\documentclass[twocolumn,amsmath,amssymb,footinbib,superscriptaddress]{revtex4-2}

\usepackage[utf8]{inputenc}
\newcommand\scalemath[2]{\scalebox{#1}{\mbox{\ensuremath{\displaystyle #2}}}}
\usepackage{xfrac}
\usepackage{amsmath}
\usepackage{mathtools}
\usepackage{float}
\usepackage{dsfont} 
\usepackage{siunitx}
\usepackage{physics}
\usepackage{soul}
\usepackage{epsfig}
\usepackage{graphicx}
\usepackage{enumitem}

\usepackage{dcolumn}
\usepackage{bm}
\usepackage{amsmath}
\usepackage{amsthm}
\usepackage{amsfonts}
\usepackage{color} 
\usepackage{booktabs}
\usepackage[hidelinks]{hyperref}

\usepackage[english]{babel}
\usepackage{blindtext}
\usepackage{balance}
\usepackage[capitalize]{cleveref}

\allowdisplaybreaks 
\allowdisplaybreaks

\usepackage{amsmath}
\usepackage{amssymb}
\usepackage{amsfonts}

\usepackage{amsthm}

\usepackage[dvipsnames]{xcolor}

\usepackage{mathtools}

\DeclareMathAlphabet\mathbfcal{OMS}{cmsy}{b}{n}

\usepackage[author=Margarida,color=pink,disable]{pdfcomment}

\makeatletter \AtEndDocument{\immediate\write\@auxout{\string\ulp@afterend}} 
\AtBeginDocument{\immediate\openout\ulp@out=\jobname.upa\relax}

\newcommand{\affvqcc}{Vigo Quantum Communication Center, University of Vigo, Vigo E-36315, Spain}
\newcommand{\affuvigo}{Escuela de Ingeniería de Telecomunicación, Department of Signal Theory and Communications, University of Vigo, Vigo E-36310, Spain}
\newcommand{\affatlantic}{atlanTTic Research Center, University of Vigo, Vigo E-36310, Spain}
\newcommand{\afftoyama}{Faculty of Engineering, University of Toyama, Gofuku 3190, Toyama 930-8555, Japan}
\newcommand{\affakihiro}{Mitsubishi Electric Corporation, Information Technology R\&D Center, 5-1-1 Ofuna, Kamakura-shi, Kanagawa, 247-8501 Japan}
\newcommand{\affkato}{National Institute of Information and Communications Technology, Nukui-kita, Koganei, Tokyo 184-8795 Japan}

\begin{document}
	\title{Modified BB84 quantum key distribution protocol robust to source imperfections}
	\author{Margarida Pereira}
	\email{mpereira@com.uvigo.es}
	\affiliation{\affvqcc} \affiliation{\affuvigo} \affiliation{\affatlantic} \affiliation{\afftoyama}
	\author{Guillermo Currás-Lorenzo}
	\affiliation{\affvqcc} \affiliation{\affuvigo} \affiliation{\affatlantic} 	\affiliation{\afftoyama}
    \author{Álvaro Navarrete}
    \affiliation{\affvqcc} \affiliation{\affuvigo} \affiliation{\affatlantic} 
    \author{Akihiro Mizutani}
    \affiliation{\affakihiro}
    \author{Go Kato}
    \affiliation{\affkato}
    \author{Marcos Curty}
    \affiliation{\affvqcc} \affiliation{\affuvigo} \affiliation{\affatlantic} 
	\author{Kiyoshi Tamaki}
    \affiliation{\afftoyama} 

	\begin{abstract}
	      The Bennett-Brassard 1984 (BB84) protocol is the most widely implemented quantum key distribution (QKD) scheme. However, despite enormous theoretical and experimental efforts in the past decades, the security of this protocol with imperfect sources has not yet been rigorously established. In this work, we address this shortcoming and prove the security of the BB84 protocol in the presence of multiple source imperfections, including state preparation flaws {and side channels, such as} Trojan-horse attacks, mode dependencies and classical correlations between the emitted pulses. To do so, we consider a modified BB84 protocol that exploits the basis mismatched events, which are often discarded in standard security analyses of this scheme; and employ the reference technique, a powerful mathematical tool to accommodate source imperfections in the security analysis of QKD. Moreover, we compare the achievable secret-key rate of the modified BB84 protocol with that of the three-state loss-tolerant protocol, and show that the addition of a fourth state, while redundant in ideal conditions, significantly improves the estimation of the leaked information in the presence of source imperfections, resulting in a better performance. This work demonstrates the relevance of the BB84 protocol in guaranteeing implementation security, taking us a step further towards closing the existing gap between theory and practice of QKD. 
	\end{abstract}

	\maketitle
	\section{Introduction}
	Quantum key distribution (QKD) enables two remote users, Alice and Bob, to securely {establish} cryptographic keys over an untrusted quantum channel \cite{lo,xu5,pirandola}. Undoubtedly, the most widely used QKD scheme is the BB84 protocol, proposed by Bennett and Brassard in 1984 \cite{bennett}. Almost four decades after its introduction, QKD has made an enormous progress both in theory and practice. However, despite its rigorous mathematical security proof, current physical implementations of QKD suffer from security loopholes due to inherent device imperfections and leakages of secret-key information.
	
	Recent years have witnessed large efforts to reduce this discrepancy between theory and practice and guarantee the implementation security of QKD. A crucial breakthrough in this direction was the proposal of measurement-device-independent QKD (MDI-QKD) \cite{lo2}, which effectively closes all security loopholes on the detector side and is practical with existing hardware \cite{rubenok,silva,liu,tang,yin}. Moreover, a variant of MDI-QKD, called twin-field QKD \cite{lucamarini2}, has been shown to provide a significant improvement on the achievable secret-key rate, allowing us to reach longer distances than ever before in fiber-based communications \cite{pittaluga,chen4,wang10}. 
	
	Having efficiently dealt with the measurement unit, the focus is now on securing the source. Essentially, source loopholes could arise from state preparation flaws (SPFs) \cite{honjo,xu,tang2, gottesman,tamaki,pereira} and {side channels, such as} Trojan-horse attacks (THAs) \cite{gisin2,vakhitov,lucamarini,tamaki3,wang,pereira}, mode dependencies \cite{nauerth,xu,tang2,pereira,dunplinskiy,bourassa}, pulse correlations {\cite{mizutani,yoshino,pereira2,grunenfelder,zapatero2,sixto}} {and through changes in, for example, electromagnetic and acoustic radiation}. Previous works have often looked at each of these imperfections individually, and developed experimental countermeasures and theoretical tools to minimise their impact on the secret-key rate and restore the security claim of QKD. For instance, the BB84 protocol has been shown to be secure in the presence of SPFs \cite{gottesman,tamaki} and THAs {\cite{lucamarini,tamaki3,wang,navarrete2}}. However, one can only guarantee the implementation security of the source if all loopholes are taken into account simultaneously in the analysis, and thus the security of this protocol with imperfect sources has not yet been rigorously established. 
	
    The importance of achieving this level of security for the BB84 protocol cannot be overstated. In particular, many existing experiments \cite{lucamarini3,yoshino2,yuan3}, field-test QKD networks {\cite{elliott,peev,chen,stucki2,sasaki,dynes2,chen5}} and satellite-based quantum communication systems \cite{liao2,liao} employ the BB84 protocol. Hence, it is crucial to ensure the practical security of their transmitting units. Moreover, while in the absence of source imperfections the achievable secret-key rate is exactly the same when using three or four states \cite{tamaki}, in the presence of imperfections the BB84 protocol may allow for a better estimation of the leaked secret information. As a consequence, this may lead to higher performances; a significant step towards attaining implementation security at an adequate level for practical QKD applications.
    
     A recently proposed analytical tool for security proofs of QKD, the reference technique (RT) \cite{pereira2}, is particularly well suited to address these issues, since it enables us to estimate the leaked secret-key information in the presence of multiple source imperfections. For this, one considers some reference states that are similar to the actual states emitted in the protocol, but whose simpler structure facilitates the estimation of some intermediate parameters. Then, since the two sets of states are close to each other, one can bound the maximum deviation between their detection probabilities using a Cauchy-Schwarz's type inequality (denoted in \cite{pereira2} as the $G$ function), and estimate the final parameters needed to guarantee the security of the actual protocol. 
     %
     %
     %
     The high flexibility and the high tolerance to source imperfections displayed by the RT with the $G$ function comes at a cost, however, as it requires the QKD protocol to be run {sequentially}, i.e.\ Alice only emits a particular pulse after Bob has measured the previous one \cite{metger}. Nonetheless, unlike other security proof approaches, the RT allows us to guarantee the security of QKD protocols with practical light sources against coherent attacks.
    
    %

    In this work, we employ the RT with the $G$ function to prove the security of the BB84 protocol in the presence of multiple source imperfections; including SPFs {and side channels, such as} THAs, mode dependencies and classical pulse correlations. To do so, we consider a modified BB84 protocol that exploits the basis mismatched events, which are usually discarded in standard implementations of the protocol. Our security proof only requires an upper bound on a few parameters that quantify the quality of the source, i.e.\ no detailed information about the side-channel states is needed, thus facilitating the work of experimentalists. Additionally, we compare the achievable secret-key rate of this modified BB84 protocol with that of the three-state loss-tolerant protocol \cite{tamaki}, and show that the emission of a fourth state, while redundant in ideal conditions, offers a significant improvement on the secret-key rate in the presence of source imperfections. This suggests that the modified BB84 protocol provides a clear performance advantage over the three-state protocol when dealing with imperfect sources.

 	\section{Description of the emitted states and assumptions}
	\label{sec:description_and_assumptions}
	In this section, we first describe in detail the state of the emitted pulses in the modified BB84 protocol (see \cref{app:protocol_description} for a full protocol description), and then list the assumptions imposed by our security proof on Alice's and Bob's devices. The security analysis is presented afterwards, in Section~\ref{sec:security}. For ease of discussion, we consider a BB84 protocol with an imperfect single-photon source. Nevertheless, our analysis could also be combined with the decoy-state method \cite{hwang,lo3,wang2} to deal with phase-randomised coherent sources. 
	
	In particular, we assume that the form of Alice's emitted states is affected by certain setting-choice-independent factors, such as temperature drifts or power fluctuations, that commonly arise in practical implementations of QKD. These factors can be modelled as a sequence of possibly-correlated random variables $\mathbfcal{G} \coloneqq \mathcal{G}_1, \hdots,\mathcal{G}_N$, where $\mathcal{G}_k$ represents all the setting-independent factors that affect the state emitted in the $k^{\rm th}$ round {and $N$ is the total number of rounds}. Since these factors are independent of Alice's encodings, for a given sequence of setting choices $j_1,\hdots,j_N$, the global state emitted is mixed over the probability distribution of $\mathbfcal{G}$. However, as shown in \cref{app:proof_fixed_g}, as long as one demonstrates that the protocol is secure for any particular outcome $\vb*{g} \coloneqq g_1,\hdots,g_N$ of $\mathbfcal{G}$, which can be assumed to have been fixed {at} the beginning of the protocol, then it is also secure for the actual case in which Alice emits mixed states. Thus, in the security proof, $\vb*{g}$ can essentially be treated as a fixed parameter that affects the form of all the emitted states. Since the latter is mathematically equivalent to considering the mixed state case, here, we take this view for simplicity of presentation. 
	
	In addition, we investigate two different scenarios for the source. In the first {one}, the state of the emitted pulse on a particular round $k$ only depends on $g_k$ and on Alice's $k^{\rm th}$ setting choice, i.e.\ it is independent of all her other setting choices. This is known as setting-independent pulse correlations and it was first modelled in \cite{mizutani}. The source model considered in this work, however, goes beyond that presented in \cite{mizutani} {even for this scenario}, as the single-mode assumption is {removed} and the effect of side channels is incorporated. In the second scenario, the state of the emitted pulse on round $k$ may not only depend on $g_k$ and on Alice's $k^{\rm th}$ setting choice, but also on Alice's previous $l_c$ setting choices, for some known correlation length $l_c$. This is often denoted as setting-dependent pulse correlations, and could arise, for instance, from memory effects in the electronic devices inside the transmitting unit \cite{grunenfelder}. The source model considered for this {latter} case is similar to that introduced in \cite{pereira2}, but here we also take into account the dependence of the emitted pulses on the setting-independent factors described above, {whose effect was disregarded in \cite{pereira2}}. In both scenarios, we assume that, given a particular sequence of setting choices $j_1, \hdots,j_N$, the global state emitted by Alice is a classical mixture of a tensor product of $N$ pure states. {That is,} we exclude the possibility of quantum correlations in which the states emitted on different rounds are entangled, {which can hardly happen in typical implementations of QKD \footnote{{Even though a QKD source is not likely to spontaneously produce entangled pulses, in principle, a powerful Eve could inject entangled light into Alice's device. Given the fragile nature of entangled states, one could expect this entanglement to almost completely vanish from the output light, especially if the source employs optical isolators. Nonetheless, one cannot be certain that all entanglement has been destroyed. In our analysis, we do not consider such attacks, although we believe that the security of QKD in the presence of these quantum correlations is an interesting problem to consider in future works.}}.}
	
	In the setting-independent scenario, for each round $k$ of the protocol, Alice chooses a setting $j \in \{0_Z, 1_Z, 0_X, 1_X\}$ and emits a state to Bob. This state can be expressed as
	\begin{align}
		&\ket{\psi_{j,{\vb*{g}}}}_{B_k,E_{k}} = \sqrt{1-\epsilon_{j,{\vb*{g}}}^{{(k)}}} \ket{{\phi_{j,{\vb*{g}}}}}_{B_k,E_{k}} + \sqrt{\epsilon_{j,{\vb*{g}}}^{{(k)}}} \ket*{\phi_{j,{\vb*{g}}}^{\perp}}_{B_k,E_{k}},
		\label{eq:general}
	\end{align}
    where $B_k$ is a two-dimensional system and $E_{k}$ includes any other systems that carry information about the $k^{\rm th}$ pulse, such as the back-reflected light from a THA (see \cref{app:THAs} for more details). 
	Note that $\ket{\psi_{j,{\vb*{g}}}}_{B_k,E_k}$ in \cref{eq:general} is a uniquely determined pure state once $j$ and $g_k$ are fixed. However, here we write ${\vb*{g}}$, rather than $g_k$, because as explained above, in our security proof the parameter ${\vb*{g}}$, which contains $g_k$ for any $k$, is fixed {at} the beginning of the protocol.
	
	From construction, Eq.~(\ref{eq:general}) is the most general description of the transmitted states within the framework of setting-independent correlations, since it is simply an expansion of the most general state $\ket*{\psi_{j,{\vb*{g}}}}_{B_k,E_{k}}$ in the basis $\{ \ket*{\phi_{j,{\vb*{g}}}}_{B_k,E_{k}}, \ket*{\phi_{j,{\vb*{g}}}^{\perp}}_{B_k,E_{k}}\}$ \cite{pereira2,navarrete}. In Eq.~(\ref{eq:general}) the parameter $\epsilon_{j,{\vb*{g}}}^{{(k)}} \in [0,1]$ quantifies the deviation of $\ket*{\psi_{j,{\vb*{g}}}}_{B_k,E_{k}}$ from the qubit state $\ket*{{\phi_{j,{\vb*{g}}}}}_{B_k,E_{k}}\coloneqq \ket*{\omega_{j,{\vb*{g}}}}_{B_k} \ket*{\lambda_{{\vb*{g}}}}_{E_{k}}$, where $\ket*{\omega_{j,{\vb*{g}}}}_{B_k}$ is the state that Alice would send to Bob in the absence of {side channels} and $\ket*{\lambda_{{\vb*{g}}}}_{E_{k}}$ is {a} setting-independent {state} for the current round. Note that the state $\ket*{\omega_{j,{\vb*{g}}}}_{B_k}$ incorporates any imperfections in a qubit space, such as SPFs and phase fluctuations. The side channels are represented in Eq.~(\ref{eq:general}) by the state $\ket*{\phi_{j,{\vb*{g}}}^{\perp}}_{B_k,E_{k}}$, which can live in a Hilbert space of arbitrary dimension and is orthogonal to $\ket*{\phi_{j,{\vb*{g}}}}_{B_k,E_{k}}$. In other words, the state $\ket*{\phi_{j,{\vb*{g}}}^{\perp}}_{B_k,E_{k}}$ corresponds to unwanted and possibly unknown modes, and it can incorporate {side channels other than setting-dependent pulse correlations, such as} THAs and mode dependencies.
	
	In the setting-dependent scenario, the emitted state for each round $k$ can instead be expressed as,
	\begin{align}
		&\scalemath{0.97}{\ket{\psi_{j,{\vb*{g}}|j_{k-1}, \hdots, j_{k-l_c}}}_{B_k,E_{k}} = \sqrt{1-\epsilon^{{(k)}}_{j,{\vb*{g}}{|j_{k-1}, \hdots, j_{k-l_c}}}} \ket{{\phi_{j,{\vb*{g}}}}}_{B_k,E_{k}}} \nonumber \\
		&\scalemath{0.97}{+ \sqrt{\epsilon^{{(k)}}_{j,{\vb*{g}}{|j_{k-1}, \hdots, j_{k-l_c}}}} \ket*{\phi_{j,{\vb*{g}}|j_{k-1}, \hdots, j_{k-l_c}}^{\perp}}_{B_k,E_{k}},}
		\label{eq:general_set_dep_corr}
	\end{align}where $j_{k-1}, \hdots, j_{k-l_c}$ represents the dependence of the $k^{\rm th}$ pulse on Alice's previous $l_c$ setting choices. As before, \cref{eq:general_set_dep_corr} is simply an expansion of the state $\ket*{\psi_{j,{\vb*{g}}|j_{k-1}, \hdots, j_{k-l_c}}}_{B_k,E_k}$ in the basis $\{\ket{{\phi_{j,{\vb*{g}}}}}_{B_k,E_k},$ $\ket*{\phi_{j,{\vb*{g}}|j_{k-1}, \hdots, j_{k-l_c}}^{\perp}}_{B_k,E_k}\}$, and within the framework of classical pulse correlations, this is the most general description of the transmitted states. Note that the state $\ket*{\phi_{j,{\vb*{g}}|j_{k-1}, \hdots, j_{k-l_c}}^{\perp}}_{B_k,E_k}$ in \cref{eq:general_set_dep_corr}, besides incorporating all the side channels in $\ket*{\phi_{j,{\vb*{g}}}^\perp}_{B_k,E_k}$, also takes into account setting-dependent pulse correlations. 
	
	Importantly, due to the form of Eqs.~(\ref{eq:general}) and (\ref{eq:general_set_dep_corr}), one can apply the RT to prove the security of the modified BB84 protocol as long as the following assumptions hold.

	\subsection{Assumptions on Alice's {transmitting unit}} 
		\label{sec:assumptions_alice}
 
		\begin{enumerate}
		\item[(A1)]	For all rounds of the protocol, Alice chooses the setting $j$ with a fixed probability $p_{j}$, {with $p_{0_Z} = p_{1_Z}$.}
		
		Alice's setting selection in a given round is independent of those of other rounds, and Eve cannot tamper with her selection probabilities.

	    \item[(A2)] {As described above, w}e consider two different {scenarios for} {the source,} which result in two security analyses {with different assumptions:}
	    
	       \begin{enumerate}[leftmargin=*,label=\alph*.]
	    \item {The emitted states do not depend on Alice's previous setting choices -- \cref{eq:general}.}
	    
	    {We assume that} an upper bound $\epsilon^{\rm U} \geq \epsilon_{j,{\vb*{g}}}^{{(k)}}$ is known for all ${k}$, $j$ {and} ${\vb*{g}}$. Note that, even in this case, the states emitted in different rounds {of the protocol} are not {necessarily} independent and identically distributed (IID) because the {random variables $\mathcal{G}_1,\hdots,\mathcal{G}_N$ that represent the setting-independent} factors {may be} correlated between consecutive rounds. We show the security analysis under this {scenario} in \cref{sec:proof_gk_independent}.

	    \item {The emitted states depend on Alice's previous $l_c$ setting choices -- \cref{eq:general_set_dep_corr}.}

	    {We assume that an upper bound $\epsilon'^{\rm U} \geq \epsilon^{{(k)}}_{j,{\vb*{g}}|j_{k-1},\hdots,j_{k-l_c}}$ is known for all ${k}$, $j$, ${\vb*{g}}$, {and} $j_{k-1},\hdots,j_{k-l_c}$. Moreover,} we assume that the state of the $k^{\rm th}$ pulse is affected by {${\vb*{g}}$ and} Alice's previous $l_c$ setting choices, {and that} $l_c$ is {a} known {parameter.} The analysis under {this scenario} is given in \cref{sec:proof_gk_dependent}. As we shall see, the {data} post-processing in this case must be done differently. In particular, one needs to divide the {sifted} key in $(l_c +1)$ groups, and then perform the parameter estimation and privacy amplification separately for each group {(see \cref{app:protocol_description}).}

	    \end{enumerate}

	     We emphasise that, while knowing the upper bound {$\epsilon^{\rm U}$} {($\epsilon'^{\rm U}$)} is a requirement to apply the RT, the characterisation of the side-channel states $\ket*{\phi_{j,{\vb*{g}}}^{\perp}}_{B_k,E_{k}}$ {($\ket*{\phi_{j,{\vb*{g}}|j_{k-1}, \hdots, j_{k-l_c}}^{\perp}}_{B_k,E_k}$)} is not needed. 
	     In other words, the inner products $\braket*{\phi_{j,{\vb*{g}}}^{\perp}}{\phi_{j',{\vb*{g}}}^{\perp}}_{B_k,E_{k}}$ {($\braket*{\phi_{j,{\vb*{g}}|j_{k-1}, \hdots, j_{k-l_c}}^{\perp}}{\phi_{{j'},{\vb*{g}}|j_{k-1}, \hdots, j_{k-l_c}}^{\perp}}_{B_k,E_k}$)} and $\braket*{\phi_{j,{\vb*{g}}}}{\phi_{{j'},{{\vb*{g}}}}^{\perp}}_{B_k,E_{k}}$  {($\braket*{\phi_{j,{\vb*{g}}}}{\phi_{{j'},{\vb*{g}}|j_{k-1}, \hdots, j_{k-l_c}}^{\perp}}_{B_k,E_k}$)} {with} $j \neq {j'}$ can be unknown. Importantly, this is not a necessary assumption but a fortunate consequence originating from the freedom to choose the reference states in the RT when using the particular inequality $G$ defined in Eq.~(\ref{eq:bound}). Since obtaining a full characterisation of the side-channel states is very challenging in practice, previous theoretical works \cite{pereira2,navarrete}, as well as this work, have exploited this advantage to consider device models that require minimal experimental characterisation. Nonetheless, {it is important to emphasise that} if any information about the side channels is available it can be incorporated in the RT framework. This would most likely lead to higher performances because a better source characterisation tends to result in a more accurate estimation of the phase-error rate. In fact, this has been recently shown for a particular time-dependent side channel {in} \cite{bourassa}.

	     \item[{(A3)}] {A partial characterisation of the qubit state $\ket{\omega_{j,{\vb*{g}}}}_{B_k}$ in \cref{eq:general,eq:general_set_dep_corr} can be obtained.}

        In the analysis presented in \cite{pereira2}, for simplicity, the qubit state $\ket*{\omega_{j,{\vb*{g}}}}_{B_k}$ in $\ket*{{\phi_{j,{\vb*{g}}}}}_{B_k,E_{k}}:= \ket*{\omega_{j,{\vb*{g}}}}_{B_k}$ $ \ket*{\lambda_{{\vb*{g}}}}_{E_{k}}$ is assumed to be perfectly characterised and stable in time. Here, we go a step further and allow $\ket*{\omega_{j,{\vb*{g}}}}_{B_k}$ to vary slightly round by round, hence its dependence on ${\vb*{g}}$. However, we assume that one can at least partially characterise {this state,} such that the upper bounds ${c_{\alpha,j}^{\rm U}}$ and $p_{\alpha_X}^{\rm (vir) U}$ {on certain quantities that are defined later} can be derived; see the discussion between \cref{eq:eph1,eq:eph2} for more details, {including the definition of these parameters.}
     
        In \cref{sec:device_model} we show {for illustration purposes} that, for a typical phase-encoding setup in which the {qubit component $\ket*{\omega_{j,{\vb*{g}}}}_{B_k}$ of all the emitted states is in a standard basis plane (such as the $XZ$ plane) and the exact} encoded phase {$\theta_{j,{\vb*{g}}}^{{(k)}}$} fluctuates over time, {this requirement translates to being able to} determine the range of these fluctuations, {i.e.\ guaranteeing that $\theta_{j,{\vb*{g}}}^{{(k)}} \in [\theta_{j}^{\rm L},\theta_{j}^{\rm U}]$ for {all $k$, $j$ and $\vb*{g}$, where} $\{\theta_{j}^{\rm L},\theta_{j}^{\rm U}\}$ {is known}.}
    	     
        \item[{(A4)}] {Alice only emits her ${k}^{\rm th}$ pulse after Bob has performed his ${k-1}^{\rm th}$ measurement \footnote{We believe that this assumption could be relaxed to the requirement that Alice's setting choices on rounds $\{k, \hdots, N\}$ {do not affect} Bob's measurement result on round $k-1$. This would allow us to take into account the maximum speed at which information about Alice's $k^{\rm th}$ setting choice could travel to Bob's measurement unit, given the distance separating the two, when determining the constraint on the maximum repetition rate of Alice's source. However, here we make the stronger assumption for conceptual simplicity.}.}\nocite{metger}

        This guarantees that the measurement operator $\hat{M}^{{(k)}}_{\gamma_X}$ in Eq.~(\ref{eq:ph_actual}) {($\hat{D}^{{(k)}}_{\gamma_X}$ in Eq.~(\ref{eq:ph_actual_dep}))} satisfies $0 \leq \hat{M}^{{(k)}}_{\gamma_X} \leq {\hat{\mathds{1}}}$ {($0 \leq \hat{D}^{{(k)}}_{\gamma_X} \leq {\hat{\mathds{1}}}$)} {(see {\cref{app:proof_measurement,app:proof_measurement_d}, respectively,} for a proof of {these} statement{s})}, which is needed to apply the RT with the $G$ function defined in Eq.~(\ref{eq:bound}). {We note that this assumption is also required when using the generalised entropy accumulation theorem \cite{metger} to prove the security of prepare-and-measure protocols against coherent attacks.}
        \end{enumerate}
        
	\subsection{Assumptions on Bob's {measurement unit}}
		\label{sec:assumptions_bob}
	\begin{enumerate}
		\item[(B1)] For all rounds of the protocol, Bob chooses a measurement basis $\beta \in \{Z,X\}$ with probabilities $p_{Z_B}$ and $p_{X_B}$, respectively.
		
		Bob's basis selection in a given round is independent of those of other rounds, and Eve cannot tamper with his selection probabilities.
		
	    \item[(B2)] Bob's measurements satisfy the basis-independent-efficiency condition.
	    
	    We assume that Bob's measurements can be represented by the positive operator-valued measures (POVMs) $\{\hat{m}_{0_\beta}, \hat{m}_{1_\beta}, \hat{m}_{f}\}$ where $\hat{m}_{0_\beta}$ ($\hat{m}_{1_\beta}$) corresponds to Bob obtaining the bit value 0 (1) when selecting the basis $\beta$, and $\hat{m}_{f}$ is associated with an inconclusive outcome. That is, the detection efficiency of Bob's unit is independent of his measurement basis choice $\beta$. This assumption is required by many security proofs of QKD to remove detector side-channel attacks exploiting channel loss \cite{lydersen,gerhardt}.
	    
	    \item[(B3)] There are no side channels on Bob's device. 

	\end{enumerate}

  \noindent All these assumptions on Bob's device can be avoided by considering a MDI-type protocol, which removes all detector loopholes and to which our analysis could easily be extended (see \cite{navarrete}).

	\section{Security Proof}
	\label{sec:security}
	Here, we show how the RT can be used to prove the security of the {modified} BB84 protocol against coherent attacks in the presence of multiple source imperfections. 
	In particular, we explain how to estimate the phase-error rate, which bounds the amount of information leakage to a potential eavesdropper, Eve, and determines the amount of privacy amplification that is needed to guarantee a secure final key. We do this for two different security analyses that consider {the two} scenarios {described previously} for the transmitting unit. Namely, in Section~\ref{sec:proof_gk_independent}, the emitted states only depend on setting-independent factors and in Section~\ref{sec:proof_gk_dependent}, in addition to {these} factors, the emitted states also depend on Alice's previous {$l_c$} setting choices. {Their} correspond{ing} assumptions {are (A2.a)} and {(A2.b)} in Section~\ref{sec:assumptions_alice}, respectively.

\subsection{{Scenario in which the emitted states} do not depend on Alice's previous setting choices}
\label{sec:proof_gk_independent}

    In 
    {this} {scenario,} for each pulse emission, Alice {sends} a state $\ket*{\psi_{j,{\vb*{g}}}}_{B_k,E_{k}}$ {given by \cref{eq:general}} through the quantum channel to Bob, who then performs his POVM measurements. The secret key is distilled from the rounds in which both Alice and Bob select the $Z$ basis. As a basic framework to prove the security {of these events}, we use the complementarity approach {\cite{koashi3,koashi2}.} First, note that from Eve's perspective, the key generation rounds are {equivalently} described by an entanglement-based scenario in which, after selecting the $Z$ basis, Alice prepares the following entangled state 
	\begin{align}
		\ket*{\Psi_{{\vb*{g}}}^{{Z}}}_{A_k,B_k,E_{k}} &= \frac{1}{\sqrt{2}} \sum_{\alpha \in \{0,1\}} \ket{\alpha_Z}_{A_k} \ket{\psi_{\alpha_Z,{\vb*{g}}}}_{B_k,E_{k}},
		\label{eq:Psi_gk_k}
	\end{align}sends systems $B_k,E_{k}$ to Bob while keeping system $A_k$ in her laboratory, and then both Alice and Bob perform $Z$-basis measurements on their local {and received} systems, {respectively}. 
	
	To prove the security of these events, we consider the number of phase errors that Alice and Bob would have obtained if they had performed their measurements in the $X$ basis instead. This virtual {scenario} is equivalent to Alice {sending} Bob the {fictitious} virtual states 
	\begin{align}
		\ket*{\psi_{\alpha_X,{\vb*{g}}}^{{\rm (vir)}}}_{B_k,E_{k}} =\frac{\ket{\psi_{0_Z,{\vb*{g}}}}_{B_k,E_{k}} + (-1)^\alpha \ket{\psi_{1_Z,{\vb*{g}}}}_{B_k,E_{k}} }{2~\sqrt{\frac{\tilde{p}^{{(k,{\rm vir})}}_{\alpha_X,{\vb*{g}}}}{p_{Z_A}} }},
		\label{eq:vir_act}
	\end{align}
	where $\alpha \in \{0,1\}$, with probabilities 
	\begin{align}
		\tilde{p}^{{(k,{\rm vir})}}_{\alpha_X,{\vb*{g}}} = \frac{1}{2} p_{Z_A}\Big[ 1+ (-1)^{\alpha} \Re \Big(\big\langle{\psi_{0_Z,{\vb*{g}}}}\big|\psi_{1_Z,{\vb*{g}}}\big\rangle {}_{B_k,E_{k}} \Big)\Big],
		\label{eq:vir_act_prob}
	\end{align}who then performs $X$-basis measurements on the received systems. In Eq.~(\ref{eq:vir_act_prob}), $\tilde{p}^{{(k,{\rm vir})}}_{\alpha_X,{\vb*{g}}}$ represents the joint probability that Alice chooses the $Z$ basis ($p_{Z_A}$) and prepares the virtual state $\ket*{\psi^{{\rm (vir)}}_{\alpha_X,{\vb*{g}}}}_{B_k,E_{k}}$. 
	
	A phase error occurs when Alice selects the virtual state associated to $1_X$ ($0_X$) and Bob obtains the bit value $0$ ($1$) in his $X$-basis measurement. {In \cref{app:proof_measurement}, we show that} the probability of obtaining a phase error on round {$k$}, conditioned on all the previous outcomes, {can be expressed as} 
	\begin{align}
		P^{{(k)}}_{{\vb*{g}}}({\rm ph|Act})&\coloneqq \sum_{\substack{\alpha, \gamma \in \{0,1\} \\ \alpha \neq \gamma}} \tilde{p}_{\alpha_X,{\vb*{g}}}^{{(k,{\rm vir})}} p_{Z_B}{\rm Tr}\big[\tilde{\sigma}_{\alpha_X,{\vb*{g}}}^{{(k,{\rm vir})}}{\hat M}_{\gamma_X}^{{(k)}}\big], 
		\label{eq:ph_actual}
	\end{align}where $\tilde{\sigma}_{\alpha_X,{\vb*{g}}}^{{(k,{\rm vir})}} \coloneqq \dyad*{\psi_{\alpha_X,{\vb*{g}}}^{{\rm (vir)}}}{\psi_{\alpha_X,{\vb*{g}}}^{{\rm (vir)}}}_{B_k,E_{k}}$, and $\hat{M}_{\gamma_X}^{{(k)}}$ with $\gamma \in \{0,1\}$ is Bob's effective POVM element for the $k^{\rm th}$ pulse after a coherent attack {(see \cref{eq:measurement_app} in \cref{app:proof_measurement} for more details).} 
	
	The detection probabilities ${\rm Tr}\big[\tilde{\sigma}_{\alpha_X,{\vb*{g}}}^{{(k,{\rm vir})}}{\hat M}_{\gamma_X}^{{(k)}} \big]$ in Eq.~(\ref{eq:ph_actual}) are {not directly observed} in the experiment because the virtual states are not actually emitted. Moreover, estimating them using the data collected in the protocol might be difficult due to the presence of multiple source imperfections. However, thanks to the RT, we can overcome this difficulty by estimating $P^{{(k)}}_{{\vb*{g}}}({\rm ph|Act})$ indirectly. For this, we first select some reference states that are similar to the actual states emitted in the protocol{, and which} allow an easy estimate of the phase-error probability that would be observed if they had been emitted: $P^{{(k)}}_{{\vb*{g}}}({\rm ph|Ref})$. Then, by evaluating the deviation between the reference and actual states, we obtain $P^{{(k)}}_{{\vb*{g}}}({\rm ph|Act})$ from $P^{{(k)}}_{{\vb*{g}}}({\rm ph|Ref})$. Finally, by applying concentration inequalities we derive an upper bound on the phase-error rate. \\
	
	\noindent \textit{Applying the reference technique} \\
	First, we define four reference states. {Even though} our choice of states is unrestricted, higher secret-key rates are achieved if they are close to the actual states. Here, we pick the set of reference states to be $\big\{\ket*{\phi_{j,{\vb*{g}}}}_{B_k,E_{k}}\big\}_{j \in \{0_Z,1_Z,0_X,1_X\}}$, which are defined in Eq.~(\ref{eq:general}) as the qubit part of the actual states. Then, by replacing the actual $Z$-basis states by their corresponding reference states in \cref{eq:Psi_gk_k}, we define
		\begin{align}
			\ket*{\Phi_{{\vb*{g}}}^{{Z}}}_{A_k,B_k,E_{k}} &= \frac{1}{\sqrt{2}} \sum_{\alpha \in \{0,1\}} \ket{\alpha_Z}_{A_k} \ket{\phi_{\alpha_Z,{\vb*{g}}}}_{B_k,E_{k}},
			\label{eq:Phi_gk_k}
		\end{align}analogous to $\ket*{\Psi_{{\vb*{g}}}^{{Z}}}_{A_k,B_k,E_{k}}$.
		Similarly, we define the virtual states $\ket*{\phi_{\alpha_X,{\vb*{g}}}^{{\rm (vir)}}}_{B_k,E_{k}}$ and the probabilities $p^{{(k,{\rm vir})}}_{\alpha_X,{\vb*{g}}}$, which are analogous to $\ket*{\psi_{\alpha_X,{\vb*{g}}}^{{\rm (vir)}}}_{B_k,E_{k}}$ and $\tilde{p}^{{(k,{\rm vir})}}_{\alpha_X,{\vb*{g}}}$, respectively. This allows us to define the quantity
	\begin{align}
		P^{{(k)}}_{{\vb*{g}}}({\rm ph|Ref})&\coloneqq \sum_{\substack{\alpha, \gamma \in \{0,1\} \\ \alpha \neq \gamma}} p_{\alpha_X,{\vb*{g}}}^{{(k,{\rm vir})}} p_{Z_B}{\rm Tr}\big[\sigma_{\alpha_X,{\vb*{g}}}^{{(k,{\rm vir})}}{\hat M}_{\gamma_X}^{{(k)}}\big],
		\label{eq:ph_ref}
	\end{align}where $\sigma_{\alpha_X,{\vb*{g}}}^{{(k,{\rm vir})}} := \dyad*{\phi_{\alpha_X,{\vb*{g}}}^{{\rm (vir)}}}_{B_k,E_{k}}$. Here, $P^{{(k)}}_{{\vb*{g}}}({\rm ph|Ref})$ could be interpreted as the probability of a phase-error on the $k^{\rm th}$ round when using the reference states.
	We emphasise that these replacements of actual states by their reference counterparts are purely mathematical. The reference states are never prepared nor sent in an actual implementation of the protocol.
	
	A {convenient} feature of \cref{eq:ph_ref} over \cref{eq:ph_actual} is that the states $\sigma_{\alpha_X,{\vb*{g}}}^{{(k,{\rm vir})}}$ live in the same qubit space as the reference states $\sigma_{j,{\vb*{g}}}^{{(k)}} \coloneqq \dyad*{\phi_{j,{\vb*{g}}}}_{B_k,E_{k}}$, and therefore one can employ the idea of the loss-tolerant protocol \cite{tamaki} to write the former states as a linear function of the latter. For simplicity, here we assume that these states all lie in the $XZ$ plane of the Bloch sphere; see {Appendix B in} \cite{lorenzo2} for a more general treatment. Then, we have that
	\begin{align}
		\sigma_{\alpha_X,{\vb*{g}}}^{{(k,{\rm vir})}} = \sum_{j} c_{\alpha,j,{\vb*{g}}}^{{(k)}} \sigma_{j,{\vb*{g}}}^{{(k)}},
		\label{eq:relationships}
	\end{align}
	 {for} {$j \in \{0_Z,1_Z,0_X,1_X\}$,} {where} $c_{\alpha,j,{\vb*{g}}}^{{(k)}}$ are real coefficients. To find these coefficients, one has to solve two systems of three linear equations with four unknowns. {These systems have} infinitely many solutions, {and therefore one can choose the solutions that provide the tightest bound on the phase-error rate.} This is the crucial difference with respect to the three-state protocol, and the reason why the {modified} BB84 protocol can provide higher secret-key rates (see \cref{sec:comparison_bb84_3state}). We note that, in the case of the three-state protocol, the $1_X$ state is not emitted and thus $c_{1,1_X,{\vb*{g}}}^{{(k)}} = c_{0,1_X,{\vb*{g}}}^{{(k)}} = 0$. This results in two systems of three linear equations with three unknowns, which have a unique solution each. 

	After substituting Eq.~(\ref{eq:relationships}) in Eq.~(\ref{eq:ph_ref}), we obtain
	\begin{align}
		\scalemath{0.96}{P^{{(k)}}_{{\vb*{g}}}({\rm ph|Ref}) = \sum_{\substack{\alpha,\gamma \in \{0,1\} \\ \alpha \neq \gamma}}  p_{\alpha_X,{\vb*{g}}}^{{(k,{\rm vir})}} p_{Z_B} \sum_{j} c_{\alpha,j,{\vb*{g}}}^{{(k)}} 
		{\rm Tr}\big[\sigma_{j,{\vb*{g}}}^{{(k)}}{\hat M}_{\gamma_X}^{{(k)}}\big],}
		\label{eq:eph1}
	\end{align}where we have used the linearity of the trace operation. Note that since the reference states $\sigma_{j,{\vb*{g}}}^{{(k)}}$ depend on ${\vb*{g}}$, the coefficients $c_{\alpha,j,{\vb*{g}}}^{{(k)}}$ and the probabilities $p_{\alpha_X,{\vb*{g}}}^{{(k,{\rm vir})}}$ also depend on ${\vb*{g}}$, and therefore their exact value is in general unknown. Nevertheless, our analysis only requires knowing upper bounds $c_{\alpha,j}^{\rm U} \geq c_{\alpha,j,{\vb*{g}}}^{{(k)}}$ and $p_{\alpha_X}^{\rm (vir)U} \geq p_{\alpha_X,{\vb*{g}}}^{{(k,{\rm vir})}}$ on each of these quantities {(see assumption (A3) in \cref{sec:assumptions_alice}).} In \cref{sec:device_model}, we show how to derive these upper bounds {in practice} for a particular device model. Substituting them in \cref{eq:eph1}, we obtain 
		\begin{align}
		P^{{(k)}}_{{\vb*{g}}}({\rm ph|Ref}) \leq \sum_{\substack{\alpha,\gamma \in \{0,1\} \\ \alpha \neq \gamma}}  p_{\alpha_X}^{{\rm (vir) U}} p_{Z_B} \sum_{j} c^{\rm U}_{\alpha,j} 
		{\rm Tr}\big[\sigma_{j,{\vb*{g}}}^{{(k)}}{\hat M}_{\gamma_X}^{{(k)}}\big].
		\label{eq:eph2}
	\end{align}
	
	The fictitious probabilities ${\rm Tr}\big[\sigma_{j,{\vb*{g}}}^{{(k)}}{\hat M}_{\gamma_X}^{{(k)}}\big]$ in Eq.~(\ref{eq:eph2}) are also unknown because, as mentioned before, the reference states are never sent in the actual protocol. However, by evaluating the deviation between the reference and the actual states we are able to bound these probabilities, and consequently estimate {an upper bound on} $P^{{(k)}}_{{\vb*{g}}}(\rm ph|Ref)$. In doing so, we may adopt a number of different inequalities; but here, we choose the following inequality \cite{pereira2}
	\begin{align}
		&G_-\big({\rm Tr}[\dyad*{A}{A}{\hat M}], |\braket{A}{R}|\big) \nonumber \\
		&\leq {\rm Tr}[\dyad*{R}{R}{\hat M}] \nonumber \\
		&\leq G_+\big({\rm Tr}[\dyad*{A}{A}{\hat M}], |\braket{A}{R}|\big),
		\label{eq:bound}
	\end{align}
	where $\ket{A}$ and $\ket{R}$ are any two normalised states, and $\hat{M}$ is a measurement operator satisfying $0 \leq \hat{M} \leq \hat{\mathds{1}}$. The latter imposes a restriction on {the repetition rate of the protocol} (see assumption {(A4)} in \cref{sec:assumptions_alice}). In Eq.~(\ref{eq:bound}), the functions $G_-(y,z)$ and $G_+(y,z)$ are defined for $0\leq y \leq 1$ and $0\leq z \leq 1$ as 
	\begin{equation}
		G_-(y,z) =
		\begin{cases}
			g_-(y,z)  & \quad \text{if } y > 1 -z^2 \\
			0  & \quad \text{otherwise},
		\end{cases}
	\end{equation}
	and
	\begin{equation}
		G_+(y,z) =
		\begin{cases}
			g_+(y,z)  & \quad \text{if } y < z^2 \\
			1  & \quad \text{otherwise},
		\end{cases}
	\end{equation}
	with
	\begin{equation}
		g_{\pm} (y,z) = y + (1-z^2)(1-2y) \pm 2z\sqrt{(1-z^2)y(1-y)}.
	\end{equation}
	
	Next, we apply the bound in Eq.~(\ref{eq:bound}) to each term in Eq.~(\ref{eq:eph2}), by selecting the function $G_-(y,z)$ for the terms whose coefficient is negative, and the function $G_+(y,z)$ for the terms whose coefficient is positive, thus maximising $P^{{(k)}}_{{\vb*{g}}}(\rm ph|Ref)$. We can then express $P^{{(k)}}_{{\vb*{g}}}(\rm ph|Ref)$ as a function of {the detection probabilities of the actual states:}
	\begin{align}
		&P^{{(k)}}_{{\vb*{g}}}({\rm ph|Ref}) \leq \sum_{\substack{\alpha,\gamma \in \{0,1\} \\ \alpha \neq \gamma}} p_{\alpha_X}^{{\rm (vir)U}} p_{Z_B} \nonumber \\
		&\left[\rule{0cm}{1cm}\right. \sum_{\substack{j  \\ c_{\alpha,j}^{\rm U} > 0}} c^{\rm U}_{\alpha,j} 
		G_+\big( {\rm Tr}\big[\tilde{\sigma}_{j,{\vb*{g}}}^{{(k)}}{\hat M}_{\gamma_X}^{{(k)}}\big],\big|\braket*{\psi_{j,{\vb*{g}}}}{\phi_{j,{\vb*{g}}}}_{{B_k,E_k}}\big| \big) + \nonumber \\
		&\sum_{\substack{j  \\ c_{\alpha,j}^{\rm U} < 0}}  c^{\rm U}_{\alpha,j} G_-\big( {\rm Tr}\big[\tilde{\sigma}_{j,{\vb*{g}}}^{{(k)}}{\hat M}_{\gamma_X}^{{(k)}}\big],\big|\braket*{\psi_{j,{\vb*{g}}}^{}}{\phi_{j,{\vb*{g}}}}_{{B_k,E_k}}\big| \big)\left.\rule{0cm}{1cm}\right],
		\label{eq:ph_ref_new}
	\end{align}
	\noindent where $\tilde{\sigma}_{j,{\vb*{g}}}^{{(k)}} \coloneqq \dyad*{\psi_{j,{\vb*{g}}}}{\psi_{j,{\vb*{g}}}}_{B_k,E_{k}}$. 
	
	Using the definitions of the actual and reference states, we find that the inner products in Eq.~(\ref{eq:ph_ref_new}) have the form $|\braket*{\psi_{j,{\vb*{g}}}}{\phi_{j,{\vb*{g}}}}_{{B_k,E_k}}| = \sqrt{\smash[b]{1-\epsilon^{{(k)}}_{j,{\vb*{g}}}}}$. For simplicity, we now use the upper bound $\epsilon^{\rm U} \geq \epsilon_{j,{\vb*{g}}}^{{(k)}}$ for all ${k}$, $j$ {and} ${\vb*{g}}$ (see assumption ({A2.a}) in \cref{sec:assumptions_alice}). Using the fact that the functions $G_{+}(y,z)$ and $-G_{-}(y,z)$ are decreasing with respect to $z$ and that $\sqrt{\smash[b]{1-\epsilon^{{(k)}}_{j,{\vb*{g}}}}} \geq \sqrt{\smash[b]{1-\epsilon^{\rm U}}}$, Eq.~(\ref{eq:ph_ref_new}) can then be upper bounded by 
	\begin{align}
		&P^{{(k)}}_{{\vb*{g}}}({\rm ph|Ref}) \leq 
		\sum_{\substack{\alpha,\gamma \in \{0,1\} \\ \alpha \neq \gamma}} p_{\alpha_X}^{\rm (vir)U} p_{Z_B} \nonumber \\
		&\left[\rule{0cm}{1cm}\right.  \sum_{\substack{j  \\ c_{\alpha,j}^{\rm U} > 0}}  c_{\alpha,j}^{\rm U}  
		G_+\bigg( \frac{P^{{(k)}}_{{\vb*{g}}}(j,\gamma_X|{\rm Act})}{p_{j} p_{X_B}},\sqrt{1-\epsilon^{\rm U}} \bigg) + \nonumber \\
		&\sum_{\substack{j  \\ c_{\alpha,j}^{\rm U} < 0}}
		c_{\alpha,j}^{\rm U} G_-\bigg( \frac{P^{{(k)}}_{{\vb*{g}}}(j,\gamma_X|{\rm Act})}{p_{j} p_{X_B}},\sqrt{1-\epsilon^{\rm U}} \bigg) \left.\rule{0cm}{1cm}\right],
			\label{eq:ph_ref_new_upper}
	\end{align}
	where $P^{{(k)}}_{{\vb*{g}}}(j,\gamma_X|{\rm Act}) \coloneqq p_{j}p_{X_B} \Tr \big[\tilde{\sigma}_{j,{\vb*{g}}}^{{(k)}} \hat{M}_{\gamma_X}^{{(k)}}\big]$ is the joint probability that Alice prepares the actual state $\ket*{\psi_{j,{\vb*{g}}}}_{B_k,E_{k}}$, Bob chooses the $X$ basis and his measurement outcome is $\gamma$. Importantly, these probabilities are related to quantities directly observed in the protocol, and all the other parameters in Eq.~(\ref{eq:ph_ref_new_upper}) are known. 
	
	Now that we have an upper bound on $P^{{(k)}}_{{\vb*{g}}}{(\rm ph|Ref)}$, the next step is to obtain an upper bound on the probability of a phase error in the actual protocol: $P^{{(k)}}_{{\vb*{g}}}({\rm ph|Act})$. For this, first note that $P^{{(k)}}_{{\vb*{g}}}{(\rm ph|Ref)}$ in Eq.~(\ref{eq:ph_ref}) can be written as
	\begin{align}
	    P^{{(k)}}_{{\vb*{g}}}{(\rm ph|Ref)} = p_{Z_A} p_{Z_B} \Tr \big[\dyad*{\Phi_{{\vb*{g}}}^{{Z}}}_{{B_k,E_k}}\hat{M}_{\rm ph}^{{(k)}}\big],
	\end{align}where 
	\begin{align}
	\hat{M}_{\rm ph}^{{(k)}} &= \hat{P}\left(\frac{\ket{0_Z}_{A_k} -  \ket{1_Z}_{A_k}}{\sqrt{2}}\right) \otimes \hat{M}_{0_X}^{{(k)}} \nonumber \\
	&+ \hat{P}\left(\frac{\ket{0_Z}_{A_k} + \ket{1_Z}_{A_k}}{\sqrt{2}}\right) \otimes \hat{M}_{1_X}^{{(k)}},
	\end{align}with $\hat{P}(\ket{\cdot}) = \dyad{\cdot}$. Similarly, $P^{{(k)}}_{{\vb*{g}}}{(\rm ph|Act)}$ in Eq.~(\ref{eq:ph_actual}) can be written as
	\begin{align}
	    P^{{(k)}}_{{\vb*{g}}}{(\rm ph|Act)} = p_{Z_A} p_{Z_B} \Tr \big[\dyad*{\Psi_{{\vb*{g}}}^{{Z}}}_{{B_k,E_k}} \hat{M}_{\rm ph}^{{(k)}} \big].
	\end{align}Hence, one can simply employ the bound in Eq.~(\ref{eq:bound}) again to obtain the following expression 
	\begin{align}
		&P^{{(k)}}_{{\vb*{g}}}({\rm ph|Act}) \nonumber \\
		&\leq p_{Z_A} p_{Z_B} G_+ \bigg(\frac{P^{{(k)}}_{{\vb*{g}}}({\rm ph|Ref})}{p_{Z_A} p_{Z_B}},\big|\braket*{\Psi^{{Z}}_{{\vb*{g}}}}{\Phi^{{Z}}_{{\vb*{g}}}}_{{B_k,E_k}} \big|\bigg) \nonumber \\
		&\leq p_{Z_A} p_{Z_B} G_+ \bigg(\frac{P^{{(k)}}_{{\vb*{g}}}({\rm ph|Ref})}{p_{Z_A} p_{Z_B}},\sqrt{1-\epsilon^{\rm U}} \bigg).
		\label{eq:phase_error_act}
	\end{align}
	In the last inequality of Eq.~(\ref{eq:phase_error_act}) we have used the fact that $G_{+}(y,z)$ is a decreasing function with respect to $z$, and the fact that $|\braket*{\Psi_{{\vb*{g}}}^{{Z}}}{\Phi^{{Z}}_{{\vb*{g}}}}_{{B_k,E_k}}| = \big(\sqrt{\smash[b]{1-\epsilon^{{(k)}}_{0_Z,{\vb*{g}}}}} + \sqrt{\smash[b]{1-\epsilon^{{(k)}}_{1_Z,{\vb*{g}}}}}\big)/2 \geq \sqrt{1-\epsilon^{\rm U}}$. 
	
	The only missing step in the security analysis is to convert Eq.~(\ref{eq:phase_error_act}) into an expression in terms of observables. We start by taking the average over all transmitted rounds $N$ on both sides of Eq.~(\ref{eq:phase_error_act}) and then applying Jensen's inequality \cite{jensen} to the right-hand side:
	\begin{align}
		&\frac{1}{N}\sum_k^N P^{{(k)}}_{{\vb*{g}}}({\rm ph|Act}) \nonumber \\
		&\leq \frac{1}{N}\sum_k^N  p_{Z_A} p_{Z_B} G_+ \bigg(\frac{P^{{(k)}}_{{\vb*{g}}}({\rm ph|Ref})}{p_{Z_A} p_{Z_B}},\sqrt{1-\epsilon^{\rm U}} \bigg) \nonumber \\
		&\leq p_{Z_A} p_{Z_B} G_+ \bigg(\frac{1}{N}\sum_k^N\frac{P^{{(k)}}_{{\vb*{g}}}({\rm ph|Ref})}{p_{Z_A} p_{Z_B}},\sqrt{1-\epsilon^{\rm U}} \bigg), 
		\label{eq:phase_error_n}
	\end{align}
	where {in the second inequality} we have used the concavity of the function $G_+(y,z)$ with respect to its argument $y$. Then, by applying Jensen's inequality to this argument we have that 
	\begin{align}
		&\frac{1}{N}\sum_k^N P^{{(k)}}_{{\vb*{g}}}({\rm ph|Ref}) 
		 \leq \sum_{\substack{\alpha,\gamma \in \{0,1\} \\ \alpha \neq \gamma}} p_{\alpha_X}^{\rm (vir)U} p_{Z_B}  \nonumber \\
		 &\left[\rule{0cm}{1cm}\right. \sum_{\substack{j  \\ c_{\alpha,j}^{\rm U} > 0}}  c_{\alpha,j}^{\rm U} G_+\bigg( \frac{1}{N}\sum_k^N \frac{P^{{(k)}}_{{\vb*{g}}}(j,\gamma_X|{\rm Act})}{p_{j} p_{X_B}},\sqrt{1-\epsilon^{\rm U}} \bigg)+ \nonumber \\
		 &\sum_{\substack{j  \\ c_{\alpha,j}^{\rm U} < 0}}  c_{\alpha,j}^{\rm U} G_-\bigg( \frac{1}{N}\sum_k^N \frac{P^{{(k)}}_{{\vb*{g}}}(j,\gamma_X|{\rm Act})}{p_{j} p_{X_B}},\sqrt{1-\epsilon^{\rm U}} \bigg)\left.\rule{0cm}{1cm}\right] ,
		\label{eq:ph_ref_bound}
		\end{align}
	where we have used Eq.~(\ref{eq:ph_ref_new_upper}) and the concavity of the functions $G_+(y,z)$ and $-G_-(y,z)$. 
	
	Now, we apply Azuma's inequality \cite{azuma} or Kato's inequality \cite{kato} to {substitute each sum of probabilities into its corresponding observable. In particular, for} {the asymptotic case where} $N \rightarrow \infty$, {we have that} $\sum_k^{N} P^{{(k)}}_{{\vb*{g}}}({\rm ph|Act}) \simeq {N_{\rm ph}}$, where ${N_{\rm ph}}$ is the number of phase errors. Similarly, for $N \rightarrow \infty$, {we find that} $\sum_{k}^{N} P_{{\vb*{g}}}^{{(k)}}(j,\gamma_X|{\rm Act}) \simeq {N_{j,\gamma_X}}$, where ${N_{j,\gamma_X}}$ is the number of {observed} events in which Alice's setting choice is $j$, Bob selects the $X$ basis and his measurement outcome is $\gamma$. Therefore, by using the fact that $G_+(y,z)$ is an increasing function with respect to $y$, we combine Eqs.~(\ref{eq:phase_error_n}) and (\ref{eq:ph_ref_bound}), then we multiply by $N$ on both sides, and apply Azuma's or Kato's inequality assuming $N \rightarrow \infty$, thus arriving at the following bound
	\begin{align}
		&{N_{\rm ph}} \leq N p_{Z_A} p_{Z_B} G_+ \left(\rule{0cm}{1cm}\right.\sum_{\substack{\alpha,\gamma \in \{0,1\} \\ \alpha \neq \gamma}} \frac{ p_{\alpha_X}^{\rm (vir)U}}{p_{Z_A}} \left[\rule{0cm}{1cm}\right. \sum_{\substack{j  \\ c_{\alpha,j}^{\rm U} > 0}} c_{\alpha,j}^{\rm U} \nonumber \\
		& G_+\bigg( \frac{{N_{j,\gamma_X}}}{{N} p_{j} p_{X_B}},\sqrt{1-\epsilon^{\rm U}} \bigg) + \sum_{\substack{j  \\ c_{\alpha,j}^{\rm U} < 0}} c_{\alpha,j}^{\rm U} \nonumber \\
		&G_-\bigg( \frac{{{N_{j,\gamma_X}}}}{N {p_{j}} p_{X_B}},\sqrt{1-\epsilon^{\rm U}}\bigg)\left.\rule{0cm}{1cm}\right],\sqrt{1-\epsilon^{\rm U}}  \left.\rule{0cm}{1cm}\right) {=:} N_{\rm ph}^{\rm U}.
		\label{eq:final}
	\end{align}
	
     {Finally}, we can calculate an upper bound on the phase-error rate of the actual protocol by using 
     {\begin{align}
         e_{\rm ph}^{\rm U} := \frac{N_{\rm ph}^{\rm U}}{N_{{\rm det}}^{(Z)}},
         \label{eq:ephU_final}
     \end{align}}where $N_{{\rm det}}^{(Z)}$
     is the number of detected rounds in which both Alice and Bob selected the $Z$ basis, i.e.\ the length of the sifted key. Importantly, both Azuma's and Kato's inequalities allow us to take into account {the fact that, under a coherent attack by Eve, the probabilities $P^{{(k)}}_{{\vb*{g}}}({\rm ph|Act})$ and $P^{{(k)}}_{{\vb*{g}}}(j,\gamma_X|{\rm Act})$ may depend on the outcomes of the previous $k-1$ rounds.}
     
	
	Note that it is straightforward to modify \cref{eq:final} to apply it to the finite-key regime by simply including the deviation terms of the concentration inequality employed, taking into account that the functions $G_{+}(y,z)$ and $G_{-}(y,z)$ are increasing with respect to $y$. Moreover, Kato's inequality has been shown to provide tight estimations for practical values of $N$ \cite{lorenzo,lorenzo2,zhou,navarrete2}, and therefore the performance of the protocol should not be significantly affected in the finite-key regime.

\subsection{{Scenario in which the emitted states} depend on Alice's previous $l_c$ setting choices}
\label{sec:proof_gk_dependent}
	
	{In this scenario, the state emitted on each round $k$ is} {$\ket*{\psi_{j,{\vb*{g}}|j_{k-1},\hdots,j_{k-l_c}}}_{B_k,{E_k}}$,} {defined in \cref{eq:general_set_dep_corr}, which} depends on setting-independent factors and on Alice's previous $l_c$ setting choices. {Note that, due to these setting-dependent pulse correlations, information about Alice's $k^{\rm th}$ setting choice is leaked to the subsequent pulses. However, this leakage of information can essentially be regarded as a side channel to the $k^{\rm th}$ pulse \cite{pereira2}. In \cref{app:derivations_corr}, we show that one can consider the states emitted in round $k$ to be $\ket*{{\tilde{\psi}}_{j,{\vb*{g}}|j_{k-1}, \hdots, j_{k-l_c}}}_{B_k,{\textit{\textbf{E}}}'_k}$ {instead,} where ${\textit{\textbf{E}}}'_k$ includes not only system $E_k$ but also the systems of all rounds after the $k^{\rm th}$ round.}
	
	{As before,} {in the key generation rounds,} {we can {then} consider that} {Alice prepares the entangled state }
		{\begin{align}
		&\ket*{\Psi^Z_{{\vb*{g}}{|j_{k-1},\hdots,j_{k-l_c}}}}_{A_k,B_k,{\textit{\textbf{E}}}'_k} \nonumber \\
		&= \frac{1}{\sqrt{2}} \sum_{\alpha \in \{0,1\}} \ket{\alpha_Z}_{A_k} \ket*{{\tilde{\psi}}_{\alpha_Z,{\vb*{g}}|j_{k-1},\hdots,j_{k-l_c}}}_{B_k,{\textit{\textbf{E}}}'_k},
		\label{eq:entangled_dep}
	\end{align} and measures her ancilla system $A_k$ in the $Z$ basis.} {To prove the security in this scenario, however, one cannot directly apply the analysis presented in \cref{sec:proof_gk_independent}. The reason is the following.} {To derive {the state $\ket*{{\tilde{\psi}}_{j,{\vb*{g}}|j_{k-1}, \hdots, j_{k-l_c}}}_{B_k,{\textit{\textbf{E}}}'_k}$} and \cref{eq:entangled_dep}, we need to assume that, if any round in $\{k-1,\hdots,k-l_c\}$ was a key-generation round, Alice measured her corresponding ancilla system in the $Z$ basis. However, in \cref{sec:proof_gk_independent}, we considered a virtual protocol in which, in all key-generation rounds, Alice measures her ancilla system in the $X$ basis. Thus, the description in \cref{eq:entangled_dep} is not valid for this virtual protocol.}

	To avoid this contradiction, we follow a similar approach to that in \cite{mizutani3}, and assume that the users assign to each round $k \in \{1, ..., N\}$ a tag $w \in \{0, ..., l_c\}$ according to the value $w = k \bmod (l_c + 1)$, and define the $w^{\rm th}$ sifted key as the subset of the total sifted key that originates from rounds with tag $w$ {(see the protocol description in  \cref{app:protocol_description}).} Then, to estimate the information leakage of the $w^\textrm{th}$ sifted key, we consider a virtual scenario in which Alice and Bob use the $X$ basis to check for phase errors only for the key rounds with tag $w$, while for the rest of the key rounds they use the $Z$ basis as in the actual protocol. We refer to this scenario as the $w^\textrm{th}$ virtual protocol, and we define the phase-error rate of the $w^\textrm{th}$ sifted key, $e_{\rm ph}^w$, as the fraction of phase errors that Alice and Bob would observe if they had run this $w^\textrm{th}$ virtual protocol. By obtaining a bound on $e_{\rm ph}^w$, Alice and Bob can determine the amount of privacy amplification that they need to apply to the $w^\textrm{th}$ sifted key to turn it into a secret key. {Importantly,} note that the $(l_c + 1)$ virtual protocols are not compatible with each other, {as} Alice and Bob could not have run them at the same time {due to the non-commutativity of the $X$ and $Z$ basis measurements.} However, {it turns out that} the $w^\textrm{th}$ virtual protocol allows us to prove the security of the $w^\textrm{th}$ key, and the security of the total key is ensured by the universal composability of each individual security proof, as shown in \cite{mizutani3}.
	
    {In the $w^{\rm th}$ virtual protocol, if any round in $\{k-1,$ $\hdots,k-l_c\}$ is a key round, Alice uses the $Z$ basis to measure her ancilla system. Thus, the description in \cref{eq:entangled_dep} is {indeed} valid for this {alternative} scenario. As shown in \cref{app:proof_measurement_d}, it follows that} the probability that, {in the $w^\textrm{th}$ virtual protocol, Alice and Bob obtain} a phase error {on some} round {$k$ with tag $w$,} conditioned on all the previous outcomes of the {$w^\textrm{th}$} virtual protocol, {can be expressed as}
		{\begin{align}
		&P^{{(k)}}_{{\vb*{g}}}({\rm ph|Act}) \coloneqq \sum_{\substack{\alpha, \gamma \in \{0,1\} \\ \alpha \neq \gamma}} \tilde{p}_{\alpha_X,{\vb*{g}}{|j_{k-1},\hdots,j_{k-l_c}}}^{{(k,{\rm vir})}} p_{Z_B} \nonumber \\
		&{\rm Tr}\big[\tilde{\sigma}_{\alpha_X,{\vb*{g}}|j_{k-1},\hdots,j_{k-l_c}}^{{(k,{\rm vir})}}{{\hat D}_{\gamma_X}^{{(k)}}}\big], 
		\label{eq:ph_actual_dep}
	\end{align}}
	{where}
{\begin{align}
	&\tilde{p}^{{(k,{\rm vir})}}_{\alpha_X,{\vb*{g}}|j_{k-1},\hdots,j_{k-l_c}} = \frac{1}{2} p_{Z_A}\Big[ 1+ (-1)^{\alpha} \nonumber \\ 
	&\Re \Big(\big\langle{{\tilde{\psi}}_{0_Z,{\vb*{g}}|j_{k-1},\hdots,j_{k-l_c}}}\big|{\tilde{\psi}}_{1_Z,{\vb*{g}}|j_{k-1},\hdots,j_{k-l_c}}\big\rangle {}_{B_k{\textit{\textbf{E}}}'_k} \Big)\Big].
	\label{eq:vir_act_prob_dep}
\end{align}}{For simplicity of notation, in \cref{eq:ph_actual_dep} we do not include a subscript $w$ to indicate that it refers to the $w^{\rm th}$ virtual protocol. Also, in this equation}, $\hat{D}_{\gamma_X}^{{(k)}}$ with $\gamma \in \{0,1\}$ is Bob's effective POVM element for the $k^{\rm th}$ pulse after Eve's coherent attack {(see \cref{eq:measurement_app_d} in \cref{app:proof_measurement_d} for more details),} {$\tilde{\sigma}_{\alpha_X,{\vb*{g}}|j_{k-1},\hdots,j_{k-l_c}}^{{(k,{\rm vir})}} \coloneqq \hat{P}\big(\ket*{\psi_{\alpha_X,{\vb*{g}}|j_{k-1},\hdots,j_{k-l_c}}^{{\rm (vir)}}}_{B_k,{\textit{\textbf{E}}}'_k}\big)$} {where $\ket*{\psi_{\alpha_X,{\vb*{g}}|j_{k-1},\hdots,j_{k-l_c}}^{{\rm (vir)}}}_{B_k,{\textit{\textbf{E}}}'_k}$ is defined as}
{\begin{align}
	\scalemath{0.95}{\frac{\ket*{{\tilde{\psi}}_{0_Z,{\vb*{g}}|j_{k-1},\hdots,j_{k-l_c}}}_{B_k,{\textit{\textbf{E}}}'_k} + (-1)^\alpha \ket*{{\tilde{\psi}}_{1_Z,{\vb*{g}}|j_{k-1},\hdots,j_{k-l_c}}}_{B_k,{\textit{\textbf{E}}}'_k} }{2~\sqrt{\frac{\tilde{p}^{{(k,{\rm vir})}}_{\alpha_X,{\vb*{g}}|j_{k-1},\hdots,j_{k-l_c}}}{p_{Z_A}} }}.}
	\label{eq:virtual_states_dep}
\end{align}}

{Importantly, n}ote that \cref{eq:ph_actual_dep} is very similar to \cref{eq:ph_actual}, and thus one can apply the RT in the same way as in  \cref{sec:proof_gk_independent} to derive an upper bound on the phase-error rate, as explained below. \\

\noindent \textit{{Applying the reference technique}} \\
{In this case, we choose the reference states to be $\{ \ket{\phi_{j,{\vb*{g}}}}_{B_k,{\textit{\textbf{E}}}'_k}\}_{j \in \{0_Z,1_Z,0_X,1_X\}}$, defined in \cref{eq:reference_new}, which live in a qubit space. Then,}
{one can follow a similar analysis to that described between} Eqs.~(\ref{eq:Phi_gk_k}) {and} (\ref{eq:phase_error_act}) {by making the following substitutions: $\ket{\psi_{j,{\vb*{g}}}}_{B_k,E_k} \to \ket*{\tilde{\psi}_{j,{\vb*{g}}|j_{k-1},\hdots,j_{k-l_c}}}_{B_k,{\textit{\textbf{E}}}'_k}$, $\ket*{\Psi_{{\vb*{g}}}^{Z}}_{A_k,B_k,E_k} \to \ket*{\Psi_{{\vb*{g}}|j_{k-1},\hdots,j_{k-l_c}}^{Z}}_{A_k,B_k,{\textit{\textbf{E}}}'_k}$, $\ket{\phi_{j,{\vb*{g}}}}_{B_k,E_k} \to \ket{\phi_{j,{\vb*{g}}}}_{B_k,{\textit{\textbf{E}}}'_k}$ and $\ket*{\Phi_{{\vb*{g}}}^Z}_{A_k,B_k,E_k} \to \ket*{\Phi_{{\vb*{g}}}^Z}_{A_k,B_k,{\textit{\textbf{E}}}'_k}$, where $\ket*{\Phi_{{\vb*{g}}}^Z}_{A_k,B_k,{\textit{\textbf{E}}}'_k}$ is defined in \cref{eq:entangled_dep_ref}.} {In the derivations, the inner products $\big|\braket*{\tilde{\psi}_{j,{\vb*{g}}|j_{k-1},\hdots,j_{k-l_c}}}{\phi_{j,{\vb*{g}}}}{}_{B_k,{\textit{\textbf{E}}}'_k}\big|$ and $\big|\braket*{\Psi^Z_{{\vb*{g}}|j_{k-1},\hdots,j_{k-l_c}}}{\Phi^Z_{{\vb*{g}}}}{}_{A_k,B_k,{\textit{\textbf{E}}}'_k}\big|$ will now appear. In {\cref{app:inner_prod_corr},} we show that these inner products are both upper bounded by $\sqrt{1-\epsilon^{\rm U}}$, where $\epsilon^{\rm U} = 1 - (1-\epsilon'^{\rm U})^{l_c+1}$. Using this result, we obtain} \cref{eq:ph_ref_new_upper,eq:phase_error_act}, which now apply to all rounds with tag $w$ in the $w^\textrm{th}$ virtual protocol.



By taking the average over all rounds with tag $w$ on both sides of \cref{eq:phase_error_act} and using the concavity of $G_+(y,z)$ with respect to $y$, we obtain 
		\begin{align}
			&\frac{1}{N_w}\sum_{k \in \mathcal{N}_w} P^{{(k)}}_{{\vb*{g}}}({\rm ph|Act}) \nonumber \\
			&\leq \frac{1}{N_w}\sum_{k \in \mathcal{N}_w}  p_{Z_A} p_{Z_B} G_+ \bigg(\frac{P^{{(k)}}_{{\vb*{g}}}({\rm ph|Ref})}{p_{Z_A} p_{Z_B}},\sqrt{1-\epsilon^{\rm U}} \bigg) \nonumber \\
			&\leq p_{Z_A} p_{Z_B} G_+ \bigg(\frac{1}{N_w}\sum_{k \in \mathcal{N}_w}\frac{P^{{(k)}}_{{\vb*{g}}}({\rm ph|Ref})}{p_{Z_A} p_{Z_B}},\sqrt{1-\epsilon^{\rm U}} \bigg), 
			\label{eq:phase_error_n_lc}
		\end{align}
		where $\mathcal{N}_w$ is the set of all rounds with tag $w$ and $N_w = \abs{\mathcal{N}_w} = \frac{N}{l_c+1}$. As in \cref{eq:ph_ref_bound}, we now apply Jensen's inequality to $\frac{1}{N_w} \sum_{k \in \mathcal{N}_w} P^{{(k)}}_{{\vb*{g}}}({\rm ph|Ref})$, using the concavity of $G_+(y,z)$ and $-G_-(y,z)$ with respect to $y$. Then, by substituting this expression back in \cref{eq:phase_error_n_lc}, and applying Azuma's or Kato's inequality {\footnote{{When we apply Azuma's or Kato's inequality to \cref{eq:phase_error_n_lc}, we consider the averaged probabilities of $P_{\vb*{g}}^{k}({\rm  ph|Act})$ and $P_{\vb*{g}}^{k}({\rm  ph|Ref})$ over all previous rounds that do not have a tag $w$.}}} under the assumption $N_w\to\infty$, we have that
		\begin{align}
			&{{N_{\textrm{ph},w}}} \leq {N_w} p_{Z_A} p_{Z_B} G_+ \left(\rule{0cm}{1cm}\right.\sum_{\substack{\alpha,\gamma \in \{0,1\} \\ \alpha \neq \gamma}} \frac{ p_{\alpha_X}^{\rm (vir)U}}{p_{Z_A}} \left[\rule{0cm}{1cm}\right. \sum_{\substack{j  \\ c_{\alpha,j}^{\rm U} > 0}} c_{\alpha,j}^{\rm U} \nonumber \\
			& G_+\bigg( \frac{{N_{j,\gamma_X,w}}}{N_w p_{j} p_{X_B}},\sqrt{1-\epsilon^{\rm U}} \bigg) + \sum_{\substack{j  \\ c_{\alpha,j}^{\rm U} < 0}} c_{\alpha,j}^{\rm U} \nonumber \\
			&G_-\bigg( \frac{{N_{j,\gamma_X,w}}}{N_w p_{j} p_{X_B}},\sqrt{1-\epsilon^{\rm U}}\bigg)\left.\rule{0cm}{1cm}\right],\sqrt{1-\epsilon^{\rm U}}  \left.\rule{0cm}{1cm}\right) =: {N_{{\rm ph}, w}^{\rm U}},
			\label{eq:final_lc}
		\end{align}
		{with} {$j\in \{0_Z,1_Z,0_X,1_X\}$,} {where} ${N_{j,\gamma_X,w}}$ is the number of events with a tag $w$ in which Alice's setting choice is $j$, Bob selects the $X$ basis and his measurement outcome is $\gamma$. 
		
		Finally, we obtain the upper bound on the phase-error rate of the $w^\textrm{th}$ sifted key,
		{
		\begin{align}
		    e_{{\rm ph},w}^{{\rm U}} := \frac{N_{{\rm ph}, w}^{\rm U}}{N_{\textrm{det},w}^{(Z)}},
		    \label{eq:ephUw_final}
		\end{align}}where $N_{{\rm det},w}^{(Z)}$ is the number of detected rounds with a tag $w$ in which both Alice and Bob selected the $Z$ basis. This means that {the users} should now apply privacy amplification to the $w^\textrm{th}$ sifted key sacrificing a fraction $h(e_{{\rm ph},w}^{{\rm U}})$ of its bits, where 
		{$h(x) = -x \log_2 x - (1-x)\log_2 (1-x)$} is the binary entropy function.
	
	The fraction of the total sifted key that {Alice and Bob} sacrifice satisfies
		\begin{equation}
			\sum_{w = 0}^{l_c} q_w h(e_{{\rm ph},w}^{{\rm U}}) \leq h\left(\sum_{w = 0}^{l_c} q_w  e_{{\rm ph},w}^{{\rm U}}\right) \leq h(e_{{\rm ph}}^{\rm U}),
			\label{eq:avg_eph}
		\end{equation}
		where $q_w \coloneqq N_{\textrm{det},w}^{(Z)}/N_{\textrm{det}}^{(Z)}$ and $e_{\rm ph}^{\rm U}$ is the bound obtained using {\cref{eq:ephU_final}} in \cref{sec:proof_gk_independent}, which in this case can be regarded as an upper bound on the average phase-error rate. The first inequality in Eq.~(\ref{eq:avg_eph}) is due to the concavity of the function $h(x)$ and the second inequality is proved in \cref{app:proof_avg_eph}. This result is useful if one is simply interested in computing a lower bound on the asymptotic secret-key rate of the protocol. However, we remark that, in practice, one cannot simply compute $e_{\rm ph}^{\rm U}$ and then apply privacy amplification to the total sifted key at once. One needs to compute each $e_{{\rm ph},w}^{{\rm U}}$, and apply privacy amplification separately to each of the $(l_c + 1)$ sifted keys {(see the protocol description in \cref{app:protocol_description}).}

	\section{Results and Discussion}
	In this section, as an example, we apply our security proof to the {modified} BB84 protocol with a phase-encoding scheme in the presence of multiple source imperfections. In particular, we show how to derive upper bounds on the coefficients $c_{\alpha,j,{\vb*{g}}}^{{(k)}}$ and on the probabilities $p_{\alpha_X,{\vb*{g}}}^{{(k,{\rm vir})}}$. Then, we simulate the secret-key rate that we would obtain in a practical implementation of the protocol. Finally, we compare its performance with that of the three-state loss-tolerant protocol \cite{tamaki} under the same parameter regimes. {We remark that the results and discussion presented here appl{y} to both {security} analyses in \cref{sec:security}.}
	
	\subsection{Particular device model} 
	\label{sec:device_model}
	In general, the state of the emitted pulses {for each round $k$} is in the form of Eqs.~(\ref{eq:general}) {and (\ref{eq:general_set_dep_corr}),} which take into account the main source imperfections. As a particular example, we assume {here} that the state $\ket*{\omega_{j,{\vb*{g}}}}_{B_k}$ in the qubit part of {these equations:} $\ket*{{\phi_{j,{\vb*{g}}}}}_{B_k,E_{k}} = \ket*{\omega_{j,{\vb*{g}}}}_{B_k}$ $ \ket*{\lambda_{{\vb*{g}}}}_{E_{k}}$, satisfies
	\begin{align}
		\ket{\omega_{j,{\vb*{g}}}}_{B_k} = \cos \Big(\frac{\theta_{j,{\vb*{g}}}^{{(k)}}}{2}\Big) \ket{0_Z}_{B_k} + \sin \Big(\frac{\theta_{j,{\vb*{g}}}^{{(k)}}}{2}\Big) \ket{1_Z}_{B_k},
		\label{eq:phase_enc}
	\end{align} 
	where $\theta_{j,{\vb*{g}}}^{{(k)}} \in [0,2\pi)$ is the encoding phase, which depends on the round $k$. Recall that the state in Eq.~(\ref{eq:phase_enc}) incorporates any imperfection in the qubit space, such as SPFs and phase fluctuations. 
	
	When applying the RT to prove the security of this protocol, we relate the virtual states $\sigma_{\alpha_X,{\vb*{g}}}^{{(k,{\rm vir})}}$ and the reference states $\sigma_{j,{\vb*{g}}}^{{(k)}}$ through Eq.~(\ref{eq:relationships}). Then, for both virtual states, we choose the unique solution of Eq.~(\ref{eq:relationships}) such that $c_{1,0_X,{\vb*{g}}}^{{(k)}} = c_{0,1_X,{\vb*{g}}}^{{(k)}} = 0$, which provides the best numerical results for our device model. We elaborate on this point in \cref{sec:comparison_bb84_3state}. Using the definition of $\ket*{\phi_{j,{\vb*{g}}}}_{B_k,E_{k}}$ and Eq.~(\ref{eq:phase_enc}), we can solve the resulting systems of linear equations to find analytical expressions for the other coefficients $c_{\alpha,j,{\vb*{g}}}^{{(k)}}$ as a function of the encoding phases, i.e.~$c_{1,j,{\vb*{g}}}^{{(k)}} (\theta_{0_Z,{\vb*{g}}}^{{(k)}},\theta_{1_Z,{\vb*{g}}}^{{(k)}},\theta_{1_X,{\vb*{g}}}^{{(k)}})$ for $j \in \{0_Z,1_Z,1_X\}$ and $c_{0,j,{\vb*{g}}}^{{(k)}} (\theta_{0_Z,{\vb*{g}}}^{{(k)}},\theta_{1_Z,{\vb*{g}}}^{{(k)}},\theta_{0_X,{\vb*{g}}}^{{(k)}})$ for $j \in \{0_Z,1_Z,0_X\}$ (see \cref{app:coefficients_bb84} for full expressions). Similarly, we can express $p_{\alpha_X,{\vb*{g}}}^{{(k,{\rm vir})}}$ as a function of $\theta_{0_Z,{\vb*{g}}}^{{(k)}}$ and $\theta_{1_Z,{\vb*{g}}}^{{(k)}}$. While the exact phases for a particular round are unknown, we assume that one can guarantee that they always fall in a known range, i.e.\
	$\theta_{j,{\vb*{g}}}^{{(k)}} \in [\theta_{j}^{\rm L},\theta_{j}^{\rm U}]$ for $j \in \{0_Z,1_Z,0_X,1_X\}$. Then, we can obtain upper bounds $c_{\alpha,j}^{\rm U}$ and $p_{\alpha_X}^{\rm (vir)U}$ on each individual coefficient $c_{\alpha,j,{\vb*{g}}}^{{(k)}}$ and probability $p_{\alpha_X,{\vb*{g}}}^{{(k,{\rm vir})}}$ by considering the worst case scenario for the encoding phases. In \cref{app:coefficients_bb84}, we provide their analytical solutions in \cref{eq:coefficients,eq:probabilities}, respectively.
	
	Finally, by substituting these bounds into {Eq.~(\ref{eq:ephU_final})}, we obtain $e_{\rm ph}^{\rm U}$. {A lower bound on the} asymptotic secret-key rate can then be expressed as \begin{equation}
		R \geq Y_Z \big[1 - h(e_{\rm ph}^{\rm U}) - fh(e_{\rm bit})\big],
		\label{eq:R}
	\end{equation}
	where $Y_Z$ is the joint probability that both Alice and Bob select the $Z$ basis and Bob obtains a detection event, $f$ is the error correction efficiency and $e_{\rm bit}$ is the bit error rate. Note that $Y_Z$ and $e_{\rm bit}$ would be directly observed in a practical implementation of the protocol. The lower bound in \cref{eq:R} applies both to the case in which {the emitted states} do not depend on Alice's previous setting choices, and to the case in which they do depend on the previous $l_c$ setting choices; the latter is due to \cref{eq:avg_eph}. {However, we emphasise that this lower bound depends on the value of $\epsilon^{\rm U}$, whose definition differs in the two cases. Namely, in the former, $\epsilon^{\rm U}$ is an upper bound on $\epsilon^{{(k)}}_{j,{\vb*{g}}}$ in \cref{eq:general}; while in the latter, $\epsilon^{\rm U} = 1 - (1-\epsilon'^{\rm U})^{l_c+1}$, where $\epsilon'^{\rm U}$ is an upper bound on $\epsilon^{{(k)}}_{j,{\vb*{g}}|j_{k-1},\hdots,j_{k-l_c}}$ in \cref{eq:general_set_dep_corr}.}
	
	\subsection{Simulation of the secret-key rate}
	To apply our analysis to this particular device model, one needs to experimentally quantify the {parameter} $\epsilon^{\rm U}$, and $\{\theta_{j}^{\rm L},\theta_{j}^{\rm U}\}$ for all $j$, which represents the uncertainty on the phase of the qubit component. In our simulations, we consider a range of values for $\epsilon^{\rm U}$ and, as an example, we take the uncertainty on the phase to be the same for all $j$, i.e.\ $\theta_{j}^{\rm L} = \hat\theta_{j} - \Delta$ and $\theta_{j}^{\rm U} = \hat\theta_{j} + \Delta$ for some known $\hat\theta_{j}$ and $\Delta$. Moreover, we assume that $\hat\theta_{j}$ deviates from the ideal encoding angles due to SPFs; in particular, we assume that $\hat\theta_{0_Z} = 0$, $\hat\theta_{1_Z} = \kappa \pi$, $\hat\theta_{0_X} = \kappa \pi/2$ and $\hat\theta_{1_X} = \kappa 3\pi/2$, where $\kappa = 1 + \delta/\pi$ and $\delta \geq 0$ quantifies this deviation.
	
	To simulate the data that one would obtain in an actual experiment, we use the channel model described in \cite{pereira}. In particular, the  model assumes that Alice sends a qubit state with an exact phase $\hat\theta_{j}$ when she selects the setting $j$. This is because $\Delta$ and $\epsilon^{\rm U}$ are both small, and do not result in significantly different experimental results compared with the ideal case. This does not contradict the assumptions of our security proof: these imperfections could still exist and allow Eve to learn some secret-key information, which our security proof takes into account.  
	
	For the simulations, we assume the following channel, device, {and protocol} parameters: $\Delta = 0.03$ \cite{xu}, $\delta= 0.063$ \cite{honjo,xu}, $f = 1.16$ and the dark count probability of Bob's detectors $p_d = 10^{-8}$ \cite{xu5,pittaluga}. Here, we do not consider a specific value for the detection efficiency of Bob's detectors because we represent the secret-key rate as a function of the overall system loss. Moreover, we assume the efficient QKD scheme \cite{lo6}, where $p_{X_A} = p_{X_B} \rightarrow 0$ in the asymptotic scenario. 
	
	The results for the {modified} BB84 protocol are shown in Fig.~1. {This figure} shows that, as $\epsilon^{\rm U}$ increases, the secret-key rate $R$ decreases. This is expected because $\epsilon^{\rm U}$ takes into account multiple side channels, and therefore, an increase in $\epsilon^{\rm U}$ corresponds to a bigger leakage of secret-key information. {In addition,} our {security proof} takes into account variable modulation flaws and {setting-independent} pulse correlations, allowing each emitted pulse to be different, i.e.\ non-IID, {even in the absence of side channels.} 
	\begin{figure}[h]
		\includegraphics[width=8.5cm]{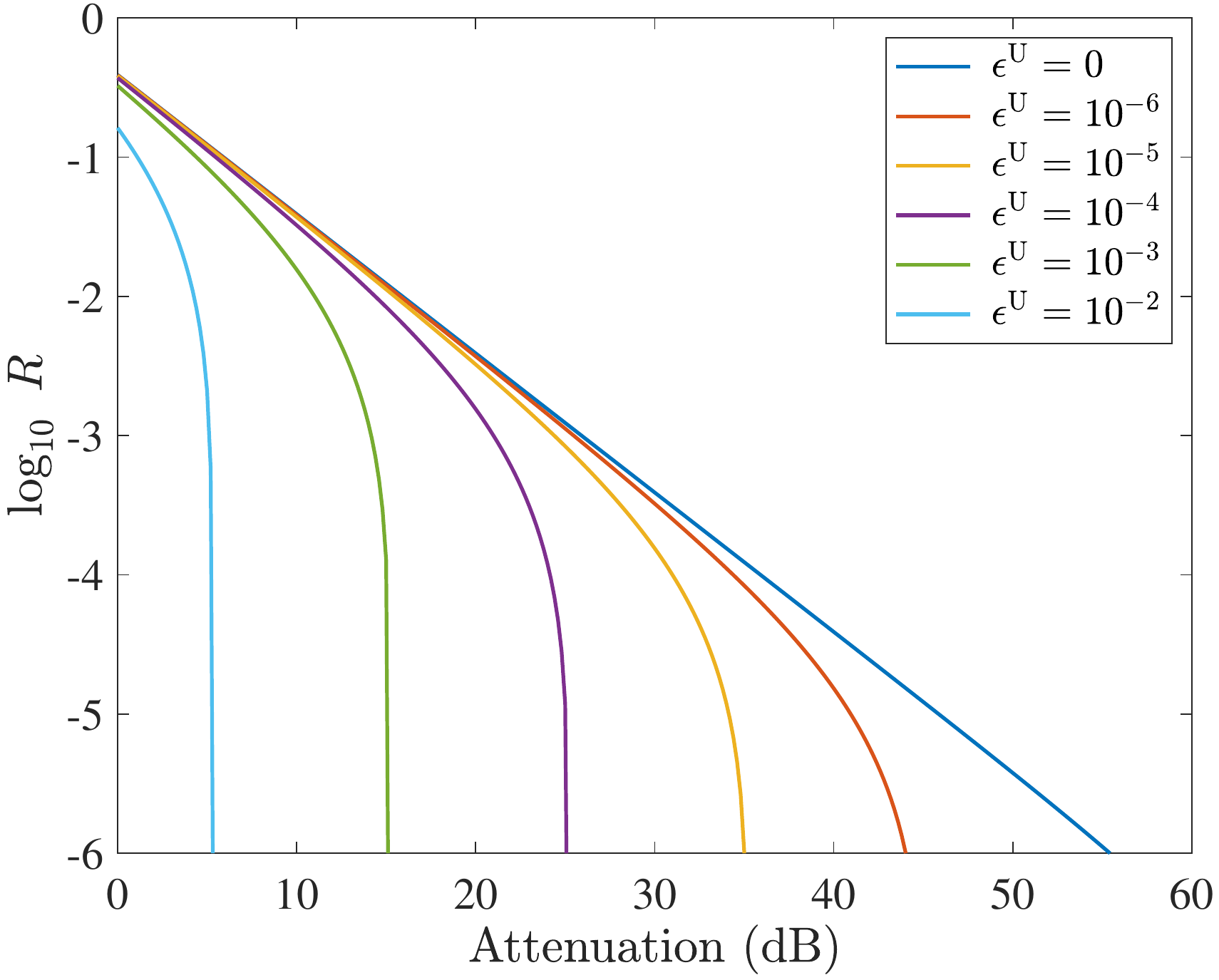} 
		\caption{Secret-key rate $R$ as a function of the overall system loss (dB) for different values of $\epsilon^{\rm U}$ when applying the RT to the {modified} BB84 protocol.}
	\end{figure}

	We note that there are other proof techniques that can also be employed to deal with source loopholes, such as those using semi-definite-programming (SDP) \cite{wang3,primaatmaja,zhou} or other convex optimisation approaches \cite{coles,winick}. However, so far, none of them is able to guarantee this level of implementation security, and therefore it is difficult to make a fair comparison. In particular, the analyses in \cite{wang3,primaatmaja,coles,winick,zhou} assume the emission of IID pulses. This assumption is unrealistic since in practice the emission of non-IID pulses is unavoidable due to fluctuations and pulse correlations.

	\subsection{Comparison between the three-state and the {modified} BB84 protocols}
	\label{sec:comparison_bb84_3state}
	For the comparison with the three-state protocol to be fair, here we consider the same particular device model (see \cref{sec:device_model}). There is, however, a difference in the security proof: for the three-state protocol, Eq.~(\ref{eq:relationships}) can only be expressed as 
		\begin{align}
		\sigma_{\alpha_X,{\vb*{g}}}^{{(k,{\rm vir})}} = \sum_{j} c_{\alpha,j,{\vb*{g}}}^{{(k)}} \sigma_{j,{\vb*{g}}}^{{(k)}},
		\label{eq:relationships3}
	\end{align}
	{for $j \in \{0_Z,1_Z,0_X\}$.} In this case, the coefficients $c_{\alpha,j,{\vb*{g}}}^{{(k)}}$ for $\alpha \in \{0,1\}$ are both functions of the same encoding phases: $c_{\alpha,j,{\vb*{g}}}^{{(k)}} (\theta_{0_Z,{\vb*{g}}}^{{(k)}},\theta_{1_Z,{\vb*{g}}}^{{(k)}},\theta_{0_X,{\vb*{g}}}^{{(k)}})$. The form of these coefficients as well as their upper bounds are defined in \cref{app:coefficients_3state}. For the simulations, we take $\epsilon^{\rm U} \in \{10^{-6},10^{-3}\}$ and $\delta \in \{0.063,0.126\}$ \cite{honjo,xu}. All the other experimental parameters are the same as above. 

	The results are presented in Fig.~2. This figure shows that the {modified} BB84 protocol achieves better secret-key rates in all parameter regimes investigated. For instance, in Fig.~2, the blue lines are significantly better than the red lines for $\epsilon^{\rm U} \in \{10^{-6},10^{-3}\}$. This indicates that using four states rather than three, while redundant in an idealised scenario \cite{tamaki}, is advantageous in the presence of multiple source imperfections. 
		\begin{figure}[h!]
		\includegraphics[width=8.5cm]{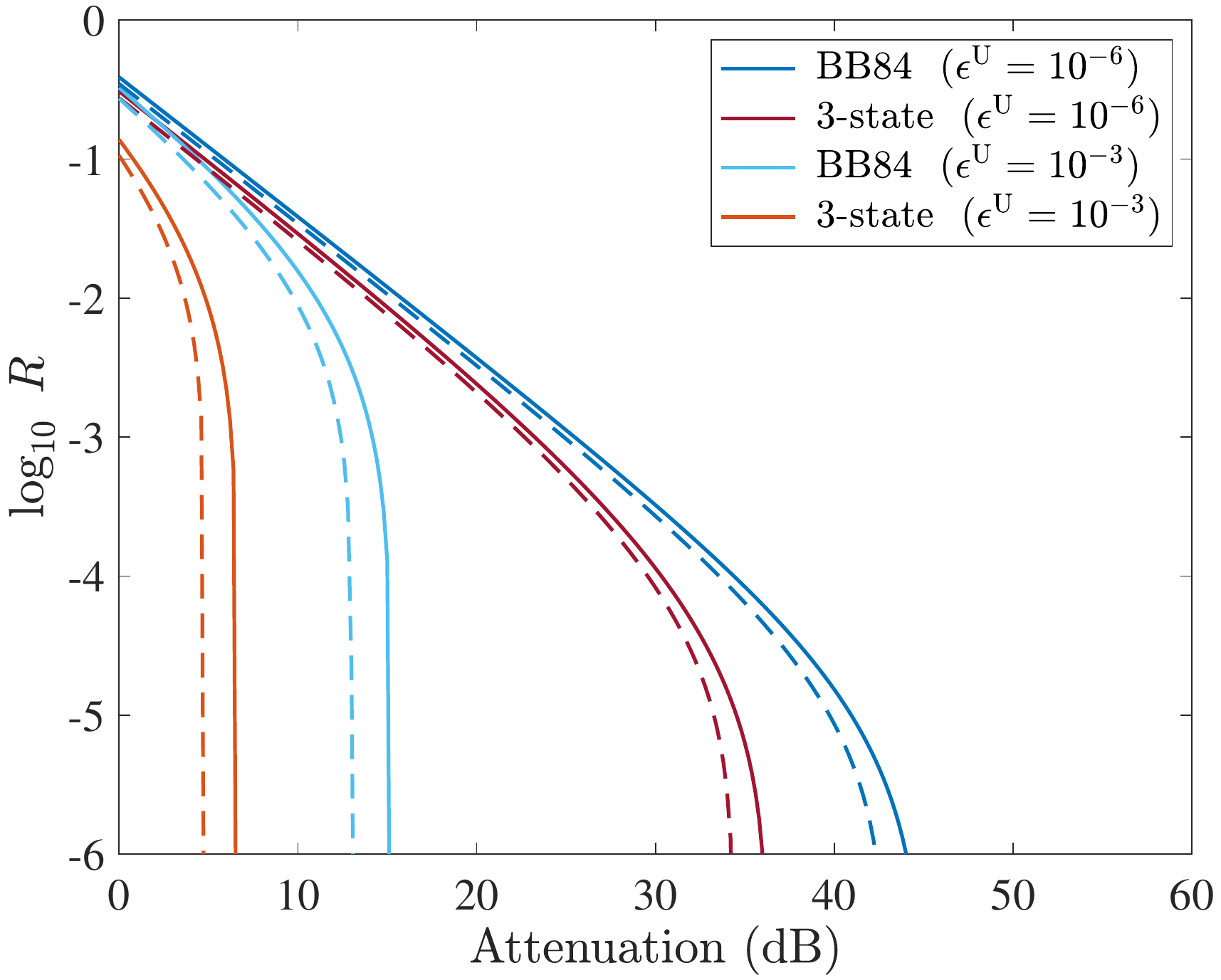} 
		\caption{Secret-key rate $R$ as a function of the overall system loss (dB) {for} different parameter regimes when applying the RT to the {modified} BB84 protocol (BB84) and to the three-state protocol (3-state). The solid lines correspond to $\delta = 0.063$ and the dashed lines correspond to $\delta = 0.126$.}
	\end{figure}

	As pointed out in \cref{sec:proof_gk_independent}, this improvement comes from the extra choice in the coefficients of \cref{eq:relationships} when adding a fourth state. Loosely speaking, for each $\alpha \in \{0,1\}$, the estimation should be tightest when $\sum_{j} \vert c_{\alpha,j,{\vb*{g}}}^{{(k)}}\vert$ is minimised, since higher absolute values for these coefficients result in increased multiplicative factors for the deviation terms introduced by the application of the bound in Eq.~(\ref{eq:bound}). Thus, since $\sigma_{1_X,{\vb*{g}}}^{{(k)}}$ and $\sigma_{1_X,{\vb*{g}}}^{{(k,{\rm vir})}}$ ($\sigma_{0_X,{\vb*{g}}}^{{(k)}}$ and $\sigma_{0_X,{\vb*{g}}}^{{(k,{\rm vir})}}$) are close in the Bloch sphere, if we take $c_{1,0_X,{\vb*{g}}}^{{(k)}} = 0$ ($c_{0,1_X,{\vb*{g}}}^{{(k)}} = 0$), then $c_{1,1_X,{\vb*{g}}}^{{(k)}}$ ($c_{0,0_X,{\vb*{g}}}^{{(k)}}$) is close to one, while $c_{1,0_Z,{\vb*{g}}}^{{(k)}}$ and $c_{1,1_Z,{\vb*{g}}}^{{(k)}}$ ($c_{0,0_Z,{\vb*{g}}}^{{(k)}}$ and $c_{0,1_Z,{\vb*{g}}}^{{(k)}}$) are close to zero, minimising the sum of absolute values. We have numerically tested all possible combinations in which $\sigma_{\alpha_X,{\vb*{g}}}^{{(k,{\rm vir})}}$ can be expressed as a function of $\sigma_{j,{\vb*{g}}}^{{(k)}}$, for $j \in \{0_Z,1_Z,0_X,1_X\}$, and confirmed that the choices $c_{1,0_X,{\vb*{g}}}^{{(k)}} = c_{0,1_X,{\vb*{g}}}^{{(k)}} = 0$ result in the highest secret-key rates. This implies that the key-rate advantage of the {modified} BB84 protocol with respect to the three-state protocol comes from the ability to express $\sigma_{1_X,{\vb*{g}}}^{{(k,{\rm vir})}}$ as a function of $\sigma_{1_X,{\vb*{g}}}^{{(k)}}$ rather than $\sigma_{0_X,{\vb*{g}}}^{{(k)}}$.
	
	We remark that, recently, a similar conclusion has been reached by employing a SDP-type security proof \cite{bourassa}. Namely, the MDI BB84 protocol has been shown to provide higher secret-key rates than the MDI three-state protocol in the presence of a particular time-dependent side channel. 
	In this work, however, we show that the {modified} BB84 protocol achieves a better performance than the three-state protocol in the presence of multiple source {imperfections,} including classical pulse correlations, when using the RT.
	
	Additionally, one can see in Fig.~2 that, as the SPFs increase, the secret-key rate is roughly the same in all cases investigated. That is, for both analyses the solid lines ({corresponding to} $\delta = 0.063$) and the dashed lines ({corresponding to} $\delta = 0.126$) are close to one another for any value of $\epsilon^{\rm U}$. In other words, as expected, the RT is robust to SPFs independently of the number of states used.

	\section{Conclusion}
	The best known and most widely implemented quantum key distribution (QKD) scheme is the Bennett-Brassard 1984 (BB84) protocol \cite{bennett}. Since its introduction, several rigorous security proofs have been {proposed} \cite{mayers,shor,biham,koashi2,gottesman}, but the security of the BB84 protocol with imperfect sources has not yet been {fully} established. In this work, we have considered a {modified} BB84 protocol that does not discard the basis mismatched events, and we have used the reference technique (RT) to prove its security in the presence of flawed, leaky and correlated sources. More precisely, our security proof accommodates multiple source imperfections, {including} state preparation flaws {and side channels, such as} Trojan-horse attacks, mode dependencies and {classical} pulse correlations. 

	
	
	{Additionally, in this paper} we have compared the achievable secret-key rates {of the modified BB84 protocol} with those obtained when {using} the three-state loss-tolerant protocol. We have shown that, in the presence of source imperfections, the addition of a fourth state results in a significant performance improvement, indicating that the {modified} BB84 protocol offers a clear advantage in guaranteeing the practical security of QKD sources.
	
	{In short,} this work proves the security of the {modified} BB84 protocol with {practical sources} and {shows its robustness to source imperfections,} taking us a step closer towards ensuring the implementation security of QKD at a level that is suitable for practical applications.

	\section{Acknowledgements}
	This work was supported by {Cisco Systems Inc.,} the Galician Regional Government (consolidation of Research Units: AtlantTIC), the Spanish Ministry of Economy and Competitiveness (MINECO), the Fondo Europeo de Desarrollo Regional (FEDER) through Grant Number PID2020-118178RB-C21 and MICIN with funding from the European Union NextGenerationEU (PRTR-C17.I1) {and the Galician Regional Government with own funding} through the ``Planes Complementarios de I+D+I con las Comunidades Autónomas" in Quantum Communication. {M.P. and G.C.-L. acknowledge support from JSPS Postdoctoral Fellowships for Research in Japan.} {A.M. acknowledges support from JST, ACT-X Grant No. JPMJAX210O.} G.K. acknowledges support from JSPS Kakenhi (C) No.20K03779 and 21K03388.  K.T. acknowledges support from JSPS KAKENHI Grant Number JP18H05237 and JST-CREST JPMJCR 1671.

	\appendix
	
	\section{{Description of the modified BB84 QKD protocol}}
	\label{app:protocol_description}
	%
        {\begin{enumerate}
	    \item \textbf{State preparation:} For each round $k \in \{1,\hdots,$ $N\}$, Alice selects the setting $j \in \{0_Z,1_Z,0_X,1_X\}$ with probability $p_{j}$, generates an encoded pulse and sends it to Bob through the quantum channel.  
	    \item \textbf{Detection:} Bob measures each incoming pulse in the basis $\beta \in \{Z,X\}$ with probabilities $p_{Z_B}$ and $p_{X_B}$, respectively. 
	    \item \textbf{Sifting:} Bob announces which rounds were detected, and Alice and Bob reveal their basis choices in {these} rounds. Let $\mathcal{K}$ be the set of detected rounds in which both users employed the $Z$ basis and $\mathcal{T}$ be the set of detected rounds in which Bob employed the $X$ basis. Then, Alice and Bob define their own sifted keys as the bit values associated with her emissions and his measurement results  on rounds $\mathcal{K}$, respectively. As for rounds $\mathcal{T}$, Alice and Bob announce their respective bit values.  
	    \item \textbf{Parameter estimation:} Alice quantifies the number of events ${N_{j,\gamma_X}}$ on rounds $\mathcal{T}$ for $j \in \{0_Z,1_Z,0_X,1_X\}$ and $\gamma \in \{0,1\}$, where $\gamma$ denotes Bob's measurement outcome. Then, using these quantities she obtains an upper bound $N^{\rm U}_{\rm ph}$ on ${N_{\rm ph}}$, the number of phase errors in her sifted key, using \cref{eq:final}.
	    \item \textbf{{Data} post-processing:} For error correction, Alice sends Bob encrypted syndrome information \footnote{{Encrypting the syndrome information ensures that the actual protocol and the virtual protocol employed in the security proof are equivalent from Eve's perspective (see \cite{koashi2,koashi3} for more information).}} about her sifted key through an authenticated public channel, which Bob uses to correct his sifted key. For error verification, Alice and Bob compute a hash of their corrected keys using a random {two-}universal hash function and check if they are identical. If not, they abort the protocol; otherwise, Alice and Bob perform privacy amplification on their corrected keys. For this, they use a random two-universal hash function to extract an identical secret key pair.
	\end{enumerate}}

{In the presence of setting-dependent pulse correlations, the protocol described above requires slight modifications. First, before step 1, Alice and Bob need to assign tags to each round:}
{\begin{enumerate}\addtocounter{enumi}{-1}
	    \item \textbf{Tag assignment:} For each round $k \in \{1,\hdots,N\}$, Alice and Bob assign a tag $w \in \{0,\hdots,l_c\}$ according to the value $w = k\bmod (l_c+1)$, where $l_c$ is the correlation length.
\end{enumerate} Then, Alice and Bob perform the steps 1-3, with $\mathcal{K}_w$ ($\mathcal{T}_w$) defined as the set of detected rounds with tag $w$ in which both users employed the $Z$ basis (Bob employed the $X$ basis). Hence, the $w^{\rm th}$ sifted key is defined as the subset of the total sifted key originating from the rounds in $\mathcal{K}_w$. After that, the users perform parameter estimation and employ {data} post-processing to obtain a final secure key: }
{\begin{enumerate}\addtocounter{enumi}{+3}
     \item \textbf{Parameter estimation:} Alice quantifies the number of events ${N_{j,\gamma_X,w}}$ on rounds $\mathcal{T}_w$ for $j \in \{0_Z,1_Z,0_X,1_X\}$ and $\gamma \in \{0,1\}$. Then, using these quantities she obtains an upper bound $N^{\rm U}_{{\rm ph},w}$ on ${N_{\textrm{ph},w}}$, the number of phase errors in her $w^{\rm th}$ sifted key, using \cref{eq:final_lc}.
    \item \textbf{{Data} post-processing:} For error correction, Alice sends Bob encrypted syndrome information about her $w^{\rm th}$ sifted key through an authenticated public channel, which Bob uses to correct his $w^{\rm th}$ sifted key. For error verification, Alice and Bob compute a hash of their $w^{\rm th}$ corrected keys using a random {two-}universal hash function and check if they are identical. If not, they discard their $w^{\rm th}$ corrected keys; otherwise, Alice and Bob perform privacy amplification. For this, they use a random two-universal hash function to extract the $w^{\rm th}$ secret key pair. 
	 \item \textbf{Key concatenation:} After repeating the steps 3-5 for all tags $w$, Alice and Bob define their final secret key pair as the concatenation of all the $w^{\rm th}$ secret keys.
\end{enumerate}}

	
\section{{Treatment of the setting-independent factors in the security proof}}
\label{app:proof_fixed_g}

{Let {${\mathbfcal{G}} = \mathcal{G}_1,\hdots,\mathcal{G}_N$} denote {the random variables that represent the setting-independent} factors affecting the form of {Alice's} emitted states, {and let ${\vb*{g}}=g_1, \hdots, g_N$ denote a particular {outcome} of these random variables. To prove the security of the actual protocol, we consider an equivalent entanglement-based scenario in which Alice prepares a entangled state with ancillary systems ${\textit{\textbf{A}}}:= A_1,\hdots,A_N$ that she can measure to learn her setting choices. Since in the actual protocol Alice's emitted states depend on ${\vb*{g}}$, this entangled state can be expressed as}}
{\begin{equation}
	\label{eq:rho_g}
	\int_{{{\vb*{g}} \in \textrm{dom}(f_{\mathbfcal{G}})}} {f_{\mathbfcal{G}}({\vb*{g}})} \ketbra{\Psi_{{\vb*{g}}}}_{\textit{\textbf{A}},\textit{\textbf{B}},\textit{\textbf{E}}} d{\vb*{g}},
\end{equation}}{where {$f_{\mathbfcal{G}}({\vb*{g}})$} is the probability density function of {${\mathbfcal{G}}$, $\textrm{dom}(f_{\mathbfcal{G}})$ is the domain of $f_{\mathbfcal{G}}({\vb*{g}})$ and {$\textit{\textbf{B}},\textit{\textbf{E}} := B_1, E_1, \hdots, B_N, E_N$}. In the setting-independent scenario, the state $\ket{\Psi_{{\vb*{g}}}}_{\textit{\textbf{A}},\textit{\textbf{B}},\textit{\textbf{E}}}$ in \cref{eq:rho_g} takes the form}}
	%
	{\begin{align}
		\ket{\Psi_{{\vb*{g}}}}_{\textit{\textbf{A}},\textit{\textbf{B}},\textit{\textbf{E}}} =\bigotimes_{k=1}^N \sum_{j_k} {\sqrt{p_{j_k}}} \ket{j_k}_{A_k} \ket*{\psi_{j_k,{\vb*{g}}}}_{B_k,E_k},
		\label{eq:running_out_of_names}
	\end{align}
	where $\ket{\psi_{j_k,{\vb*{g}}}}_{B_k,E_k}$ is defined in \cref{eq:general}. Note that, for ease of discussion, in this appendix we explicitly write the setting choice on the $k^{\rm th}$ round as $j_k$, rather than $j$. Alternatively, in the setting-dependent scenario, the state $\ket{\Psi_{{\vb*{g}}}}_{\textit{\textbf{A}},\textit{\textbf{B}},\textit{\textbf{E}}}$ in \cref{eq:rho_g} takes the form }
	{\begin{align}
	\ket{\Psi_{{\vb*{g}}}}_{\textit{\textbf{A}},\textit{\textbf{B}},\textit{\textbf{E}}} =\bigotimes_{k=1}^N \sum_{j_k} {\sqrt{p_{j_k}}} \ket{j_k}_{A_k} \ket*{\psi_{j_k,{\vb*{g}}|j_{k-1},\hdots,j_{k-l_c}}}_{B_k,E_k},
		\label{eq:Psi_g1g2gn}
	\end{align}}{where $\ket*{\psi_{j_k,{\vb*{g}}|j_{k-1},\hdots,j_{k-l_c}}}_{B_k,E_k}$ is defined in \cref{eq:general_set_dep_corr}.}

	{However, {note that Eve cannot distinguish the scenario in which Alice prepares the mixed state in \cref{eq:rho_g} from the scenario in which she} prepares its purification, {
	\begin{equation}
			\label{eq:rho_g2 }
			\int_{{\vb*{g}} \in \textrm{dom}(f_{\mathbfcal{G}})} \sqrt{f_{\mathbfcal{G}}({\vb*{g}})} \ket{{\vb*{g}}}_G  \ket{\Psi_{{\vb*{g}}}}_{\textit{\textbf{A}},\textit{\textbf{B}},\textit{\textbf{E}}} d{\vb*{g}},
	\end{equation}}and {then} measures system $G$ in the very beginning of the protocol, obtaining some outcome ${\vb*{g}}$ and post-measurement state $\ket{\Psi_{{\vb*{g}}}}_{\textit{\textbf{A}},\textit{\textbf{B}},\textit{\textbf{E}}}$.} 
	{{For simplicity of presentation, in} our security proof, {we consider this equivalent scenario, in which the value of} ${\vb*{g}}$ has {effectively become} fixed {before Alice emits any pulses to Bob.} {Then, in \cref{sec:proof_gk_independent}, we} derive a bound {$N_{\rm ph}^{\rm U}$ on the number of phase errors $N_{\rm ph}$ that is in fact} conditional on {this value of} ${\vb*{g}}$, {such that
	\begin{equation}
		\Pr[N_{\rm ph} > N_{\rm ph}^{\rm U} \vert {\mathbfcal{G}}={\vb*{g}}] \leq \varepsilon,
		\label{eq:bound_set_ind}
	\end{equation}		
  	where $\varepsilon$ is the failure probability. {Importantly}, however, this bound is valid for all possible {outcomes} ${\vb*{g}}$, {as} neither $N_{\rm ph}^{\rm U}$ nor $\varepsilon$ depend on ${\vb*{g}}$. This implies that the bound in \cref{eq:bound_set_ind} is also valid for the scenario in which Alice prepares the mixed state in \cref{eq:rho_g}, since 
  	\begin{align}
  		\Pr[N_{\rm ph} > N_{\rm ph}^{\rm U}] = &\int_{{\vb*{g}} \in \textrm{dom}(f_{\mathbfcal{G}})} \!\!\! f_{\mathbfcal{G}}({\vb*{g}}) \Pr[N_{\rm ph} > N_{\rm ph}^{\rm U} \vert {\mathbfcal{G}}={\vb*{g}}] \nonumber \\ 
  		\leq &\int_{{\vb*{g}} \in \textrm{dom}(f_{\mathbfcal{G}})} \!\!\! f_{\mathbfcal{G}}({\vb*{g}}) \, \varepsilon = \varepsilon.
  	\end{align}
  By a similar argument, we conclude that the phase-error rate bounds derived in \cref{sec:proof_gk_dependent}} also {apply} to the scenario in which Alice prepares the mixed state {in \cref{eq:rho_g}.}}

\section{{Trojan-horse attacks}}
\label{app:THAs} 

{Here, we explicitly show how to incorporate THAs in our security analysis. In particular, we assume that Eve's injected light does not alter Alice's prepared signals, other than adding extra modes of light that contain information about Alice's setting choices. That is, in the presence of a THA, the total emitted state consists of a tensor product between the state that Alice would have emitted in Eve's absence and the back-reflected light from the THA. Moreover, we assume that the back-reflected light on round $k$ only contains information about the $k^{\rm th}$ setting choice, and is in tensor product form with the back-reflected light from all the other rounds. In this context, we consider two cases for the THA: (1) the back-reflected light is a pure state, and (2) the back-reflected light is a mixed state. For simplicity of discussion, in what follows we assume that THAs are the only side channel present. However, one can easily combine these results with the other side channels considered in this work, as shown below.}
    {\subsection{Pure output light}}
{In this case, the state emitted on round $k$ can be written as $\ket{\psi_{j,\vb*{g}}}_{B_k,E_k} = \ket{\omega_{j,\vb*{g}}}_{B_k} \otimes \ket{E_{j,\vb*{g}}}_{E_k}$, where $\ket{\omega_{j,\vb*{g}}}_{B_k}$ is a qubit state and} 
{\begin{equation}
\ket{E_{j,\vb*{g}}}_{E_k} = \sqrt{1-\epsilon^{{(k)}}} \ket{v}_{E_k} + \sqrt{\epsilon^{{(k)}}} \ket*{e_{j,\vb*{g}}}_{E_k},
\label{eq:pure_Eve}
\end{equation}is the back-reflected light from the THA. Here, $\epsilon^{{(k)}} \in [0,1]$ quantifies the deviation of $\ket{E_{j,\vb*{g}}}_{E_k}$ from the vacuum state $\ket{v}_{E_k}$, and $\ket*{e_{j,\vb*{g}}}_{E_k}$ is a setting-dependent state orthogonal to $\ket{v}_{E_k}$, i.e.\ it {belongs to} the Fock subspace $\{\ket{1},\ket{2},\hdots\}$. Note that $\ket{E_{j,\vb*{g}}}_{E_k}$ could also represent other side channels, such as electromagnetic radiation. Using \cref{eq:pure_Eve}, it is straightforward to write the $k^{\rm th}$ emitted state in the form of \cref{eq:general}:}
{\begin{equation}
\ket{\psi_{j,\vb*{g}}}_{B_k,E_k} = \sqrt{1-\epsilon^{{(k)}}} \ket{\phi_{j,\vb*{g}}}_{B_k,E_k} + \sqrt{\epsilon^{{(k)}}} \ket*{\phi_{j,\vb*{g}}^\perp}_{B_k,E_k},
\label{eq:pure_emitted}
\end{equation} where the qubit state $\ket{\phi_{j,\vb*{g}}}_{B_k,E_k} := \ket{\omega_{j,\vb*{g}}}_{B_k} \ket{v}_{E_k}$ and the side-channel state $\ket*{\phi_{j,\vb*{g}}^\perp}_{B_k,E_k} := \ket{\omega_{j,\vb*{g}}}_{B_k}\ket*{e_{j,\vb*{g}}}_{E_k} $; and consequently, apply our security analysis, given that an upper bound on $\epsilon^{(k)}$ is known.}

{We remark that the inclusion of side channels other than THAs can be readily accommodated as follows. Suppose that, in the absence of a THA, the state of Alice's prepared signal on round $k$ {is the following:}
\begin{align}
    \ket{\Omega_{j,\vb*{g}}}_{B_k} = \sqrt{1-\tilde{\epsilon}^{(k)}_{j,\vb*{g}}} \ket{\omega_{j,\vb*{g}}}_{B_k} + \sqrt{\tilde{\epsilon}^{(k)}_{j,\vb*{g}}} \ket*{\omega_{j,\vb*{g}}^\perp}_{B_k},
    \label{eq:mode_dep}
\end{align}where $\ket*{\omega_{j,\vb*{g}}^\perp}_{B_k}$ is a side-channel state that lives in any Hilbert space orthogonal to $\ket{\omega_{j,\vb*{g}}}_{B_k}$. Note that the state in \cref{eq:mode_dep} includes any kind of mode dependencies with $\ket*{\omega_{j,\vb*{g}}}_{B_k}$ indicating the desired mode. Then, in the presence of a THA, the $k^{\rm th}$ emitted state would {simply} become $\ket{\psi_{j,\vb*{g}}}_{B_k,E_k} = \ket{\Omega_{j,\vb*{g}}}_{B_k} \otimes \ket{E_{j,\vb*{g}}}_{E_k}$, which can be written in the form of \cref{eq:general}:
\begin{align}
    &\ket{\psi_{j,\vb*{g}}}_{B_k,E_k} = \sqrt{1-\tilde{\epsilon}^{(k)}_{j,\vb*{g}}} \sqrt{1-{\epsilon}^{(k)}} \ket{\phi_{j,\vb*{g}}}_{B_k,E_k} \nonumber \\
    &+ \sqrt{1-(1-\tilde{\epsilon}^{(k)}_{j,\vb*{g}})(1-{\epsilon}^{(k)})} \ket*{\phi_{j,\vb*{g}}^\perp}_{B_k,E_k},
\end{align}where $\ket{\phi_{j,\vb*{g}}}_{B_k,E_k} := \ket{\omega_{j,\vb*{g}}}_{B_k} \ket{v}_{E_k}$ and 
\begin{align}
    &\ket*{\phi_{j,\vb*{g}}^\perp}_{B_k,E_k} := \Big[\sqrt{1-\tilde{\epsilon}^{(k)}_{j,\vb*{g}}} \sqrt{{\epsilon}^{(k)}} \ket{\omega_{j,\vb*{g}}}_{B_k} \ket{e_{j,\vb*{g}}}_{E_k} \nonumber \\
    &+ \sqrt{\tilde{\epsilon}^{(k)}_{j,\vb*{g}}} \ket*{\omega_{j,\vb*{g}}^{\perp}}_{B_k} \ket{E_{j,\vb*{g}}}_{E_k}\Big]/\sqrt{1-(1-\tilde{\epsilon}^{(k)}_{{j,\vb*{g}}})(1-{\epsilon}^{(k)})}.
\end{align}

In a similar way, if in the absence of a THA, Alice's prepared signal on round $k$ {is the following:}
\begin{align}
       &\ket{\Omega_{j,\vb*{g}|j_{k-1},\hdots,j_{k-l_c}}}_{B_k} = \sqrt{1-\tilde{\epsilon}^{(k)}_{j,\vb*{g}|j_{k-1},\hdots,j_{k-l_c}}} \ket{\omega_{j,\vb*{g}}}_{B_k} \nonumber \\
       &+ \sqrt{\tilde{\epsilon}^{(k)}_{j,\vb*{g}|j_{k-1},\hdots,j_{k-l_c}}} \ket*{\omega_{j,\vb*{g}|j_{k-1},\hdots,j_{k-l_c}}^\perp}_{B_k},
    \label{eq:mode_dep_corr}
\end{align}
due to setting-dependent pulse correlations and mode dependencies; then, in the presence of a THA, the $k^{\rm th}$ emitted state would become $\ket*{\psi_{j,\vb*{g}|j_{k-1},\hdots,j_{k-l_c}}}_{B_k,E_k} = \ket*{\Omega_{j,\vb*{g}|j_{k-1},\hdots,j_{k-l_c}}}_{B_k} \otimes \ket{E_{j,\vb*{g}}}_{E_k}$, which can be written in the form of \cref{eq:general_set_dep_corr}. \\} 

\noindent {\textit{{E}xperimental method to upper bound $\epsilon^{{(k)}}$}} \\
{Here, we show how one {could upper} bound $\epsilon^{{(k)}}$ {by} using only information about the output light intensity (average number of photons). To obtain this information, one {needs to} first determine the maximum light intensity that Eve could inject into Alice's source without being detected. Then, by characterising the value of the attenuation in Alice's setup \cite{lucamarini}, one can bound the maximum intensity of the output light $\nu_{\rm max}$. Mathematically, this bound can be expressed as $\ev{\hat{N}}{E_{j,\vb*{g}}}_{E_k} \leq \nu_{\rm max}$ where $\hat{N} = \sum_n n \ketbra{n}$ is the photon-number operator. Then, using \cref{eq:pure_Eve}, we have that}
{\begin{equation}
	\nu_{\rm max} \geq \ev{\hat{N}}{E_{j,\vb*{g}}}_{E_k} = \epsilon^{{(k)}} \ev{\hat{N}}{e_{j,\vb*{g}}}_{E_k} \geq \epsilon^{{(k)}} ,
	\label{eq:pure_vmax}
\end{equation}with equality if and only if $\ket{e_{j,\vb*{g}}}_{E_k}$ is a single-photon state. }
{We note that one could obtain slightly tighter bounds if one had more information on the back-reflected light, such as it being a coherent state \cite{navarrete2}.}
{\subsection{Mixed output light}}
{In this case, we assume that the back-reflected light system $\tilde{E}_k$ is purified by an ancillary system $E'_k$ that is in Eve's hands \footnote{{In practice, even if Eve injects half of a maximally entangled state into the source, the joint state of the back-reflected light and Eve's ancilla will become mixed as the light is attenuated in Alice's setup. Mathematically, this mixedness could be purified by another ancillary system that Eve has no access to. However, for simplicity, here we consider the worst case scenario in which she has full access to {an} ancillary system that purifies the back-reflected light.}}. We can then express the joint state of the back-reflected light and this ancillary system as}
{\begin{align}
\ket{E_{j,\vb*{g}}}_{E_k} &= \sum_c \sqrt{p_c} \ket{c}_{E'_k} \Big(\sqrt{1-\epsilon^{{(k)}}_c} \ket{v}_{\tilde{E}_k} \nonumber \\
&+ \sqrt{\epsilon^{{(k)}}_c} \ket*{e_{j,\vb*{g},c}}_{\tilde{E}_k}\Big),
\label{eq:mixed_Eve}
\end{align}{where $E_k := E'_k,\tilde{E}_k$ and $\{\ket{c}\}$ forms an orthonormal basis}. If $\epsilon^{{(k)}}_c = \epsilon^{{(k)}}$ for all $c$, it is straightforward to write the emitted state on round $k$ in the form of \cref{eq:pure_emitted}, with $\ket{\phi_{j,\vb*{g}}}_{B_k,E_k} = \ket{\omega_{j,\vb*{g}}}_{B_k} \ket{\lambda}_{E_k}$, where $\ket{\lambda}_{E_k} := \sum_c \sqrt{p_c} \ket{c}_{E'_k} \ket{v}_{\tilde{E}_k}$, and $\ket*{\phi_{j,\vb*{g}}^\perp}_{B_k,E_k} := \ket{\omega_{j,\vb*{g}}}_{B_k}$ $\sum_c \sqrt{p_c} \ket{c}_{E'_k} \ket*{e_{j,\vb*{g},c}}_{\tilde{E}_k} $. If 
$\epsilon^{{(k)}}_c \leq \epsilon^{{(k)}}$ for all $c$, then the worst case scenario is such that the bound is saturated for all $c$, and one can assume that $\epsilon^{{(k)}}_c = \epsilon^{{(k)}}$. In both of these cases, one can directly use the experimental bound $\nu_{\rm max} \geq \epsilon^{{(k)}}$ in \cref{eq:pure_vmax} and apply our security analysis.}

{However, even if the conditions above do not hold, our security analysis can still be applied as long as we have an upper bound on $\sum_c p_c \epsilon^{{(k)}}_c$, which, as we will show below, can be related to $\nu_{\rm max}$.} {To use the RT in this case, one {can} select the reference states {as follows}}
{\begin{align}
\ket{\phi_{j,\vb*{g}}}_{B_k,E_k} &= \ket{\omega_{j,\vb*{g}}}_{B_k} \ket{\lambda}_{E_k} \nonumber \\
& := \frac{ \ket{\omega_{j,\vb*{g}}}_{B_k} \sum_c \sqrt{p_c} \ket{c}_{E'_k} \sqrt{1-\epsilon^{{(k)}}_c} \ket{v}_{\tilde{E}_k}}{\sqrt{\sum_c p_c (1-\epsilon^{{(k)}}_c)}},
\label{eq:mixed_reference}
\end{align} for $j \in \{0_Z,1_Z,0_X,1_X\}$. Then, the required inner products $|\braket*{\psi_{j,{\vb*{g}}}}{\phi_{j,{\vb*{g}}}}_{B_k,E_k}|$ can be calculated as} 
{\begin{align}
|\braket*{\psi_{j,{\vb*{g}}}}{\phi_{j,{\vb*{g}}}}_{B_k,E_k}|  = \sqrt{1 - \sum_c p_c \epsilon^{{(k)}}_c},
\label{eq:mixed_inner_prod}
\end{align}where we have used the fact that $\braket*{\phi_{j,{\vb*{g}}}^\perp}{\phi_{j,{\vb*{g}}}}_{B_k,E_k} = 0$ for any $j$. \\} 

\noindent {\textit{{E}xperimental method to upper bound $\sum_c p_c \epsilon^{{(k)}}_c$}} \\
{Similarly to the pure state case, we can obtain a bound on $\sum_c p_c \epsilon^{{(k)}}_c$ using only information about the output light intensity. {In particular, from} \cref{eq:mixed_Eve}, we have that}
{\begin{align}
	\nu_{\rm max} &\geq \ev{{{\hat{\mathds{1}}}_{E'_k} \otimes \hat{N}_{\tilde{E}_k}}}{E_{j,\vb*{g}}}_{E_k} \nonumber \\
	&= \sum_c p_c \epsilon^{{(k)}}_c \ev{\hat{N}_{\tilde{E}_k}}{e_{j,\vb*{g},c}}_{\tilde{E}_k} \geq \sum_c p_c \epsilon^{{(k)}}_c,
	\label{eq:mixed_vmax}
\end{align}with equality if $\ket{e_{j,\vb*{g},c}}_{\tilde{E}_k}$ is a single-photon state.}\\ \\

    \section{{Proof of \cref{eq:ph_actual}}}
    \label{app:proof_measurement} 
   {Here, we prove \cref{eq:ph_actual} and show that {when the} QKD {protocol is run sequentially (see assumption (A4) in \cref{sec:assumptions_alice})}, the operator $\hat{M}^{{(k)}}_{\gamma_X}$ satisfies $0 \leq \hat{M}^{{(k)}}_{\gamma_X} \leq {\hat{\mathds{1}}}$. {To do so, we} consider an entanglement-based scenario {that} is equivalent to the actual protocol. For a fixed ${\vb*{g}}$ {(see \cref{app:proof_fixed_g})}, the transmission of $N$ pulses {can then be} described by Alice first preparing the following entangled state}
    {\begin{align}
       \ket{\Psi_{{\vb*{g}}}}_{\textit{\textbf{A}},\textit{\textbf{B}},\textit{\textbf{E}}} &=  \bigotimes_{k=1}^{N} \sum_{j_{k}} \sqrt{p_{j_{k}}} \ket{j_{k}}_{A_k} \ket{\psi_{j_{k},{\vb*{g}}}}_{B_k,E_k} \nonumber \\
       &=: \bigotimes_{k=1}^{N} \ket*{\Psi'_{{\vb*{g}}}}_{A_k,B_k,E_k},
       \label{eq:full_stateB}
    \end{align}}{then keeping system ${\textit{\textbf{A}}}$ in her laboratory and sending systems ${\textit{\textbf{B}},\textit{\textbf{E}}}$ through the quantum channel. Here, ${\textit{\textbf{A}}:=} A_1,\hdots,A_N$ and ${\textit{\textbf{B}},\textit{\textbf{E}}:=} B_1,E_1,\hdots,B_N,E_N$ refers to the composite systems of Alice's ancillae and to the pulses sent to Bob, respectively, where $A_k$ and $B_k$ {for $k \in \{1,\hdots,N\}$} denote Alice's and Bob's $k^{\rm th}$ systems and $E_k$ denotes any other systems that are emitted by Alice on round $k$, such as the back-reflected light from a THA {(see \cref{app:THAs} for more details)}.} {Also, in} {Eq.~(\ref{eq:full_stateB}), $\{\ket*{j_{k}}_{A_k}\}_{j_{k} \in \{0_Z,1_Z,0_X,1_X\}}$ is a set of {orthonormal} states {and $\ket{\psi_{j_k,{\vb*{g}}}}_{B_k,E_k}$ is defined in \cref{eq:general}. Note that, for ease of discussion, in this appendix we explicitly write the setting choice on the $k^{\rm th}$ round as $j_k$, rather than $j$.}} 
    
\begin{figure}[h]
		\includegraphics[width=8.7cm]{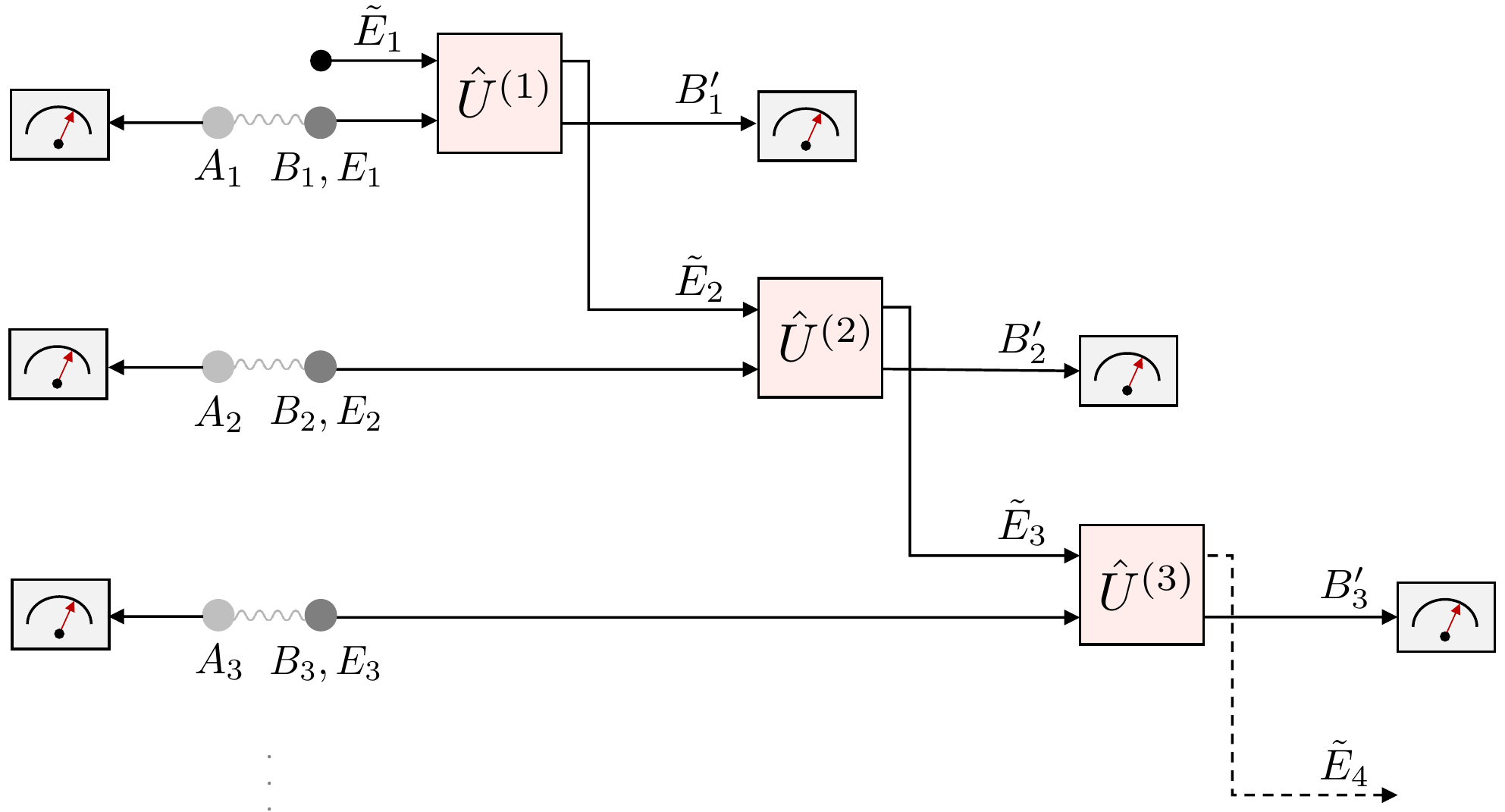}
		\caption{{Diagram of the entanglement-based scenario and Eve's most general coherent attack when the {QKD} protocol is run {sequentially}. {Eve must perform her attack on the {$k^{\rm th}$} pulse before she learns information about systems $B_{{k+1}},E_{{k+1}}$, and therefore $\hat{U}^{{({k})}}$ cannot take systems $B_{{k+1}},E_{{k+1}}$ as an input. Conversely,} Eve's attack on the {$k^{\rm th}$} pulse can depend on information {that} she has learned {about the} systems $B_{{k-1}},E_{{k-1}}$. {This} is why {$\hat{U}^{{({k})}}$} takes Eve's updated ancilla $\tilde{E}_{{k}}$ as an input. }} 
		\label{fig:eve_attack_slow}
	\end{figure}

   {As already discussed, i}n our analysis we assume that the QKD protocol is run {sequentially} such that Alice only generates the ${k}^{\rm th}$ pulse once Bob has obtained his ${k-1}^{\rm th}$ measurement result. Under this restriction, Eve's most general coherent attack can be described as the application of a sequence of $N$ unitary operators $\hat{U}_{B_N,E_N,\tilde{E}_N}^{{(N)}} \hdots \hat{U}_{B_1,E_1,\tilde{E}_1}^{{(1)}}$, where $\hat{U}_{B_k,E_k,\tilde{E}_k}^{{(k)}}$ acts on the $k^{\rm th}$ photonic system $B_k,E_k$ and on Eve's updated ancilla $\tilde{E}_k$, {resulting in systems $B'_k$ and $\tilde{E}_k$, respectively, as shown in} \cref{fig:eve_attack_slow}.
    
    {After Eve's attack on each round,} {Alice and Bob measure their local systems. {First, Alice measures system $A_k$ to know if $j_{k} \in \{0_Z, 1_Z\}$ or if $j_{k} \in \{0_X, 1_X\}$, whose respective outcomes are denoted by $a = Z$ and  $a = X$, and Bob selects the measurement basis $\beta \in \{Z,X\}$ with probability $p_{\beta_B}$.} In the modified BB84 protocol, the secret key is generated from the rounds in which {$(a,\beta) = (Z,Z)$.} As explained in \cref{sec:proof_gk_independent}, to prove the security of these rounds, one needs to consider the number of phase errors that Alice and Bob would have observed if they had performed their local measurements in the {phase} basis instead, {i.e.\ $\{{\ket*{\alpha_X^{\rm (vir)}}_{A_k}} = (\ket{0_Z}_{A_k} + (-1)^\alpha \ket{1_Z}_{A_k})/\sqrt{2}\}_{\alpha \in \{0,1\}}$ and $\{\hat{m}_{0_X},\hat{m}_{1_X},\hat{m}_{f}\}$, respectively.} We can then define a virtual protocol in which, in the key generation rounds, Alice and Bob perform {these} measurements, and in all the other rounds they perform measurements in their selected basis {({$a,\beta$)}.}}
    
    {In this virtual protocol, let us denote all possible outcomes for round $k$ as {$o_k \in \{(a,\beta,\alpha,\gamma),(f)\}$,} where $\alpha,$ $\gamma \in \{0,1\}$ are Alice's and Bob's observed bit value{s}, respectively, {and $f$ is associated with an inconclusive outcome.} Now, let us define the POVM element associated with obtaining the outcome $o_k$ by $\hat{F}^{o_k}_{A_k,B'_k}$. Then, we can summarise the possible POVM elements as}   
    {\begin{align}
    &{\hat{F}_{A_k,B'_k}^{(Z,Z,\alpha,\gamma)} = {\dyad*{\alpha_X^{\rm (vir)}}_{A_k}} \otimes p_{Z_B} \hat{m}_{\gamma_X},} \nonumber \\
   &\hat{F}_{A_k,B'_k}^{(Z,X,\alpha,\gamma)} = \dyad{\alpha_Z}_{A_k} \otimes p_{X_B} \hat{m}_{\gamma_X}, \nonumber \\
    &{\hat{F}_{A_k,B'_k}^{(X,\beta,\alpha,\gamma)} = \dyad{\alpha_X}_{A_k} \otimes p_{\beta_B} \hat{m}_{\gamma_\beta},} \nonumber \\  
    &{\hat{F}_{A_k,B'_k}^{(f)} = {\hat{\mathds{1}}}_{A_k} \otimes  \hat{m}_{f}.}
    \label{eq:combined_measurements}
    \end{align}}
    
    {Note that $\hat{F}_{A_k,B'_k}^{(Z,Z,\alpha,\gamma)}$ is the POVM element associated with the estimation of the number of phase errors. A phase error on the $k^{\rm th}$ round can then be defined as obtaining the outcome $o_{{k}}=(Z,Z,0,1)$ or $o_{{k}}=(Z,Z,1,0)$. Therefore, the probability of a phase error on the $k^{\rm th}$ round conditional on all the previous outcomes  $o_{{\overrightarrow{k-1}}}$ can be expressed as}
    {\begin{align}
       \scalemath{0.98}{P^{{(k)}}_{{\vb*{g}}}({\rm ph|Act})} &\scalemath{0.98}{= P\big(o_{{k}} = (Z,Z,0,1) {~\rm or~} (Z,Z,1,0) |o_{{\overrightarrow{k-1}}}\big)} \nonumber \\
       &\scalemath{0.98}{=\frac{P\big((o_{{k}} = (Z,Z,0,1) {~\rm or~} (Z,Z,1,0)) {,} o_{{\overrightarrow{k-1}}}\big)}{P\big(o_{{\overrightarrow{k-1}}}\big)},} 
        \label{eq:phase_error_prob_appendix}
   \end{align}}where $P\big(o_{{\overrightarrow{k-1}}}\big)$ is the probability of obtaining Alice's and Bob's previous {$k-1$} outcomes. Note that this conditional probability on all the previous outcomes is required to take into account any correlations between the measurement outcomes of different rounds of the protocol, which could arise due to Eve's coherent attack. Because of this dependence, Azuma's or Kato's inequality {can be employed} in order to estimate the total number of phase errors in the protocol. 
   
   Now, using the form of the full emitted state in \cref{eq:full_stateB}, the description of Eve's coherent attack {when the} QKD protocol {is run sequentially}, and the POVM elements in \cref{eq:combined_measurements}, we can mathematically express {the time evolution illustrated in} \cref{fig:eve_attack_slow} and calculate the numerator of \cref{eq:phase_error_prob_appendix} as follows
{ \begin{widetext}\begin{align}
       &\scalemath{0.90}{P\big((o_{{k}} = (Z,Z,0,1) {~\rm or~} (Z,Z,1,0)) {,} o_{{\overrightarrow{k-1}}}\big) =  \sum_{\substack{\alpha, \gamma \in \{0,1\} \\ \alpha \neq \gamma}} {\rm Tr}~ \Bigg[\hat{F}_{A_k,B'_k}^{(Z,Z,\alpha,\gamma)}} \nonumber \\
       & \scalemath{0.90}{\hat{P}\left(\hat{U}^{{(k)}}_{B_k,E_k,\tilde{E}_k}\prod^{k-1}_{n=1} \sqrt{\hat{F}^{o_{{k-n}}}_{A_{k-n},B'_{k-n}}} \hat{U}^{{(k-n)}}_{B_{k-n},E_{k-n},\tilde{E}_{k-n}} \ket{0}_{\tilde{E}_1}  \bigotimes_{n=1}^{k-1} \ket*{\Psi'_{{\vb*{g}}}}_{A_n,B_n,E_n} \ket*{\Psi'_{{\vb*{g}}}}_{A_k,B_k,E_k} \bigotimes_{n=k+1}^{N} \ket*{\Psi'_{{\vb*{g}}}}_{A_n,B_n,E_n}  \right) \Bigg]} \nonumber \\
       & \scalemath{0.90}{= \sum_{\substack{\alpha, \gamma \in \{0,1\} \\ \alpha \neq \gamma}} \Tr [\hat{F}_{A_k,{B'_k}}^{(Z,Z,\alpha,\gamma)} \hat{P}\left(\hat{U}^{{(k)}}_{B_k,E_k,\tilde{E}_k}\prod^{k-1}_{n=1} \sqrt{\hat{F}^{o_{k-n}}_{A_{k-n},B'_{k-n}}} \hat{U}^{{(k-n)}}_{B_{k-n},E_{k-n},\tilde{E}_{k-n}} \ket{0}_{\tilde{E}_1} \bigotimes_{n=1}^{k-1} \ket*{\Psi'_{{\vb*{g}}}}_{A_n,B_n,E_n} \ket*{\Psi'_{{\vb*{g}}}}_{A_k,B_k,E_k} \right)]} \nonumber \\
       & \scalemath{0.90}{= \sum_{\substack{\alpha, \gamma \in \{0,1\} \\ \alpha \neq \gamma}} \Tr \left[{}_{{\textit{\textbf{A}}}_{\leq k-1},{\textit{\textbf{B}}'}_{\leq k-1},{\textit{\textbf{E}}}_{\leq k-1},{\tilde{E}_k}}\bra*{\Psi_{{\vb*{g}}}^{\overrightarrow{k-1}}} \hat{U}^{{(k)}\dagger}_{B_k,E_k,\tilde{E}_k} \hat{F}_{A_k,{B'_k}}^{(Z,Z,\alpha,\gamma)} \hat{U}^{{(k)}}_{B_k,E_k,\tilde{E}_k} \ket*{\Psi_{{\vb*{g}}}^{\overrightarrow{k-1}}}_{{\textit{\textbf{A}}}_{\leq k-1},{\textit{\textbf{B}}'}_{\leq k-1},{\textit{\textbf{E}}}_{\leq k-1},{\tilde{E}_k}} \dyad*{\Psi'_{{\vb*{g}}}}_{A_k,B_k,E_k} \right]} \nonumber \\
        & \scalemath{0.90}{= \sum_{\substack{\alpha, \gamma \in \{0,1\} \\ \alpha \neq \gamma}} \Tr \left[{}_{{\textit{\textbf{A}}}_{\leq k-1},{\textit{\textbf{B}}'}_{\leq k-1},{\textit{\textbf{E}}}_{\leq k-1},{\tilde{E}_k}}\bra*{\Psi_{{\vb*{g}}}^{\overrightarrow{k-1}}} \hat{U}^{{(k)}\dagger}_{B_k,E_k,\tilde{E}_k} \hat{m}_{\gamma_X} \hat{U}^{{(k)}}_{B_k,E_k,\tilde{E}_k} \ket*{\Psi_{{\vb*{g}}}^{\overrightarrow{k-1}}}_{{\textit{\textbf{A}}}_{\leq k-1},{\textit{\textbf{B}}'}_{\leq k-1},{\textit{\textbf{E}}}_{\leq k-1},{\tilde{E}_k}} \tilde{p}^{{(k,{\rm vir})}}_{\alpha_X,{\vb*{g}}} p_{Z_B} \dyad*{\psi_{\alpha_X,{\vb*{g}}}^{\rm (vir)}}_{B_k,E_k}\right],}
        \label{eq:prob_derivation}
   \end{align}\end{widetext}}\noindent {where}
   {\begin{align}
       &\ket*{\Psi_{{\vb*{g}}}^{\overrightarrow{k-1}}}_{{\textit{\textbf{A}}}_{\leq k-1},{\textit{\textbf{B}}'}_{\leq k-1},{\textit{\textbf{E}}}_{\leq k-1},{\tilde{E}_k}} \nonumber \\
       &:= \prod^{k-1}_{n=1} \sqrt{\hat{F}^{o_{k-n}}_{A_{k-n},B'_{k-n}}} \hat{U}^{{(k-n)}}_{B_{k-n},E_{k-n},\tilde{E}_{k-n}} \ket{0}_{\tilde{E}_1} \nonumber \\
       &\bigotimes_{n=1}^{k-1} \ket*{\Psi'_{{\vb*{g}}}}_{A_n,B_n,E_n},
       \label{eq:new_state}
   \end{align}}{with ${\textit{\textbf{A}}}_{\leq k-1} := A_1, \hdots, A_{k-1}$ and ${\textit{\textbf{B}}'}_{\leq k-1},{\textit{\textbf{E}}}_{\leq k-1} := {B'_1},E_1,\hdots, {B'_{k-1}},E_{k-1}$,} {and {where} $\ket*{\psi_{\alpha_X,{\vb*{g}}}^{\rm (vir)}}_{B_k,E_k}$ and $\tilde{p}^{{(k,{\rm vir})}}_{\alpha_X,{\vb*{g}}}$ are defined in Eqs.~(\ref{eq:vir_act}) and (\ref{eq:vir_act_prob}), respectively. In the second equality of \cref{eq:prob_derivation}, we took the partial trace over systems ${\textit{\textbf{A}}}_{\geq k+1},{\textit{\textbf{B}}}_{\geq k+1},{\textit{\textbf{E}}}_{\geq k+1}$. In the third equality of \cref{eq:prob_derivation}, we defined the state $\ket*{\Psi_{{\vb*{g}}}^{\overrightarrow{k-1}}}_{{\textit{\textbf{A}}}_{\leq k-1},{\textit{\textbf{B}}'}_{\leq k-1},{\textit{\textbf{E}}}_{\leq k-1},{\tilde{E}_k}}$ using \cref{eq:new_state} and used the cyclic property of the trace operation. Finally, in the last equality of \cref{eq:prob_derivation}, we used {\cref{eq:Psi_gk_k} and} the definition of $\hat{F}_{A_k,B'_k}^{(Z,Z,\alpha,\gamma)}$ in \cref{eq:combined_measurements} for $\alpha,\gamma \in \{0,1\}$ such that $\alpha \neq \gamma$.}

    {By substituting \cref{eq:prob_derivation} in \cref{eq:phase_error_prob_appendix}, we can now express $P^{{(k)}}_{{\vb*{g}}}({\rm ph|Act})$ as}
    {\begin{align}
        P^{{(k)}}_{{\vb*{g}}}({\rm ph|Act}) = \sum_{\substack{\alpha, \gamma \in \{0,1\} \\ \alpha \neq \gamma}} \tilde{p}^{{(k,{\rm vir})}}_{\alpha_X,{\vb*{g}}} p_{Z_B} \Tr \left[ \tilde{\sigma}^{{(k,{\rm vir})}}_{\alpha_X,{\vb*{g}}} \hat{M}_{\gamma_X}^{{(k)}} \right], 
   \end{align}}{{with} $\tilde{\sigma}^{{(k,{\rm vir})}}_{\alpha_X,{\vb*{g}}} := \dyad*{\psi_{\alpha_X,{\vb*{g}}}^{\rm (vir)}}_{B_k,E_k}$ and}
    {\begin{align}
       \hat{M}_{\gamma_X}^{{(k)}} &:= \frac{\bra*{\Psi_{{\vb*{g}}}^{\overrightarrow{k-1}}}\hat{U}^{{(k)}\dagger}_{B_k,E_k,\tilde{E}_k} \hat{m}_{\gamma_X} \hat{U}^{{(k)}}_{B_k,E_k,\tilde{E}_k} \ket*{\Psi_{{\vb*{g}}}^{\overrightarrow{k-1}}}}{P(o_{{\overrightarrow{k-1}}})},
       \label{eq:measurement_app}
   \end{align}}{where} {we have omitted the mode subscripts in the quantum states for simplicity of notation. This concludes the derivation of \cref{eq:ph_actual}.} 
   
    {Next, we prove that in this case, the operator $\hat{M}_{\gamma_X}^{{(k)}}$ in \cref{eq:measurement_app} satisfies $0 \leq \hat{M}^{{(k)}}_{\gamma_X} \leq {\hat{\mathds{1}}}$. For this, first note that $\hat{M}_{\gamma_X}^{{(k)}} \geq 0$ {holds because for any state $\ket{\varphi}$ we have that $\bra{\varphi}\hat{M}_{\gamma_X}^{{(k)}} \ket{\varphi} = \big|\big| \sqrt{\hat{m}_{\gamma_X}} \hat{U}^{{(k)}}_{B_k,E_k,\tilde{E}_k} \ket*{\Psi_{{\vb*{g}}}^{\overrightarrow{k-1}}} \ket{\varphi} {P(o_{\overrightarrow{k-1}})^{-1/2}} \big| \big|^2 \geq 0$. Then, since} $\hat{m}_{\gamma_X}$ is a POVM element, {we have that} $\hat{m}_{\gamma_X}\leq {\hat{\mathds{1}}}$, {and thus} we can upper bound $\hat{M}_{\gamma_X}^{{(k)}}$ as }
    
    {\begin{align}
        \hat{M}_{\gamma_X}^{{(k)}} &\leq \frac{\bra*{\Psi_{{\vb*{g}}}^{\overrightarrow{k-1}}} \hat{U}^{{{(k)}}\dagger}_{B_k,E_k,\tilde{E}_k} {\hat{\mathds{1}}} \hat{U}^{{(k)}}_{B_k,E_k,\tilde{E}_k} \ket*{\Psi_{{\vb*{g}}}^{\overrightarrow{k-1}}}}{P(o_{{\overrightarrow{k-1}}})} \nonumber \\ 
        &= \frac{\left|\left| \ket*{\Psi_{{\vb*{g}}}^{\overrightarrow{k-1}}} \right|\right|^2 {\hat{\mathds{1}}}_{{B_k,E_k}}}{P(o_{{\overrightarrow{k-1}}})} = \frac{P(o_{{\overrightarrow{k-1}}}) {\hat{\mathds{1}}}_{{B_k,E_k}}}{P(o_{{\overrightarrow{k-1}}})} = {\hat{\mathds{1}}}_{{B_k,E_k}}.
    \end{align}}
    {Therefore, by combining these two arguments, we have that, {when the} QKD protocol {is run sequentially}, $0 \leq \hat{M}^{{(k)}}_{\gamma_X} \leq {\hat{\mathds{1}}}$, as required.} 

	
	\section{{Results used in \cref{sec:proof_gk_dependent}}}
	\label{app:Eq1_derivation}
	
	\subsection{{Derivation of ${\ket*{\tilde{\psi}_{j_k,\vb*{g}|j_{k-1},\hdots,j_{k-l_c}}}_{B_k,\vb*{E}'_k}}$ and \cref{eq:entangled_dep}}}
	\label{app:derivations_corr}
	
	{Here, we show how to derive { $\ket*{\tilde{\psi}_{j_k,\vb*{g}|j_{k-1},\hdots,j_{k-l_c}}}_{B_k,\vb*{E}'_k}$} and \cref{eq:entangled_dep}, which are required to prove the security of the modified BB84 protocol in the presence of setting-dependent pulse correlations. To do so, we consider an entanglement-based picture of the protocol.} {In this case, for a fixed ${\vb*{g}}$ {(see \cref{app:proof_fixed_g})}, the transmission of $N$ pulses is described by Alice first preparing $N$ ancilla systems and $N$ pulses in the state} 
	{\begin{align}
    &\ket{\Psi_{{\vb*{g}}}}_{\textit{\textbf{A}},\textit{\textbf{B}},\textit{\textbf{E}}} = \sum_{j_1} {\sqrt{p_{j_1}}} \ket*{j_1}_{A_1} \ket{\psi_{j_1,{\vb*{g}}}}_{B_1,E_1} \nonumber \\ &\otimes \sum_{j_2} {\sqrt{p_{j_2}}} \ket{j_2}_{A_2} \ket*{\psi_{j_2,{\vb*{g}}|j_1}}_{B_2,E_2} {\otimes} \hdots \nonumber \\
    &\otimes \sum_{j_N} {\sqrt{p_{j_N}}} \ket{j_N}_{A_N} \ket*{\psi_{j_N,{\vb*{g}}|j_{N-1},\hdots,j_{N-l_c}}}_{B_N,E_N}, 
    \label{eq:full_entangled_state}
        \end{align}}{and then sending systems ${\textit{\textbf{B}},\textit{\textbf{E}}}$ through the quantum channel. {Here, ${\textit{\textbf{A}}:=} A_1,\hdots,A_N$ and ${\textit{\textbf{B}},\textit{\textbf{E}}:=} B_1,E_1,\hdots,B_N,E_N$ refers to the composite systems of Alice's ancillae and to the pulses sent to Bob, respectively, where $A_k$ and $B_k$ for $k \in \{1,2,\hdots,N\}$ denote Alice's and Bob's $k^{\rm th}$ systems, and $E_k$ denotes any other systems sent by Alice through the quantum channel.} In Eq.~(\ref{eq:full_entangled_state}), $\{\ket*{j_k}_{A_k}\}_{j_k \in \{0_Z,1_Z,0_X,1_X\}}$ is a set of {orthonormal} states and $\ket*{\psi_{j_k,{{\vb*{g}}}|j_{k-1},\hdots,j_{k-l_c}}}_{B_k,E_k}$ {is defined in \cref{eq:general_set_dep_corr}.}} {Note that, for ease of discussion, in this appendix we explicitly write the setting choice on the $k^{\rm th}$ round as $j_k$, rather than $j$. }

    {We are interested in the post-measurement state after Alice has measured her ancillae $\textit{\textbf{A}}_{\leq k-1} := A_1,\hdots,A_{k-1}$. Once we trace out the systems ${\textit{\textbf{A}}}_{\leq k-1},{\textit{\textbf{B}}}_{\leq k-1},{\textit{\textbf{E}}}_{\leq k-1}$, this post-measurement state can be expressed as}  
    {\begin{widetext}\begin{align}
    & \scalemath{0.98}{{\ket{\Psi_{{\vb*{g}}|j_{k-1},\hdots,j_{k-l_c}}}_{A_k,B_k,{\textit{\textbf{E}}}'_k}}}\nonumber \\
    &\scalemath{0.98}{:=\sum_{j_k} {\sqrt{p_{j_k}}} \ket{j_k}_{A_k} \ket*{\psi_{j_k,{\vb*{g}}|j_{k-1},\hdots,j_{k-l_c}}}_{B_k,E_k} \bigotimes^{l_c}_{n=1} \sum_{j_{k+n}} {\sqrt{p_{j_{k+n}}}} \ket*{j_{k+n}}_{A_{k+n}} \ket*{\psi_{j_{k+n},{\vb*{g}}|j_{k+n-1},\hdots,j_{k+n-l_c}}}_{B_{k+n},E_{k+n}}} \nonumber \\
    &\scalemath{0.98}{\otimes \sum_{j_{k+l_c+1}} {\sqrt{p_{j_{k+l_c+1}}}} \ket*{j_{k+l_c+1}}_{A_{k+l_c+1}} \ket*{\psi_{j_{k+l_c+1},{\vb*{g}}|j_{k+l_c},\hdots,j_{k+1}}}_{B_{k+l_c+1},E_{k+l_c+1}} \hdots \sum_{j_N} {\sqrt{p_{j_{N}}}}  \ket*{j_N}_{A_N} \ket*{\psi_{j_N,{\vb*{g}}|j_{N-1},\hdots,j_{N-l_c}}}_{B_N,E_N},}
    \label{eq:after_meas}
    \end{align}\end{widetext}
    {where $j_{k-1},\hdots,j_{k-l_c}$ denotes Alice's measurement outcomes on systems $A_{k-1},\hdots,A_{k-l_c}${,} and ${\textit{\textbf{E}}}'_k := {\textit{\textbf{A}}}_{\geq k+1},{\textit{\textbf{B}}}_{\geq k+1},{\textit{\textbf{E}}}_{\geq k}$ with ${\textit{\textbf{A}}}_{\geq k+1}$ and ${\textit{\textbf{B}}}_{\geq k+1},{\textit{\textbf{E}}}_{\geq k}$ defined as $A_{k+1},\hdots,A_N$ and $E_k, B_{k+1}, E_{k+1}, \hdots, B_{N}, E_{N}$, respectively. Note that, to derive \cref{eq:after_meas}, we need to assume that the outcomes $j_{k-1},\hdots,j_{k-l_c} \in \{0_Z,1_Z,0_X, 1_X\}$, i.e.\ that Alice has measured her ancillae $A_{k-1},\hdots,A_{k-l_c}$ in the basis $\{\ket{0_Z},\ket{1_Z},\ket{0_X},\ket{1_X}\}$.  This has important consequences when using the complementary approach to prove the security of the protocol; see discussion after \cref{eq:entangled_dep} in the main text. 
    }}
    
     {In Eq.~(\ref{eq:after_meas}), one can {explicitly} see that some information about the $k^{\rm th}$ {setting choice} {is} leaked to the subsequent {$l_c$} pulses. However, the pulses $k+l_c+1$ to $N$ are independent of $j_k$.} 
    {To simplify the {description} in Eq.~(\ref{eq:after_meas}), we now introduce two definitions. First, we define a state that is independent of $j_k$ as}
	{\begin{align}
	    &\scalemath{0.98}{\ket*{\psi_{j_{k+1},\hdots,j_{k+l_c},{\vb*{g}}}}_{{\textit{\textbf{A}}}_{\ge k+1},{\textit{\textbf{B}}}_{\ge k+l_c+1},{\textit{\textbf{E}}}_{\ge k+l_c+1}}} \nonumber \\
	    &\scalemath{0.98}{:= \ket{j_{k+1}}_{A_{k+1}} \hdots \ket*{j_{k+l_c}}_{A_{k+l_c}}} \sum_{j_{k+l_c+1}} {\sqrt{p_{j_{k+l_c+1}}}} \nonumber \\ 
	    &\scalemath{0.98}{  \ket*{j_{k+l_c+1}}_{A_{k+l_c+1}} \ket*{\psi_{j_{k+l_c+1},{\vb*{g}}|j_{k+l_c},\hdots,j_{k+1}}}_{B_{k+l_c+1},E_{k+l_c+1}} \hdots} \nonumber \\
	    &\scalemath{0.98}{\sum_{j_N} {\sqrt{p_{j_{N}}}} \ket{j_N}_{A_N} \ket*{\psi_{j_N,{\vb*{g}}|j_{N-1},\hdots,j_{N-l_c}}}_{B_N,E_N},}
	    \label{eq:def1}
	\end{align}}{\noindent which forms a set of orthogonal {states} as $\big\{\ket*{\psi_{j_{k+1},\hdots,j_{k+l_c},{\vb*{g}}}}_{{\textit{\textbf{A}}}_{\ge k+1},{\textit{\textbf{B}}}_{\ge k+l_c+1},{\textit{\textbf{E}}}_{\ge k+l_c+1}}\big\}_{j_{k+1},\hdots,j_{k+l_c} {\in} \{0_Z,}$ ${}_{1_Z,0_X,1_X\}}$. Next, using Eq.~(\ref{eq:def1}), we define the following state}
	{\begin{align}
	    &\ket*{\lambda_{j_{k},{\hdots, j_{k+1-l_c},\vb*{g}}}}_{{\textit{\textbf{A}}}_{\ge k+1},{\textit{\textbf{B}}}_{\ge k+1},{\textit{\textbf{E}}}_{\ge k+1}} := \sum_{j_{k+1}} \sqrt{p_{j_{k+1}}} \hdots  \nonumber \\ &\sum_{j_{k+l_c}} \sqrt{p_{j_{k+l_c}}} \ket*{\psi_{j_{k+1},\hdots,j_{k+l_c},{\vb*{g}}}}_{{\textit{\textbf{A}}}_{\ge k+1},{\textit{\textbf{B}}}_{\ge k+l_c+1},{\textit{\textbf{E}}}_{\ge k+l_c+1}} \nonumber \\ 
	    &\bigotimes^{l_c}_{n=1} \ket*{\psi_{j_{k+n},{\vb*{g}}|j_{k+n-1},\hdots,j_{k+n-l_c}}}_{B_{k+n},E_{k+n}}.
	    \label{eq:def2}
	\end{align}}{\noindent Note that $\ket{\lambda_{j_{k},{\hdots, j_{k+1-l_c},\vb*{g}}}}_{{\textit{\textbf{A}}}_{\ge k+1},{\textit{\textbf{B}}}_{\ge k+1},{\textit{\textbf{E}}}_{\ge k+1}}$ depends on $j_k$ even though it does not include system $B_k$. This means that it is effectively a side channel to the $k^{\rm th}$ pulse. Finally, using Eq.~(\ref{eq:def2}), we can rewrite Eq.~(\ref{eq:after_meas}) as}
	{\begin{align}
    &\scalemath{0.97}{\ket{\Psi_{{\vb*{g}}|j_{k-1},\hdots,j_{k-l_c}}}_{A_k,B_k,{\textit{\textbf{E}}}'_k} =  \sum_{j_k} {\sqrt{p_{j_{k}}}} \ket{j_k}_{A_k}} \nonumber \\
    & \scalemath{0.97}{\ket*{\psi_{j_k,{\vb*{g}}|j_{k-1},\hdots,j_{k-l_c}}}_{B_k,E_k} \ket{\lambda_{j_{k},{\hdots, j_{k+1-l_c},\vb*{g}}}}_{{\textit{\textbf{A}}}_{\geq k+1},{\textit{\textbf{B}}}_{\geq k+1},{\textit{\textbf{E}}}_{\geq k+1}.}}
    \label{eq:protocol2}
    \end{align}}

    {Now, it is useful to {decompose} $\ket*{\psi_{j_k,{\vb*{g}}|j_{k-1},\hdots,j_{k-l_c}}}_{B_k,E_k}$ $\ket*{\lambda_{j_{k},{\hdots, j_{k+1-l_c},\vb*{g}}}}_{{\textit{\textbf{A}}}_{\geq k+1},{\textit{\textbf{B}}}_{\ge k+1},{\textit{\textbf{E}}}_{\ge k+1}}$ in \cref{eq:protocol2}.}
   {{For this, we {first combine \cref{eq:protocol2} and \cref{eq:def2}, and then} use Eq.~(\ref{eq:general_set_dep_corr}) such that }}
	{\begin{widetext}\begin{align}
	&\scalemath{0.99}{\ket*{\psi_{j_k,{\vb*{g}}|j_{k-1},\hdots,j_{k-l_c}}}_{B_k,E_k} \sum_{j_{k+1}} \sqrt{p_{j_{k+1}}} \hdots \sum_{j_{k+l_c}} \sqrt{p_{j_{k+l_c}}} \ket*{\psi_{j_{k+1},\hdots,j_{k+l_c},{\vb*{g}}}}_{{\textit{\textbf{A}}}_{\geq k+1},{\textit{\textbf{B}}}_{\geq k+l_c+1},{\textit{\textbf{E}}}_{\geq k+l_c+1}}} \nonumber \\
	&\scalemath{0.99}{\bigotimes^{l_c}_{n=1} \ket*{\psi_{j_{k+n},{\vb*{g}}|j_{k+n-1},\hdots,j_{k+n-l_c}}}_{B_{k+n},E_{k+n}}} \nonumber \\
	& \scalemath{0.99}{= \left(\sqrt{1- {\epsilon^{{(k)}}_{j_k,{\vb*{g}}|j_{k-1},\hdots,j_{k-l_c}}}}\ket{\phi_{j_k,{\vb*{g}}}}_{B_k,E_k} + \sqrt{{\epsilon^{{(k)}}_{j_k,{\vb*{g}}|j_{k-1},\hdots,j_{k-l_c}}}} \ket*{\phi^\perp_{j_k,{\vb*{g}}|j_{k-1},\hdots,j_{k-l_c}}}_{B_k,E_k}\right)} \nonumber \\
	&\scalemath{0.99}{\sum_{j_{k+1}} \sqrt{p_{j_{k+1}}} \hdots \sum_{j_{k+l_c}} \sqrt{p_{j_{k+l_c}}} \ket*{\psi_{j_{k+1},\hdots,j_{k+l_c},{\vb*{g}}}}_{{\textit{\textbf{A}}}_{\geq k+1},{\textit{\textbf{B}}}_{\geq k+l_c+1},{\textit{\textbf{E}}}_{\geq k+l_c+1}}}  \nonumber \\
	& \scalemath{0.99}{\bigotimes^{l_c}_{n=1} \Big(\sqrt{1- {\epsilon^{{(k)}}_{j_{k+n},{\vb*{g}}|j_{k+n-1},\hdots,j_{k+n-l_c}}}}\ket*{\phi_{j_{k+n},{\vb*{g}}}}_{B_{k+n},E_{k+n}} + \sqrt{{\epsilon^{{(k)}}_{j_{k+n},{\vb*{g}}|j_{k+n-1},\hdots,j_{k+n-l_c}}}} \ket*{\phi^\perp_{j_{k+n},{\vb*{g}}|j_{k+n-1},\hdots,j_{k+n-l_c}}}_{B_{k+n},E_{k+n}}\Big)} \nonumber \\
	&\scalemath{0.99}{=: \sqrt{1- {\tilde{\epsilon}^{{(k)}}_{j_k,{\vb*{g}}|j_{k-1},\hdots,j_{k-l_c}}}} \ket*{\tilde{\phi}_{j_k,{\vb*{g}|j_{k-1},\hdots,j_{k+1-l_c}}}}_{{B_k,{\textit{\textbf{E}}}'_k}} + \sqrt{{\tilde{\epsilon}^{{(k)}}_{j_k,{\vb*{g}}|j_{k-1},\hdots,j_{k-l_c}}}} \ket*{\tilde{\phi}^{\perp}_{j_k,{\vb*{g}}|j_{k-1},\hdots,j_{k-l_c}}}_{{B_k{\textit{\textbf{E}}}'_k}}} \nonumber \\
	&\scalemath{0.99}{{=: \ket*{\tilde{\psi}_{j_k,{\vb*{g}}|j_{k-1},\hdots,j_{k-{l_c}}}}_{B_k{\textit{\textbf{E}}}'_k}.}}
	\label{eq:full_derivation_last_state}
	\end{align}\end{widetext}
	{In this equation,} we have defined
	\begin{align}
	&\tilde{\epsilon}^{{(k)}}_{j_k,{\vb*{g}}|j_{k-1},\hdots,j_{k-l_c}} := 1 - (1 - \epsilon^{{(k)}}_{j_k,{\vb*{g}}|j_{k-1},\hdots,j_{k-l_c}}) \nonumber \\
	&\prod^{l_c}_{n=1} \sum_{j_{k+n}} p_{j_{k+n}} (1- \epsilon^{{(k)}}_{j_{k+n},{\vb*{g}}|j_{k+n-1},\hdots,j_{k+n-l_c}}),
	\label{eq:epsilon_def}
	\end{align}}{and have used the fact that $\braket{j_k}{j'_k}_{A_k} = \delta_{j_k,j'_k}$. {Moreover}, the normalised state $\ket*{\tilde{\phi}_{j_k,{\vb*{g}|j_{k-1},\hdots,j_{k+1-l_c}}}}_{{B_k,{\textit{\textbf{E}}}'_k}}$ is expressed as 
	\begin{widetext}\begin{align}
	& \scalemath{0.85}{\frac{\ket{\phi_{j_k,{\vb*{g}}}}_{B_k,E_k} \sum_{j_{k+1}} \sqrt{p_{j_{k+1}}} \hdots \sum_{j_{k+l_c}} \sqrt{p_{j_{k+l_c}}} \ket*{\psi_{j_{k+1},\hdots,j_{k+l_c},{\vb*{g}}}}_{{\textit{\textbf{A}}}_{\geq k+1},{\textit{\textbf{B}}}_{\geq k+l_c+1},{\textit{\textbf{E}}}_{\geq k+l_c+1}}\bigotimes^{l_c}_{n=1} \sqrt{1- {\epsilon^{{(k)}}_{j_{k+n},{\vb*{g}}|j_{k+n-1},\hdots,j_{k+n-l_c}}}} \ket*{\phi_{j_{k+n},{\vb*{g}}}}_{B_{k+n},E_{k+n}}}{{\prod^{l_c}_{n=1} \sqrt{\sum_{j_{k+n}} p_{j_{k+n}} (1- \epsilon^{{(k)}}_{j_{k+n},{\vb*{g}}|j_{k+n-1},\hdots,j_{k+n-l_c}})}}}} \nonumber \\
	& \scalemath{0.85}{=: \ket{\phi_{j_k,{\vb*{g}}}}_{B_k,E_k} \ket{\Lambda_{{j_k},{\hdots,j_{k+1-l_c},\vb*{g}}}}_{{\textit{\textbf{A}}}_{\geq k+1},{\textit{\textbf{B}}}_{\geq k+1},{\textit{\textbf{E}}}_{\geq k+1}} =: \ket*{\tilde{\phi}_{j_k,{\vb*{g}|j_{k-1},\hdots,j_{k+1-l_c}}}}_{{B_k,{\textit{\textbf{E}}}'_k}},}
	\label{eq:state_new_def}
	\end{align}\end{widetext}}
%
%
\noindent {and $\ket*{\tilde{\phi}^\perp_{j_k,{\vb*{g}}|j_{k-1},\hdots,j_{k-l_c}}}_{{B_k,{\textit{\textbf{E}}}'_k}}$ is a state orthogonal to $\ket*{\tilde{\phi}_{j_k,{\vb*{g}|j_{k-1},\hdots,j_{k+1-l_c}}}}_{{B_k,{\textit{\textbf{E}}}'_k}}$, living in a Hilbert space of any dimension, and whose explicit form is omitted {here} for simplicity.}
%
{Importantly, substituting \cref{eq:full_derivation_last_state} into \cref{eq:protocol2} we obtain
\begin{align}
		&\ket*{\Psi_{{\vb*{g}}|j_{k-1},\hdots,j_{k-l_c}}}_{A_k,B_k,{\textit{\textbf{E}}}'_k} \nonumber \\
		&= \sum_{j_k} \sqrt{p_{j_k}} \ket{j_k}_{A_k} \ket*{\tilde{\psi}_{j_k,{\vb*{g}}|j_{k-1},\hdots,j_{k-l_c}}}_{B_k,{\textit{\textbf{E}}}'_k},
		\label{eq:entangled_dep_gen_app}
	\end{align} 
{which} implies that we can regard the emitted states on round $k$ to be $\ket*{\tilde{\psi}_{j_k,{\vb*{g}}|j_{k-1},\hdots,j_{k-l_c}}}_{B_k,{\textit{\textbf{E}}}'_k}$.}

{Now, suppose that Alice performs a measurement on system $A_k$ such that she learns if $j_{k} \in \{0_Z, 1_Z\}$ or if $j_{k} \in \{0_X, 1_X\}$, i.e.\ her choice of basis for the $k^{\rm th}$ round. If she obtains the former outcome, her post-measurement state can be expressed as 
\begin{align}
		&\ket*{\Psi^Z_{{\vb*{g}}|j_{k-1},\hdots,j_{k-l_c}}}_{A_k,B_k,{\textit{\textbf{E}}}'_k} \nonumber \\
		&= \frac{1}{\sqrt{2}} \sum_{\alpha \in \{0,1\}} \ket{\alpha_Z}_{A_k} \ket*{{\tilde{\psi}}_{\alpha_Z,{\vb*{g}}|j_{k-1},\hdots,j_{k-l_c}}}_{B_k,{\textit{\textbf{E}}}'_k},
		\label{eq:entangled_dep_app}
	\end{align}
	where we have used $p_{0_Z} = p_{1_Z}$ (see assumption (A1) in \cref{sec:assumptions_alice}). This concludes the derivation of \cref{eq:entangled_dep}.
}

\subsection{{Proof of \cref{eq:ph_actual_dep}}}
\label{app:proof_measurement_d}
    {Here, we prove \cref{eq:ph_actual_dep} and show that {when the} QKD {protocol is run sequentially (see assumption (A4) \cref{sec:assumptions_alice})}, the operator $\hat{D}^{(k)}_{\gamma_X}$ satisfies $0 \leq \hat{D}^{(k)}_{\gamma_X} \leq {\hat{\mathds{1}}}$. To do so, we consider again an entanglement-based scenario that is equivalent to the actual protocol. For a fixed $\vb*{g}$, the transmission of $N$ pulses can then be described by Alice first preparing the entangled state $\ket{\Psi_{\vb*{g}}}_{\textit{\textbf{A}},\textit{\textbf{B}},\textit{\textbf{E}}}$ in \cref{eq:full_entangled_state}, keeping system $\textit{\textbf{A}}$ in her laboratory and sending systems $\textit{\textbf{B}},\textit{\textbf{E}}$ through the quantum channel. In our analysis we assume that the QKD protocol is run {sequentially}. As explained in \cref{app:proof_measurement}, under this restriction, Eve's most general attack is described by $\hat{U}_{B_N,E_N,\tilde{E}_N}^{(N)} \hdots \hat{U}_{B_1,E_1,\tilde{E}_1}^{(1)}$, where $\hat{U}_{B_k,E_k,\tilde{E}_k}^{(k)}$ acts on the $k^{\rm th}$ photonic system $B_k,E_k$ and on Eve's updated ancilla $\tilde{E}_k$, resulting in systems $B'_k$ and $\tilde{E}_k$ (see \cref{fig:eve_attack_slow} for more details). After Eve's attack on each round, Alice and Bob measure their local systems to obtain the experimental data. The secret key is generated from the rounds in which both have selected the $Z$ basis, i.e.~$(a,\beta) = (Z,Z)$, where $a$ ($\beta$) denotes Alice's (Bob's) basis selection.}
    
    {As before, to prove the security of these rounds, one needs to consider a virtual protocol to estimate the number of phase errors that Alice and Bob would have obtained if they had performed their local measurements in the phase basis instead, i.e.~$\{\ket*{\alpha_X^{\rm (vir)}}_{A_k} = (\ket{0_Z}_{A_k} + (-1)^\alpha \ket{1_Z}_{A_k})/\sqrt{2}\}_{\alpha \in \{0,1\}}$ and $\{\hat{m}_{0_X},\hat{m}_{1_X},\hat{m}_{f}\}$, respectively. Unlike in \cref{app:proof_measurement}, however, each emitted pulse $k$ now depends on the previous $l_c$ setting choices $\{j_{k-1},\hdots,j_{k-l_c}\}$. As explained in \cref{sec:proof_gk_dependent}, this dependence does not allow us to naively consider a virtual protocol in which Alice and Bob perform phase basis measurements on all key generation rounds. Instead, we consider that Alice and Bob assign a tag $w \in \{0,\hdots,l_c\}$ to each round $k$ according to the value $w = k \bmod (l_c +1)$ and construct $(l_c +1)$ virtual protocols whose respective $w^{\rm th}$ sifted keys are subsets of the total sifted key originating from the rounds with a tag $w$. In each $w^{\rm th}$ virtual protocol, Alice's and Bob's measurements on the $w$-tagged rounds are described by the POVM $\{\hat{F}_{A_k,B_k'}^{o_k}\}_{{o_k}}$, whose elements are defined in \cref{eq:combined_measurements}, and on the other rounds, Alice's and Bob's measurements are described by the POVM $\{\hat{J}_{A_k,B_k'}^{o_k}\}_{{o_k}}$, whose elements are defined as $\hat{J}_{A_k,B_k'}^{(a,\beta,\alpha,\gamma)} = \dyad{\alpha_a}_{A_k} \otimes p_{\beta_B} \hat{m}_{\gamma_\beta}$ and $\hat{J}_{A_k,B'_k}^{(f)} = {\hat{\mathds{1}}}_{A_k} \otimes \hat{m}_{f}$, where $\alpha,\gamma \in \{0,1\}$ are Alice's and Bob's observed bit value respectively, and $f$ is associated with an inconclusive outcome. Note that each of the these virtual protocols is indistinguishable from the actual protocol.}
    
    {The probability that, in the $w^{\rm th}$ virtual protocol, a phase error is obtained on some round $k$ with a tag $w$, conditioned on all the previous outcomes of the $w^{\rm th}$ virtual protocol, can then be expressed as \cref{eq:phase_error_prob_appendix}, rewritten here for convenience: 
     \begin{align}
       \scalemath{0.98}{P^{(k)}_{\vb*{g}}({\rm ph|Act})}
       &\scalemath{0.98}{=\frac{P\big((o_{k} = (Z,Z,0,1) {~\rm or~} (Z,Z,1,0)) , o_{\overrightarrow{k-1}}\big)}{P\big(o_{\overrightarrow{k-1}}\big)},} 
        \label{eq:phase_error_prob_appendix_corr}
   \end{align} where $P\big(o_{\overrightarrow{k-1}}\big)$ is the probability of obtaining Alice's and Bob's previous outcomes. Let us now introduce the POVM $\{\hat{L}_{A_k,B_k'}^{o_k}\}_{{o_k}}$, which is defined as 
   $\{\hat{F}_{A_k,B_k'}^{o_k}\}_{{o_k}}$ when {a} pulse $k$ has {the} tag $w$ and defined as $\{\hat{J}_{A_k,B_k'}^{o_k}\}_{{o_k}}$ when it has {another} tag. Therefore, by using this POVM, the form of the emitted state in \cref{eq:full_entangled_state} as well as the description of Eve's coherent attack {when the} QKD protocol {is run sequentially}, we have that 
    \begin{widetext}\begin{align}
       &P\big((o_k = (Z,Z,0,1) ~{\rm or}~ (Z,Z,1,0)) , o_{\overrightarrow{k-1}}\big) \nonumber \\
       & = \sum_{\substack{\alpha, \gamma \in \{0,1\} \\ \alpha \neq \gamma}} {\rm Tr}~ \Bigg[\hat{F}_{A_k,B'_k}^{(Z,Z,\alpha,\gamma)} \hat{P}\left(\hat{U}^{(k)}_{B_k,E_k,\tilde{E}_k}\prod^{k-1}_{n=1} \sqrt{\hat{L}^{o_{k-n}}_{A_{k-n},B'_{k-n}}} \hat{U}^{(k-n)}_{B_{k-n},E_{k-n},\tilde{E}_{k-n}} \ket{0}_{\tilde{E}_1} \ket{\Psi_{\vb*{g}}}_{\textit{\textbf{A}},\textit{\textbf{B}},\textit{\textbf{E}}}   \right) \Bigg] \nonumber \\
       & = \sum_{\substack{\alpha, \gamma \in \{0,1\} \\ \alpha \neq \gamma}} \Tr [\hat{F}_{A_k,B'_k}^{(Z,Z,\alpha,\gamma)} \hat{P}\left(\hat{U}^{(k)}_{B_k,E_k,\tilde{E}_k} \Bigg[\prod^{k-1}_{n=1} \sqrt{\hat{L}^{o_{k-n}^B}_{B'_{k-n}}} \hat{U}^{(k-n)}_{B_{k-n},E_{k-n},\tilde{E}_{k-n}}\Bigg] \ket{0}_{\tilde{E}_1} \Bigg[\prod^{k-1}_{n=1} \sqrt{\hat{L}^{o_{k-n}^A}_{A_{k-n}}} \ket{\Psi_{\vb*{g}}}_{\textit{\textbf{A}},\textit{\textbf{B}},\textit{\textbf{E}}} \Bigg] \right)] \nonumber \\
        & = \sum_{\substack{\alpha, \gamma \in \{0,1\} \\ \alpha \neq \gamma}} \Tr \Bigg[\hat{F}_{A_k,B'_k}^{(Z,Z,\alpha,\gamma)} \hat{P}\Bigg(\hat{U}^{(k)}_{B_k,E_k,\tilde{E}_k} \Bigg[\prod^{k-1}_{n=1} \sqrt{\hat{L}^{o_{k-n}^B}_{B'_{k-n}}} \hat{U}^{(k-n)}_{B_{k-n},E_{k-n},\tilde{E}_{k-n}} \Bigg] \ket{0}_{\tilde{E}_1} \nonumber \\ 
        &\Bigg[\bigotimes_{n=1}^{k-1} \sqrt{p_{j_n}} \ket{j_n}_{A_n} \ket{\psi_{j_n,\vb*{g}|j_{n-1},\hdots,j_{n-l_c}}}_{B_{n}}\Bigg] \ket{\Psi_{\vb*{g}|j_{k-1},\hdots,j_{k-l_c}}}_{A_k,B_k,\textit{\textbf{E}}'_k}  \Bigg)\Bigg] \nonumber \\
       & = \sum_{\substack{\alpha, \gamma \in \{0,1\} \\ \alpha \neq \gamma}} \Tr \Bigg[{}_{\textit{\textbf{A}}_{\leq k-1},\textit{\textbf{B}}'_{\leq k-1},\textit{\textbf{E}}_{\leq k-1},\tilde{E}_k}\bra*{\Psi_{\vb*{g}}^{\overrightarrow{k-1}}} \hat{U}^{(k)\dagger}_{B_k,E_k,\tilde{E}_k} \hat{F}_{A_k,B'_k}^{(Z,Z,\alpha,\gamma)} \hat{U}^{(k)}_{B_k,E_k,\tilde{E}_k}  \ket*{\Psi_{\vb*{g}}^{\overrightarrow{k-1}}}_{\textit{\textbf{A}}_{\leq k-1},\textit{\textbf{B}}'_{\leq k-1},\textit{\textbf{E}}_{\leq k-1},\tilde{E}_k} \nonumber \\
       &\dyad*{\Psi_{\vb*{g}|j_{k-1},\hdots,j_{k-l_c}}}_{A_k,B_k,\textit{\textbf{E}}'_k} \Bigg] \nonumber \\
        & = \sum_{\substack{\alpha, \gamma \in \{0,1\} \\ \alpha \neq \gamma}} \Tr \Bigg[{}_{\textit{\textbf{A}}_{\leq k-1},\textit{\textbf{B}}'_{\leq k-1},\textit{\textbf{E}}_{\leq k-1},\tilde{E}_k}\bra*{\Psi_{\vb*{g}}^{\overrightarrow{k-1}}} \hat{U}^{(k)\dagger}_{B_k,E_k,\tilde{E}_k} \hat{m}_{\gamma_X} \hat{U}^{(k)}_{B_k,E_k,\tilde{E}_k} \ket*{\Psi_{\vb*{g}}^{\overrightarrow{k-1}}}_{\textit{\textbf{A}}_{\leq k-1},\textit{\textbf{B}}'_{\leq k-1},\textit{\textbf{E}}_{\leq k-1},\tilde{E}_k} \nonumber \\
        &~~~\tilde{p}^{(k,{\rm vir})}_{\alpha_X,\vb*{g}|j_{k-1},\hdots,j_{k-l_c}} p_{Z_B} \dyad*{\psi_{\alpha_X,\vb*{g}|j_{k-1},\hdots,j_{k-l_c}}^{\rm (vir)}}_{B_k,\vb*{E}'_k}\Bigg],
        \label{eq:prob_derivation_corr}
   \end{align}\end{widetext}where
   \begin{align}
       &\ket*{\Psi_{\vb*{g}}^{\overrightarrow{k-1}}}_{\textit{\textbf{A}}_{\leq k-1},\textit{\textbf{B}}'_{\leq k-1},\textit{\textbf{E}}_{\leq k-1},\tilde{E}_k} \nonumber \\
       &:= \Bigg[\prod^{k-1}_{n=1} \sqrt{\hat{L}^{o_{k-n}^B}_{B'_{k-n}}} \hat{U}^{(k-n)}_{B_{k-n},E_{k-n},\tilde{E}_{k-n}}\Bigg] \ket{0}_{\tilde{E}_1} \nonumber \\
       &\Bigg[ \bigotimes_{n=1}^{k-1} \sqrt{p_{j_n}} \ket{j_n}_{A_n} \ket{\psi_{j_n,\vb*{g}|j_{n-1},\hdots,j_{n-l_c}}}_{B_k}\Bigg],
       \label{eq:new_state_corr}
   \end{align}with $\textit{\textbf{A}}_{\leq k-1} := A_1, \hdots, A_{k-1}$ and $\textit{\textbf{B}}'_{\leq k-1},\textit{\textbf{E}}_{\leq k-1} := B'_1,E_1,\hdots, B'_{k-1},E_{k-1}$, and {where} $\ket*{\psi_{\alpha_X,\vb*{g}|j_{k-1},\hdots,j_{k-l_c}}^{\rm (vir)}}_{B_k,\vb*{E}'_k}$ and $\tilde{p}^{(k,{\rm vir})}_{\alpha_X,\vb*{g}|j_{k-1},\hdots,j_{k-l_c}}$ are defined in Eqs.~(\ref{eq:virtual_states_dep}) and (\ref{eq:vir_act_prob_dep}), respectively. In the second equality of \cref{eq:prob_derivation_corr}, we have used the fact that Alice's measurements on her ancillae $\textit{\textbf{A}}_{\leq k-1}$ commute with any measurements or operations on systems $B_1,E_1,\tilde{E}_1, \hdots,B_{k-1},E_{k-1},\tilde{E}_{k-1}$, and the fact that $\hat{L}^{o_{k}}_{A_{k},B'_{k}}$ can be decomposed as $\hat{L}^{o_{k}^A}_{A_{k}} \otimes \hat{L}^{o_{k}^B}_{B'_{k}}$, where Alice's outcome $o_{k}^A$ and Bob's outcome $o_{k}^B$ on round $k$ are defined as $(a,\alpha)$ and $(\beta,\gamma)$, respectively. In the third equality of \cref{eq:prob_derivation_corr}, we have used the definition of the post-measurement state $\ket*{\Psi_{\vb*{g}|j_{k-1},\hdots,j_{k-l_c}}}_{A_k,B_k,\textit{\textbf{E}}'_k}$ in \cref{eq:after_meas}, and in the fourth equality of \cref{eq:prob_derivation_corr}, we have defined the state $\ket*{\Psi_{\vb*{g}}^{\overrightarrow{k-1}}}_{\textit{\textbf{A}}_{\leq k-1},\textit{\textbf{B}}'_{\leq k-1},\textit{\textbf{E}}_{\leq k-1},\tilde{E}_k}$ using \cref{eq:new_state_corr} and {we have taken advantage of} the cyclic property of the trace operation. Finally, in the last equality of \cref{eq:prob_derivation_corr}, we have used \cref{eq:entangled_dep_gen_app}, \cref{eq:entangled_dep} and the definition of $\hat{F}_{A_k,B'_k}^{(Z,Z,\alpha,\gamma)}$ in \cref{eq:combined_measurements} for $\alpha,\gamma \in \{0,1\}$ such that $\alpha \neq \gamma$.}
  
 {By substituting \cref{eq:prob_derivation_corr} in \cref{eq:phase_error_prob_appendix_corr}, we can now express $P^{(k)}_{\vb*{g}}({\rm ph|Act})$ as
    \begin{align}
        &P^{(k)}_{\vb*{g}}({\rm ph|Act}) \coloneqq \sum_{\substack{\alpha, \gamma \in \{0,1\} \\ \alpha \neq \gamma}} \tilde{p}_{\alpha_X,\vb*{g}|j_{k-1},\hdots,j_{k-l_c}}^{(k,{\rm vir})} p_{Z_B} \nonumber \\
		&{\rm Tr}\big[\tilde{\sigma}_{\alpha_X,\vb*{g}|j_{k-1},\hdots,j_{k-l_c}}^{(k,{\rm vir})}{\hat D_{\gamma_X}^{(k)}}\big], 
   \end{align}with $\tilde{\sigma}_{\alpha_X,\vb*{g}|j_{k-1},\hdots,j_{k-l_c}}^{(k,{\rm vir})} \coloneqq \hat{P}\big(\ket*{\psi_{\alpha_X,\vb*{g}|j_{k-1},\hdots,j_{k-l_c}}^{\rm (vir)}}_{B_k,\textit{\textbf{E}}'_k}\big)$ and
    \begin{align}
       \hat{D}_{\gamma_X}^{(k)} := \frac{\bra*{\Psi_{\vb*{g}}^{\overrightarrow{k-1}}}\hat{U}^{(k)\dagger}_{B_k,E_k,\tilde{E}_k} \hat{m}_{\gamma_X} \hat{U}^{(k)}_{B_k,E_k,\tilde{E}_k} \ket*{\Psi_{\vb*{g}}^{\overrightarrow{k-1}}}}{P(o_{\overrightarrow{k-1}})},
       \label{eq:measurement_app_d}
   \end{align}where $P(o_{\overrightarrow{k-1}}) = \big|\big| \ket*{\Psi_{\vb*{g}}^{\overrightarrow{k-1}}} \big|\big|^2$, and where we have omitted the mode subscripts in the quantum states for simplicity of notation. This concludes the derivation of \cref{eq:ph_actual_dep}.}
   
   {Finally, note that the operator $\hat{D}_{\gamma_X}^{(k)}$ in  \cref{eq:measurement_app_d} has a very similar form to the operator $\hat{M}_{\gamma_X}^{(k)}$ in \cref{eq:measurement_app}. Therefore, we can use an analogous argument to that in \cref{app:proof_measurement} to show that {when the} QKD {protocol is run sequentially}, the operator $\hat{D}_{\gamma_X}^{(k)}$ satisfies $0 \leq \hat{D}^{(k)}_{\gamma_X} \leq {\hat{\mathds{1}}}$.}

\subsection{{Upper bounds on $\big|\braket*{\tilde{\psi}_{j_k,{\vb*{g}}|j_{k-1},\hdots,j_{k-l_c}}}{\phi_{j_k,{\vb*{g}}}}{}_{B_k,{\textit{\textbf{E}}}'_k}\big|$ and $\big|\braket*{\Psi_{{\vb*{g}}|j_{k-1},\hdots,j_{k-l_c}}}{\Phi_{{\vb*{g}}}}{}_{A_k,B_k,{\textit{\textbf{E}}}'_k}\big|$}}
\label{app:inner_prod_corr}

{When employing the RT to prove the security of the protocol, we define the reference states}
{\begin{align}
    &\ket*{\phi_{j_k,{\vb*{g}}}}_{B_k,{\textit{\textbf{E}}}'_k} = \ket*{\phi_{j_k,{\vb*{g}}}}_{B_k,E_k} \sum_{j_{k+1}} \sqrt{p_{j_{k+1}}} \hdots \sum_{j_{k+l_c}} \sqrt{p_{j_{k+l_c}}} \nonumber \\
    &\ket*{\psi_{j_{k+1},\hdots,j_{k+l_c},{\vb*{g}}}}_{{\textit{\textbf{A}}}_{\geq k+1},{\textit{\textbf{B}}}_{\geq k+l_c+1},{\textit{\textbf{E}}}_{\geq k+l_c+1}} \nonumber \\ &\bigotimes^{l_c}_{n=1}  \ket*{\phi_{j_{k+n},{\vb*{g}}}}_{B_{k+n},E_{k+n}} \nonumber \\
    &{=: \ket{\phi_{j_k,{\vb*{g}}}}_{B_k,E_k} \ket*{{\Lambda}_{{\vb*{g}}}}_{{\textit{\textbf{A}}}_{\geq k+1},{\textit{\textbf{B}}}_{\geq k+1},{\textit{\textbf{E}}}_{\geq k+1}},}
    \label{eq:reference_new}
\end{align}and
\begin{align}
    \ket*{\Phi^Z_{{\vb*{g}}}}_{A_k,B_k,{\textit{\textbf{E}}}'_k} = \frac{1}{\sqrt{2}} \sum_{\alpha \in \{0,1\}} \ket{\alpha_Z}_{A_k} \ket{\phi_{\alpha_Z,{\vb*{g}}}}_{B_k,{\textit{\textbf{E}}}'_k},
   \label{eq:entangled_dep_ref}
\end{align} which are analogous to $\ket*{\tilde{\psi}_{j_k,{\vb*{g}}|j_{k-1},\hdots,j_{k-l_c}}}_{B_k,{\textit{\textbf{E}}}'_k}$ and $\ket*{\Psi^Z_{{\vb*{g}}{|j_{k-1},\hdots,j_{k-l_c}}}}_{A_k,B_k,{\textit{\textbf{E}}}'_k}$, respectively. Note that, unlike $\ket*{\tilde{\phi}_{j_k,{\vb*{g}|j_{k-1},\hdots,j_{k+1-l_c}}}}_{{B_k,{\textit{\textbf{E}}}'_k}}$ in \cref{eq:state_new_def}, $\ket*{\phi_{j_k,{\vb*{g}}}}_{B_k, {\textit{\textbf{E}}}_k'}$ in \cref{eq:reference_new} is a qubit state, i.e.~$\{\ket{\phi_{j_k,{\vb*{g}}}}_{B_k, {\textit{\textbf{E}}}'_k}\}_{j_k=0_Z,1_Z,0_X,1_X}$ spans a qubit space, because $\ket{{\Lambda}_{{\vb*{g}}}}_{{\textit{\textbf{A}}}_{\geq k+1},{\textit{\textbf{B}}}_{\geq k+1},{\textit{\textbf{E}}}_{\geq k+1}}$ is a state independent of $j_k$.}

{Using Eqs.~(\ref{eq:full_derivation_last_state}) and (\ref{eq:reference_new}), we can then calculate an upper bound on the inner product $\big|\braket*{{\tilde{\psi}}_{j_k,{\vb*{g}}|j_{k-1},\hdots,j_{k-l_c}}}{\phi_{j_k,{\vb*{g}}}}{}_{B_k,{\textit{\textbf{E}}}'_k}\big|$ such that}
{\begin{widetext}\begin{align}
    &\scalemath{0.99}{\left|\braket*{{\tilde{\psi}}_{j_k,{\vb*{g}}|j_{k-1},\hdots,j_{k-l_c}}}{\phi_{j_k,{\vb*{g}}}}{}_{B_k{\textit{\textbf{E}}}'_k}\right|} \nonumber \\
    & \scalemath{0.99}{= \left|\sqrt{1- \tilde{\epsilon}^{{(k)}}_{j_k,{\vb*{g}}|j_{k-1},\hdots,j_{k-l_c}}} \braket*{\tilde{\phi}_{j_k,{\vb*{g}|j_{k-1},\hdots,j_{k+1-l_c}}}}{\phi_{j_k,{\vb*{g}}}}{}_{B_k{\textit{\textbf{E}}}'_k} + \sqrt{{\tilde{\epsilon}^{{(k)}}_{j_k,{\vb*{g}}|j_{k-1},\hdots,j_{k-l_c}}}}\braket*{\tilde{\phi}^{\perp}_{j_k,{\vb*{g}}|j_{k-1},\hdots,j_{k-l_c}}}{\phi_{j_k,{\vb*{g}}}}{}_{B_k{\textit{\textbf{E}}}'_k}\right|}\nonumber \\
    &\scalemath{0.99}{= \left|\sqrt{1- {\tilde{\epsilon}^{{(k)}}_{j_k,{\vb*{g}}|j_{k-1},\hdots,j_{k-l_c}}}} \braket*{\phi_{j_k,{\vb*{g}}}}{\phi_{j_k,{\vb*{g}}}}{}_{B_kE_k} \braket{\Lambda_{j_k,{\hdots,j_{k+1-l_c},\vb*{g}}}}{\Lambda_{{\vb*{g}}}}{}_{{\textit{\textbf{A}}}_{\geq k+1},{\textit{\textbf{B}}}_{\geq k+1},{\textit{\textbf{E}}}_{\geq k+1}}\right|} \nonumber \\
    & \scalemath{0.99}{= \sqrt{1- {\tilde{\epsilon}^{{(k)}}_{j_k,{\vb*{g}}|j_{k-1},\hdots,j_{k-l_c}}}} \frac{\prod_{n=1}^{l_c} \sum_{j_{k+n}}p_{j_{k+n}} \sqrt{1- {\epsilon^{{(k)}}_{j_{k+n},{\vb*{g}}|j_{k+n-1},\hdots,j_{k+n-l_c}}}}}{\prod_{n=1}^{l_c} \sqrt{\sum_{j_{k+n}}p_{j_{k+n}}(1- {\epsilon^{{(k)}}_{j_{k+n},{\vb*{g}}|j_{k+n-1},\hdots,j_{k+n-l_c}}})}}}  \nonumber \\
    &\scalemath{0.99}{{= \sqrt{1- {\epsilon^{{(k)}}_{j_k,{\vb*{g}}|j_{k-1},\hdots,j_{k-l_c}}}} {\prod_{n=1}^{l_c} \sqrt{\sum_{j_{k+n}}p_{j_{k+n}}(1- {\epsilon^{{(k)}}_{j_{k+n},{\vb*{g}}|j_{k+n-1},\hdots,j_{k+n-l_c}}})}} \frac{\prod_{n=1}^{l_c} \sum_{j_{k+n}}p_{j_{k+n}} \sqrt{1- {\epsilon^{{(k)}}_{j_{k+n},{\vb*{g}}|j_{k+n-1},\hdots,j_{k+n-l_c}}}}}{\prod_{n=1}^{l_c} \sqrt{\sum_{j_{k+n}}p_{j_{k+n}}(1- {\epsilon^{{(k)}}_{j_{k+n},{\vb*{g}}|j_{k+n-1},\hdots,j_{k+n-l_c}}})}}}} \nonumber \\
    &\scalemath{0.99}{= \sqrt{1- {\epsilon^{{(k)}}_{j_k,{\vb*{g}}|j_{k-1},\hdots,j_{k-l_c}}}} \prod_{n=1}^{l_c} \sum_{j_{k+n}} p_{j_{k+n}} \sqrt{1- {\epsilon^{{(k)}}_{j_{k+n},{\vb*{g}}|j_{k+n-1},\hdots,j_{k+n-l_c}}}} \geq \prod_{n=1}^{l_c+1} \sqrt{1-{\epsilon'^{\rm U}}} {=\left(\sqrt{1-{\epsilon'^{\rm U}}}\right)^{l_c+1} } =: \sqrt{1-\epsilon^{\rm U}},}
    \label{eq:bound1}
\end{align}\end{widetext}}
\noindent {where $\epsilon^{\rm U} := 1 - {\left(1-{\epsilon'^{\rm U}}\right)^{l_c+1}}$. In the second equality of \cref{eq:bound1}, we have used the fact that $\braket*{\tilde{\phi}^{\perp}_{j_k,{\vb*{g}}|j_{k-1},\hdots,j_{k-l_c}}}{\phi_{j_k,{\vb*{g}}}}{}_{B_k,{\textit{\textbf{E}}}'_k} = 0$ for any $j_k$, {in the third equality we have used the definition in \cref{eq:state_new_def}, and in} the forth equality, we have used the definition in \cref{eq:epsilon_def}. {Finally,} in the inequality of \cref{eq:bound1}, we have used that $\sum_{j_k} p_{j_k} = 1$ for any round $k$ and that $\epsilon^{{(k)}}_{j_k,{\vb*{g}}|j_{k-1},\hdots,j_{k-l_c}} \leq {\epsilon'^{\rm U}}$ for all ${k},j_k,{\vb*{g}},$ and $j_{k-1},\hdots,j_{k-l_c}$ {(see assumption (A2.b) in \cref{sec:assumptions_alice}).}} 

{Similarly, using Eqs.~(\ref{eq:entangled_dep_app}) and  (\ref{eq:entangled_dep_ref}), we can calculate an upper bound on the inner product $\big|\braket*{\Psi^{Z}_{{\vb*{g}}|j_{k-1},\hdots,j_{k-l_c}}}{\Phi^{Z}_{{\vb*{g}}}}{}_{A_k,B_k,{\textit{\textbf{E}}}'_k}\big|$ such that}
{\begin{align}
    &\left|\braket*{\Psi^{Z}_{{\vb*{g}}|j_{k-1},\hdots,j_{k-l_c}}}{\Phi^{Z}_{{\vb*{g}}}}{}_{A_k,B_k,{\textit{\textbf{E}}}'_k}\right| \nonumber \\
    &= \frac{1}{2} \Big|\braket*{{\tilde{\psi}}_{0_Z,{\vb*{g}}|j_{k-1},\hdots,j_{k-1}}}{\phi_{0_Z,{\vb*{g}}}}{}_{B_k,{\textit{\textbf{E}}}'_k} \nonumber \\
    &+ \braket*{{\tilde{\psi}}_{1_Z,{\vb*{g}}|j_{k-1},\hdots,j_{k-1}}}{\phi_{1_Z,{\vb*{g}}}}{}_{B_k,{\textit{\textbf{E}}}'_k} \Big| \nonumber \\
    &\leq \frac{1}{2}(\sqrt{1-\epsilon} + \sqrt{1-\epsilon}) = \sqrt{1-\epsilon},
    \label{eq:bound2}
\end{align}where we have used the fact that $\braket{j_k}{j'_k}_{A_k} = \delta_{j_k,j'_k}$. Also, in the inequality of \cref{eq:bound2} we have used the upper bound in \cref{eq:bound1}. This concludes the calculation of the upper bounds on the inner products $\big|\braket*{\tilde{\psi}_{j_k,{\vb*{g}}|j_{k-1},\hdots,j_{k-l_c}}}{\phi_{j_k,{\vb*{g}}}}{}_{B_k,{\textit{\textbf{E}}}'_k}\big|$ and $\big|\braket*{\Psi_{{\vb*{g}}|j_{k-1},\hdots,j_{k-l_c}}}{\Phi_{{\vb*{g}}}}{}_{A_k,B_k,{\textit{\textbf{E}}}'_k}\big|$.}

	\section{Proof of \cref{eq:avg_eph}}
	\label{app:proof_avg_eph}
    
    Here, we provide a proof of the second inequality in \cref{eq:avg_eph}. From {\cref{eq:ephU_final}}, we can express $e_{\rm ph}^{\rm U}$ as
    \begin{equation}
        e_{\rm ph}^{\rm U} = \frac{N_{\rm ph}^{\rm U}}{N_{\rm det}^{(Z)}} = \frac{N}{N_{\rm det}^{(Z)}} f\left(\frac{\vectorbold{x}}{N}\right),
        \label{eq:ephU_f}
    \end{equation}
	where $\vectorbold{x}$ is a tuple whose elements are ${N_{j,\gamma_X}}$ for all $j, \gamma$, and $f$ is a multivariate function that is concave with respect to all of its arguments due to the concavity of $G_{+}(y,z)$ and $-G_{-}(y,z)$ with respect to $y$. Similarly, from {\cref{eq:ephUw_final}}, we can express $e_{\textrm{ph},w}^{\rm U}$ as
	\begin{equation}
	   e_{\textrm{ph},w}^{\rm U} = \frac{N_{\textrm{ph},w}^{\rm U}}{N_{\textrm{det},w}^{(Z)}} = \frac{N_w}{N_{\textrm{det},w}^{(Z)}} f\left(\frac{\vectorbold{x}_w}{N_w}\right),
	   \label{eq:ephUw_f}
	\end{equation}
	where $\vectorbold{x}_w$ is a tuple whose components are ${N_{j,\gamma_X,w}}$ for all $j, \gamma$, and $f$ is the same function as in \cref{eq:ephU_f}. We have that
	\begin{align}
	    \sum_{w = 0}^{l_c} q_w  e_{{\rm ph},w}^{{\rm U}} &= \sum_{w = 0}^{l_c} \frac{N_{\textrm{det},w}^{(Z)}}{N_{\rm det}^{(Z)}}  \frac{N_w}{N_{\textrm{det},w}^{(Z)}} f\left(\frac{\vectorbold{x}_w}{N_w}\right) \nonumber \\ 
	    &= \sum_{w = 0}^{l_c}\frac{N_w}{N_{\textrm{det}}^{(Z)}} f\left(\frac{\vectorbold{x}_w}{N_w}\right) \nonumber \\
	    &= \frac{N}{N_{\textrm{det}}^{(Z)}} \sum_{w = 0}^{l_c} \frac{N_w}{N} f\left(\frac{\vectorbold{x}_w}{N_w}\right) \nonumber \\
	    &\leq  \frac{N}{N_{\textrm{det}}^{(Z)}} f \left(\sum_{w = 0}^{l_c}  \frac{N_w}{N} \frac{\vectorbold{x}_w}{N_w}\right) \nonumber \\
	    &= \frac{N}{N_{\textrm{det}}^{(Z)}} f \left(\frac{\vectorbold{x}}{N}\right) = e_{\rm ph}^{\rm U},
	\end{align}
	where the inequality is due to the concavity of $f$, and the {second to} last equality is due to $\sum_w \vectorbold{x}_w = \vectorbold{x}$, since $\sum_w {N_{j,\gamma_X,w}} = {N_{j,\gamma_X}}$.
	
	
	\section{Parameters $c_{\alpha,j,{\vb*{g}}}^{{(k)}}$,  $c_{\alpha,j}^{\rm U}$ and $p_{\alpha_X}^{{\rm (vir)U}}$ for the {modified} BB84 protocol}
	\label{app:coefficients_bb84}
	
	In this appendix, we provide the full expressions of the coefficients $c_{\alpha,j,{\vb*{g}}}^{{(k)}}$, with $\alpha \in \{0,1\}$ and $j \in \{0_Z,1_Z,$ $0_X,1_X\}$, associated with the {modified} BB84 protocol, for the particular phase-encoding scheme considered in \cref{sec:device_model}. Direct calculations show that these coefficients are given by
		\begin{widetext}\begin{align}
			c_{1,0_Z,{\vb*{g}}}^{{(k)}} &= \frac{\sin(\frac{\theta_{0_Z,{\vb*{g}}}^{{(k)}}}{2} - \frac{\theta_{1_X,{\vb*{g}}}^{{(k)}}}{2}) - \sin(\frac{\theta_{1_Z,{\vb*{g}}}^{{(k)}}}{2} - \frac{\theta_{1_X,{\vb*{g}}}^{{(k)}}}{2})}{\sin(\frac{\theta_{1_Z,{\vb*{g}}}^{{(k)}}}{2} - \theta_{0_Z,{\vb*{g}}}^{{(k)}} + \frac{\theta_{1_X,{\vb*{g}}}^{{(k)}}}{2}) + 2\sin(\frac{\theta_{0_Z,{\vb*{g}}}^{{(k)}}}{2} - \frac{\theta_{1_X,{\vb*{g}}}^{{(k)}}}{2}) - \sin(\frac{\theta_{1_Z,{\vb*{g}}}^{{(k)}}}{2} - \frac{\theta_{1_X,{\vb*{g}}}^{{(k)}}}{2})}, \nonumber \\
			c_{1,1_Z,{\vb*{g}}}^{{(k)}} &= \frac{-\sin(\frac{\theta_{0_Z,{\vb*{g}}}^{{(k)}}}{2} - \frac{\theta_{1_X,{\vb*{g}}}^{{(k)}}}{2}) + \sin(\frac{\theta_{1_Z,{\vb*{g}}}^{{(k)}}}{2} - \frac{\theta_{1_X,{\vb*{g}}}^{{(k)}}}{2})}{\sin(\frac{\theta_{0_Z,{\vb*{g}}}^{{(k)}}}{2} - \theta_{1_Z,{\vb*{g}}}^{{(k)}} + \frac{\theta_{1_X,{\vb*{g}}}^{{(k)}}}{2}) - \sin(\frac{\theta_{0_Z,{\vb*{g}}}^{{(k)}}}{2} - \frac{\theta_{1_X,{\vb*{g}}}^{{(k)}}}{2}) + 2\sin(\frac{\theta_{1_Z,{\vb*{g}}}^{{(k)}}}{2} - \frac{\theta_{1_X,{\vb*{g}}}^{{(k)}}}{2})}, \nonumber \\
			c_{1,1_X,{\vb*{g}}}^{{(k)}} &=  \frac{\cos(\theta_{0_Z,{\vb*{g}}}^{{(k)}} - \theta_{1_Z,{\vb*{g}}}^{{(k)}}) - 1}{\left[\splitfrac{\cos(\theta_{0_Z,{\vb*{g}}}^{{(k)}} - \theta_{1_Z,{\vb*{g}}}^{{(k)}}) - \cos(\theta_{0_Z,{\vb*{g}}}^{{(k)}} - \theta_{1_X,{\vb*{g}}}^{{(k)}}) - \cos(\theta_{1_Z,{\vb*{g}}}^{{(k)}} - \theta_{1_X,{\vb*{g}}}^{{(k)}})}{ + 2\cos(\frac{\theta_{0_Z,{\vb*{g}}}^{{(k)}}}{2} + \frac{\theta_{1_Z,{\vb*{g}}}^{{(k)}}}{2} - \theta_{1_X,{\vb*{g}}}^{{(k)}}) - 2\cos(\frac{\theta_{0_Z,{\vb*{g}}}^{{(k)}}}{2} - \frac{\theta_{1_Z,{\vb*{g}}}^{{(k)}}}{2}) + 1}\right]}, \nonumber \\
			c_{0,0_Z,{\vb*{g}}}^{{(k)}} &= \frac{\sin(\frac{\theta_{0_Z,{\vb*{g}}}^{{(k)}}}{2} - \frac{\theta_{0_X,{\vb*{g}}}^{{(k)}}}{2}) + \sin(\frac{\theta_{1_Z,{\vb*{g}}}^{{(k)}}}{2} - \frac{\theta_{0_X,{\vb*{g}}}^{{(k)}}}{2})}{2\sin(\frac{\theta_{0_Z,{\vb*{g}}}^{{(k)}}}{2} - \frac{\theta_{0_X,{\vb*{g}}}^{{(k)}}}{2}) - \sin(\frac{\theta_{1_Z,{\vb*{g}}}^{{(k)}}}{2} - \theta_{0_Z,{\vb*{g}}}^{{(k)}} + \frac{\theta_{0_X,{\vb*{g}}}^{{(k)}}}{2}) + \sin(\frac{\theta_{1_Z,{\vb*{g}}}^{{(k)}}}{2} - \frac{\theta_{0_X,{\vb*{g}}}^{{(k)}}}{2})}, \nonumber \\
			c_{0,1_Z,{\vb*{g}}}^{{(k)}} &= \frac{\sin(\frac{\theta_{0_Z,{\vb*{g}}}^{{(k)}}}{2} - \frac{\theta_{0_X,{\vb*{g}}}^{{(k)}}}{2}) + \sin(\frac{\theta_{1_Z,{\vb*{g}}}^{{(k)}}}{2} - \frac{\theta_{0_X,{\vb*{g}}}^{{(k)}}}{2})}{\sin(\frac{\theta_{0_Z,{\vb*{g}}}^{{(k)}}}{2} - \frac{\theta_{0_X,{\vb*{g}}}^{{(k)}}}{2}) - \sin(\frac{\theta_{0_Z,{\vb*{g}}}^{{(k)}}}{2} - \theta_{1_Z,{\vb*{g}}}^{{(k)}} + \frac{\theta_{0_X,{\vb*{g}}}^{{(k)}}}{2}) + 2\sin(\frac{\theta_{1_Z,{\vb*{g}}}^{{(k)}}}{2} - \frac{\theta_{0_X,{\vb*{g}}}^{{(k)}}}{2})}, \nonumber \\
			c_{0,0_X,{\vb*{g}}}^{{(k)}} &=  \frac{\cos(\theta_{0_Z,{\vb*{g}}}^{{(k)}} - \theta_{1_Z,{\vb*{g}}}^{{(k)}}) - 1}{\left[ \splitfrac{\cos(\theta_{0_Z,{\vb*{g}}}^{{(k)}} - \theta_{1_Z,{\vb*{g}}}^{{(k)}}) - \cos(\theta_{0_Z,{\vb*{g}}}^{{(k)}} - \theta_{0_X,{\vb*{g}}}^{{(k)}}) - \cos(\theta_{1_Z,{\vb*{g}}}^{{(k)}} - \theta_{0_X,{\vb*{g}}}^{{(k)}})}{ - 2\cos(\frac{\theta_{0_Z,{\vb*{g}}}^{{(k)}}}{2} + \frac{\theta_{1_Z,{\vb*{g}}}^{{(k)}}}{2} - \theta_{0_X,{\vb*{g}}}^{{(k)}}) + 2\cos(\frac{\theta_{0_Z,{\vb*{g}}}^{{(k)}}}{2} - \frac{\theta_{1_Z,{\vb*{g}}}^{{(k)}}}{2}) + 1}\right]}.
			\label{eq:coefficients_4state}
		\end{align}\end{widetext}

These coefficients are constrained by the fact that $\theta_{j,{\vb*{g}}}^{{(k)}} \in [\theta^{\rm L}_{j},\theta^{\rm U}_{j}]$ for some known $\theta^{\rm L}_{j}$ {and} $\theta^{\rm U}_{j}$. While one can in principle find numerical upper bounds on these coefficients regardless of the value of $\theta^{\rm L}_{j}$ {and} $\theta^{\rm U}_{j}$, here we obtain analytical bounds under the following assumption: $-\pi/6 \leq \theta_{0_Z}^{\rm L} \leq \theta_{0_Z}^{\rm U} \leq \pi/6$, $5\pi/6 \leq \theta_{1_Z}^{\rm L} \leq \theta_{1_Z}^{\rm U} \leq 7\pi/6$, $\pi/3 \leq \theta_{0_X}^{\rm L} \leq \theta_{0_X}^{\rm U} \leq 2\pi/3$ and $4\pi/3 \leq \theta_{1_X}^{\rm L} \leq \theta_{1_X}^{\rm U} \leq 5\pi/3$. Note that this assumption is reasonable because a deviation of $\pm \pi/6$ from the ideal phase value is much larger than the modulation errors characterised in recent experiments \cite{xu}. 

To derive the analytical upper bounds, we consider the partial differential equations of $c_{\alpha,j,{\vb*{g}}}^{{(k)}}$ with respect to each $\theta_{j,{\vb*{g}}}^{{(k)}}$ and then select the values of $\theta_{j,{\vb*{g}}}^{{(k)}} \in [\theta_{j}^{\rm L},\theta_{j}^{\rm U}]$ that maximise them. For example, since $\partial_{\theta_{0_Z,{\vb*{g}}}} {c_{1,0_Z,{\vb*{g}}}^{{(k)}}} > 0$ when $\theta_{0_Z,{\vb*{g}}}^{{(k)}} \in [-\pi/6, \pi/6]$, $\theta_{1_Z,{\vb*{g}}}^{{(k)}} \in [5 \pi/6, 7\pi/6]$ and $\theta_{0_X,{\vb*{g}}}^{{(k)}} \in [\pi/3, 2\pi/3]$, $c_{1,0_Z,{\vb*{g}}}^{{(k)}}$ is maximised when $\theta_{0_Z,{\vb*{g}}}^{{(k)}} = \theta_{0_Z}^{\rm U}$. Alternatively, if the function is not always increasing or decreasing with respect to a particular argument, but given the ranges stated above it is convex with respect to that argument, e.g.\ $\partial^2_{\theta_{0_Z,{\vb*{g}}}} {c_{1,1_X,{\vb*{g}}}^{{(k)}}} > 0$, then we conclude that $c_{1,1_X,{\vb*{g}}}$ is maximised either when $\theta_{0_Z,{\vb*{g}}}^{{(k)}} = \theta_{0_Z}^{\rm L}$ or when $\theta_{0_Z,{\vb*{g}}}^{{(k)}} = \theta_{0_Z}^{\rm U}$, and we select the appropriate value by inspection. The overall solution is the following:
\begin{align}
		&c_{1,0_Z}^{\rm U} \rightarrow c_{1,0_Z,{\vb*{g}}}^{{(k)}} (\theta_{0_Z}^{\rm U},\theta_{1_Z}^{\rm U},\theta_{1_X}^{\rm L}), \nonumber \\
		&c_{1,1_Z}^{\rm U} \rightarrow c_{1,1_Z,{\vb*{g}}}^{{(k)}} (\theta_{0_Z}^{\rm L},\theta_{1_Z}^{\rm L},\theta_{1_X}^{\rm U}), \nonumber \\
		&c_{1,1_X}^{\rm U} \rightarrow \max_{x_1,y_1,z_1 \in \{{\rm L,U}\}}c_{1,1_X,{\vb*{g}}}^{{(k)}} (\theta_{0_Z}^{x_1},\theta_{1_Z}^{y_1},\theta_{1_X}^{z_1}), \nonumber \\
		&c_{0,0_Z}^{\rm U} \rightarrow c_{0,0_Z,{\vb*{g}}}^{{(k)}} (\theta_{0_Z}^{\rm L},\theta_{1_Z}^{\rm L},\theta_{1_X}^{\rm U}), \nonumber \\
		&c_{0,1_Z}^{\rm U} \rightarrow c_{0,1_Z,{\vb*{g}}}^{{(k)}} (\theta_{0_Z}^{\rm U},\theta_{1_Z}^{\rm U},\theta_{1_X}^{\rm L}), \nonumber \\
		&c_{0,0_X}^{\rm U} \rightarrow \max_{x_0,y_0,z_0 \in \{{\rm L,U}\}}c_{0,0_X,{\vb*{g}}}^{{(k)}} (\theta_{0_Z}^{x_0},\theta_{1_Z}^{y_0},\theta_{1_X}^{z_0}).
		\label{eq:coefficients}
	\end{align}
	
	Similarly, we can obtain upper bounds on the probabilities $p_{\alpha_X,{\vb*{g}}}^{{(k,{\rm vir})}}$. For this, first recall that
	\begin{align}
	    p_{\alpha_X,{\vb*{g}}}^{{(k,{\rm vir})}} = \frac{1}{2} p_{Z_A} \big[1+ (-1)^{\alpha} \Re \big(\big\langle{\phi_{0_Z,{\vb*{g}}}}\big|\phi_{1_Z,{\vb*{g}}}\big\rangle {}_{B_k,E_{k}} \big)\big].
	\end{align}Then, by using the definition of $\ket*{\phi_{j,{\vb*{g}}}}_{B_k,E_{k}}$ and Eq.~(\ref{eq:phase_enc}), and by considering the partial differential equations of $p_{\alpha_X,{\vb*{g}}}^{{(k,{\rm vir})}}$ with respect to each phase $\theta_{j,{\vb*{g}}}^{{(k)}}$ we have that 
	\begin{align}
		&p_{1_X}^{\rm (vir)U} \rightarrow \frac{1}{2} p_{Z_A} \left[ 1 - \cos (\frac{\theta_{0_Z}^{\rm L} - \theta_{1_Z}^{\rm U}}{2})\right], \nonumber \\
		&p_{0_X}^{\rm (vir)U} \rightarrow \frac{1}{2} p_{Z_A} \left[ 1 + \cos (\frac{\theta_{0_Z}^{\rm U} - \theta_{1_Z}^{\rm L}}{2})\right].
		\label{eq:probabilities}
	\end{align} 
		

	\section{Parameters $c_{\alpha,j,{\vb*{g}}}^{{(k)}}$, $c_{\alpha,j}^{\rm U}$ and $p_{\alpha_X}^{{\rm (vir)U}}$ for the three-state protocol}
	\label{app:coefficients_3state}
	Here, we provide the full expressions of the coefficients $c_{1,j,{\vb*{g}}}^{{(k)}}$ with $j \in \{0_Z,1_Z,0_X\}$ associated with the three-state protocol for the particular phase-encoding scheme considered in \cref{sec:device_model}, as well as their upper bounds. Direct calculations show that $c_{1,j,{\vb*{g}}}^{{(k)}}$ are given by
		\begin{widetext}\begin{align}
			c_{1,0_Z,{\vb*{g}}}^{{(k)}} &= \frac{\sin(\frac{\theta_{0_Z,{\vb*{g}}}^{{(k)}}}{2} - \frac{\theta_{0_X,{\vb*{g}}}^{{(k)}}}{2}) - \sin(\frac{\theta_{1_Z,{\vb*{g}}}^{{(k)}}}{2} - \frac{\theta_{0_X,{\vb*{g}}}^{{(k)}}}{2})}{\sin(\frac{\theta_{1_Z,{\vb*{g}}}^{{(k)}}}{2} - \theta_{0_Z,{\vb*{g}}}^{{(k)}} + \frac{\theta_{0_X,{\vb*{g}}}^{{(k)}}}{2}) + 2\sin(\frac{\theta_{0_Z,{\vb*{g}}}^{{(k)}}}{2} - \frac{\theta_{0_X,{\vb*{g}}}^{{(k)}}}{2}) - \sin(\frac{\theta_{1_Z,{\vb*{g}}}^{{(k)}}}{2} - \frac{\theta_{0_X,{\vb*{g}}}^{{(k)}}}{2})}, \nonumber \\
			c_{1,1_Z,{\vb*{g}}}^{{(k)}} &= \frac{- \sin(\frac{\theta_{0_Z,{\vb*{g}}}^{{(k)}}}{2} - \frac{\theta_{0_X,{\vb*{g}}}^{{(k)}}}{2}) + \sin(\frac{\theta_{1_Z,{\vb*{g}}}^{{(k)}}}{2} - \frac{\theta_{0_X,{\vb*{g}}}^{{(k)}}}{2})}{\sin(\frac{\theta_{0_Z,{\vb*{g}}}^{{(k)}}}{2} - \theta_{1_Z,{\vb*{g}}}^{{(k)}} + \frac{\theta_{0_X,{\vb*{g}}}^{{(k)}}}{2}) - \sin(\frac{\theta_{0_Z,{\vb*{g}}}^{{(k)}}}{2} - \frac{\theta_{0_X,{\vb*{g}}}^{{(k)}}}{2}) + 2\sin(\frac{\theta_{1_Z,{\vb*{g}}}^{{(k)}}}{2} - \frac{\theta_{0_X,{\vb*{g}}}^{{(k)}}}{2})}, \nonumber \\
			c_{1,0_X,{\vb*{g}}}^{{(k)}} &=  \frac{\cos(\theta_{0_Z,{\vb*{g}}}^{{(k)}} - \theta_{1_Z,{\vb*{g}}}^{{(k)}}) - 1}{\left[\splitfrac{\cos(\theta_{0_Z,{\vb*{g}}}^{{(k)}} - \theta_{1_Z,{\vb*{g}}}^{{(k)}}) - \cos(\theta_{0_Z,{\vb*{g}}}^{{(k)}} - \theta_{0_X,{\vb*{g}}}^{{(k)}}) - \cos(\theta_{1_Z,{\vb*{g}}}^{{(k)}} - \theta_{0_X,{\vb*{g}}}^{{(k)}})}{ + 2\cos(\frac{\theta_{0_Z,{\vb*{g}}}^{{(k)}}}{2} + \frac{\theta_{1_Z,{\vb*{g}}}^{{(k)}}}{2} - \theta_{0_X,{\vb*{g}}}^{{(k)}}) - 2\cos(\frac{\theta_{0_Z,{\vb*{g}}}^{{(k)}}}{2} - \frac{\theta_{1_Z,{\vb*{g}}}^{{(k)}}}{2}) + 1}\right]}.
		\end{align}\end{widetext}
	Their upper bounds can be derived by following a similar approach as in Appendix~\ref{app:coefficients_bb84} and are summarised below \cite{mizutani}:
		\begin{widetext}\begin{align}
			&c_{1,0_Z}^{\rm U} \rightarrow c_{1,0_Z,{\vb*{g}}}^{{(k)}} (\theta_{0_Z}^{\rm U},\theta_{1_Z}^{\rm L},\theta_{0_X}^{\rm L}), \nonumber \\
			&c_{1,1_Z}^{\rm U} \rightarrow c_{1,1_Z,{\vb*{g}}}^{{(k)}} (\theta_{0_Z}^{\rm U},\theta_{1_Z}^{\rm L},\theta_{0_X}^{\rm U}), \nonumber \\
			&c_{1,0_X}^{\rm U} \rightarrow 
			\begin{cases}
				c_{1,0_X,{\vb*{g}}}^{{(k)}} (\theta_{0_Z}^{\rm L},\theta_{1_Z}^{\rm U},\theta_{0_X}^{\rm U}) & \text{if} ~~\theta_{0_X}^{\rm U} < \frac{\theta_{0_Z}^{\rm L} + \theta_{1_Z}^{\rm U}}{2},  \\
				c_{1,0_X,{\vb*{g}}}^{{(k)}} \left(\theta_{0_Z}^{\rm L},\theta_{1_Z}^{\rm U},\frac{\theta_{0_Z}^{\rm L} + \theta_{1_Z}^{\rm U}}{2}\right) & \text{if}~~\frac{\theta_{0_Z}^{\rm L} + \theta_{1_Z}^{\rm U}}{2} \in \left[\theta_{0_X}^{\rm L}, \theta_{0_X}^{\rm U}\right], \\
			 	c_{1,0_X,{\vb*{g}}}^{{(k)}} (\theta_{0_Z}^{\rm L},\theta_{1_Z}^{\rm U},\theta_{0_X}^{\rm L}) & \text{if}~~\frac{\theta_{0_Z}^{\rm L} + \theta_{1_Z}^{\rm U}}{2} < \theta_{0_X}^L. 
			\end{cases}
		\end{align}\end{widetext}
	The full expressions for the coefficients $c_{0,j,{\vb*{g}}}^{{(k)}}$ with $j \in \{0_Z,1_Z,0_X\}$ as well as their upper bounds, and the probabilities $p_{\alpha_X}^{\rm (vir)U}$ are the same as in \cref{app:coefficients_bb84}.

	
	
%

\end{document}